\documentclass[prd,eqsecnum,twocolumn,amsfonts,amssymb]{revtex4}

\usepackage{graphicx}

\usepackage{bm}

\newcommand{\beq}{\begin{equation}}
\newcommand{\eeq}{\end{equation}}
\newcommand{\beqs}{\begin{eqnarray}}
\newcommand{\eeqs}{\end{eqnarray}}
\newcommand{\lsim}{\mathrel{\raisebox{-
.6ex}{$\stackrel{\textstyle<}{\sim}$}}}
\newcommand{\gsim}{\mathrel{\raisebox{-
.6ex}{$\stackrel{\textstyle>}{\sim}$}}}

\begin{document}

\title{Higher-Order Scheme-Independent Series Expansions of 
$\gamma_{\bar\psi\psi,IR}$ and $\beta'_{IR}$ in Conformal Field Theories}

\author{Thomas A. Ryttov$^a$ and Robert Shrock$^b$}

\affiliation{(a) \ CP$^3$-Origins and Danish Institute for Advanced Study \\
Southern Denmark University, Campusvej 55, Odense, Denmark}

\affiliation{(b) \ C. N. Yang Institute for Theoretical Physics and 
Department of Physics and Astronomy, \\
Stony Brook University, Stony Brook, NY 11794, USA }

\begin{abstract}

  We study a vectorial asymptotically free gauge theory, with gauge group $G$
  and $N_f$ massless fermions in a representation $R$ of this group, that
  exhibits an infrared (IR) zero in its beta function, $\beta$, at the coupling
  $\alpha=\alpha_{IR}$ in the non-Abelian Coulomb phase.  For general $G$ and
  $R$, we calculate the scheme-independent series expansions of (i) the
  anomalous dimension of the fermion bilinear, $\gamma_{\bar\psi\psi,IR}$, to
  $O(\Delta_f^4)$ and (ii) the derivative $\beta' = d\beta/d\alpha$, to
  $O(\Delta_f^5)$, both evaluated at $\alpha_{IR}$, where $\Delta_f$ is
  an $N_f$-dependent expansion variable.  These are the highest orders to which
  these expansions have been calculated. We apply these general results to
  theories with $G={\rm SU}(N_c)$ and $R$ equal to the fundamental, adjoint,
  and symmetric and antisymmetric rank-2 tensor representations. It is shown
  that for all of these representations, $\gamma_{\bar\psi\psi,IR}$, calculated
  to the order $\Delta_f^p$, with $1 \le p \le 4$, increases monotonically with
  decreasing $N_f$ and, for fixed $N_f$, is a monotonically increasing function
  of $p$. Comparisons of our scheme-independent calculations of
  $\gamma_{\bar\psi\psi,IR}$ and $\beta'_{IR}$ are made with our earlier higher
  $n$-loop values of these quantities, and with lattice measurements.  For
  $R=F$, we present results for the limit $N_c \to \infty$ and $N_f \to \infty$
  with $N_f/N_c$ fixed. We also present expansions for $\alpha_{IR}$ calculated
  to $O(\Delta_f^4)$.

\end{abstract}

\maketitle

% =======================================================================

% section 1
\section{Introduction}
\label{intro_section}

An important advance in the understanding of quantum field theory was the
realization that the properties of a theory depend on the Euclidean
energy/momentum scale $\mu$ at which they are measured.  This is of particular
interest in an asymptotically free non-Abelian gauge theory, in which the
running gauge coupling $g(\mu)$ and the associated quantity $\alpha(\mu) =
g(\mu)^2/(4\pi)$ approach zero at large $\mu$ in the deep ultraviolet (UV).  We
shall consider a theory of this type, with gauge group $G$ and $N_f$ massless
fermions $\psi_j$, $j=1,..., N_f$, in a representation $R$ of $G$.  The
dependence of $\alpha(\mu)$ on $\mu$ is described by the renormalization-group
(RG) \cite{rg} beta function, $\beta = d\alpha(\mu)/dt$, where $dt=d\ln\mu$.
The condition that the theory be asymptotically free implies that $N_f$ be less
than a certain value, $N_u$, given below in Eq. (\ref{nfb1z}). 
Since $\alpha(\mu)$ is small at large $\mu$, one can self-consistently
calculate $\beta$ as a power series in $\alpha(\mu)$.  As $\mu$ decreases from
large values in the UV to small values in the infrared (IR), $\alpha(\mu)$
increases.  A situation of special interest occurs if the beta function has a
zero at some value away from the origin. For a given $G$ and $R$, this can
happen for sufficiently large $N_f$, while still in the asymptotically free
regime.  In this case, as $\mu$ decreases from large values in the UV toward
$\mu=0$ in the IR, the coupling increases, but approaches the value of $\alpha$
at this zero in the beta function, which is thus denoted $\alpha_{IR}$.  Since
$\beta=0$ at $\alpha=\alpha_{IR}$, the resultant theory in this IR limit is
scale-invariant, and generically also conformally invariant \cite{scalecon,fm}.
A fundamental question concerns the properties of the interacting theory at
such an IR fixed point (IRFP) of the renormalization group. There is convincing
evidence that if $\alpha_{IR}$ is small enough, then the IR theory is in a
(deconfined) non-Abelian Coulomb phase (NACP), also called the conformal window
\cite{nacp_comment}. In terms of $N_f$, this phase occurs if $N_f$ is in the
interval $N_{f,cr} < N_f < N_u$, where $N_u$ and $N_{f,cr}$ depend on $G$ and
$R$.  Here, $N_{f,cr}$ denotes the value of $N_f$ below which the running 
$\alpha(\mu)$ becomes large enough to cause spontaneous chiral symmetry
breaking and dynamical fermion mass generation. 

Physical quantities in the IR-limit theory at $\alpha_{IR}$ cannot depend on
the scheme used for the regularization and subtraction procedure in
renormalization.  In conventional computations of these quantities, first, one
expresses them as series expansions in powers of $\alpha$, calculated to
$n$-loop order; second, one computes the IR zero of the beta function at the
$n$-loop ($n\ell$) level, denoted $\alpha_{IR,n\ell}$; and third, one sets
$\alpha=\alpha_{IR,n\ell}$ in the series expansion for the given quantity to
obtain its value at the IR zero of the beta function to this $n$-loop order.
However, these conventional series expansions in powers of $\alpha$, calculated
to a finite order, are scheme-dependent beyond the leading one or two terms.
Specifically, the terms in the beta function are scheme-dependent at loop order
$\ell \ge 3$ and the terms in an anomalous dimension are scheme-dependent at
loop order $\ell \ge 2$ \cite{gross75}. Indeed, as is well-known, the presence
of scheme-dependence in higher-order perturbative calculations is a general
property in quantum field theory.

It is therefore of great value to use a complementary approach in which one
expresses these physical quantities at $\alpha_{IR}$ as an expansion in powers
of a variable such that, at every order in this expansion, the result is
scheme-independent.  A very important property is that one can recast the
expressions for physical quantities in a manner that is scheme-independent. A
crucial point here is that, for a given gauge group $G$ and fermion
representation $R$, as $N_f$ (formally generalized from non-negative integers
to the real numbers) approaches the upper limit allowed by asymptotic freedom,
denoted $N_u$ (given by Eq.  (\ref{nfb1z}) below), the resultant value of
$\alpha_{IR}$ approaches zero. This means that one can equivalently express a
physical quantity in a scheme-independent manner as a series in powers of the
variable
\beq
\Delta_f = N_u - N_f = \frac{11C_A}{4T_f} - N_f \ , 
\label{deltaf}
\eeq
where $C_A$ is the quadratic Casimir invariant for the adjoint representation,
and $T_f$ is the trace invariant for the fermion representation $R$
\cite{casimir}. Here, $\alpha_{IR} \to 0 \ \Longleftrightarrow \ \Delta_f \to
0$. Hence, for $N_f$ less than, but close to $N_u$, this expansion variable
$\Delta_f$ is reasonably small, and one can envision reliable perturbative
calculations of physical quantities at this IR fixed point in powers of
$\Delta_f$. Following the original calculations of the one- and
two-loop coefficients of the beta function \cite{b1}-\cite{gw2}, some early
work on this was reported in \cite{bz,grunberg92}.

In this paper we consider a vectorial, asymptotically free gauge theory and
present scheme-independent calculations, for a general gauge group $G$ and
fermion representation $R$, of two physical quantities in the IR theory at
$\alpha_{IR}$ of considerable importance, namely (i) the anomalous dimension,
denoted $\gamma_{\bar\psi\psi,IR}$, of the (gauge-invariant) fermion bilinear
$\bar\psi\psi = \sum_{j=1}^{N_f} \bar\psi_j \psi_j$ to $O(\Delta_f^4)$ and
(ii) the derivative $\beta'_{IR} = d\beta/d\alpha$ to $O(\Delta_f^5)$, both
evaluated at $\alpha=\alpha_{IR}$.  These are the highest orders in powers of
$\Delta_f$ to which these quantities have been calculated.  We give explicit
expressions for these quantities in the special cases where $G={\rm SU}(N_c)$
and the fermion representation $R$ is the fundamental ($F$), adjoint ($adj$),
and symmetric and antisymmetric rank-2 tensors, ($S_2$, \ $A_2$).  Our results
extend our previous scheme-independent calculations of
$\gamma_{\bar\psi\psi,IR}$ to $O(\Delta_f^3)$ in \cite{gtr} and of the 
derivative $\beta'_{IR}$ to $O(\Delta_f^4)$ in \cite{dex} 
for general $G$ and $R$, and 
our scheme-independent calculation of $\gamma_{\bar\psi\psi,IR}$ to
$O(\Delta_f^4)$ for $G={\rm SU}(3)$ and $R=F$ in \cite{gsi} (see also
\cite{flir}).  A brief report on some of our results was given in \cite{dexs}.

Scheme-independent series expansions of $\gamma_{\bar\psi\psi,IR}$ and
$\beta'_{IR}$ can be written as 
\beq
\gamma_{\bar\psi\psi,IR} = \sum_{j=1}^\infty \kappa_j \, \Delta_f^j
\label{gamma_ir_Deltaseries}
\eeq
and
\beq
\beta'_{IR} = \sum_{j=1}^\infty d_j \, \Delta_f^j \ ,
\label{betaprime_ir_Deltaseries}
\eeq
where $d_1=0$ for all $G$ and $R$ \cite{gtr,gsi,dex}.  In general, the
calculation of the coefficient $\kappa_j$ in Eq. (\ref{gamma_ir_Deltaseries})
requires, as inputs, the values of the $b_\ell$ for $1 \le \ell \le j+1$ and
the $c_\ell$ for $1 \le \ell \le j$.  The calculation of the coefficient $d_j$
in Eq. (\ref{betaprime_ir_Deltaseries}) requires, as inputs the values of the
$b_\ell$ for $1 \le \ell \le j$.  We refer the reader to \cite{gtr} and
\cite{dex} for discussions of the procedure for calculating the coefficients
$\kappa_j$ and $d_j$.  We denote the truncation of these series to maximal
power $j=p$ as $\gamma_{\bar\psi\psi,IR,\Delta_f^p}$ and
$\beta'_{IR,\Delta_f^p}$, respectively. Where it is necessary for
clarity, we will also indicate the fermion representation $R$ in the subscript.

Our main new results here include the general expressions, for arbitrary gauge
group $G$ and fermion representation $R$, for the coefficient, $\kappa_4$ in
Eq. (\ref{kappa4}) below, and for the coefficient $d_5$, given in
Eq. (\ref{d5}) below, as well as reductions of these formulas for special cases
and, for $R=F$, calculations in the LNN limit (\ref{lnn}).  As will be
discussed further below, the derivative $\beta'_{IR}$ is equivalent to the
anomalous dimension of the non-Abelian field strength squared, ${\rm
  Tr}(F_{\mu\nu}F^{\mu\nu})$.  Our present calculations make use of the newly
computed five-loop coefficient in the beta function for this gauge theory for
general $G$ and $R$ in \cite{b5}, as our work in \cite{flir,gsi} made use of
the calculation of this five-loop coefficient for the case $G={\rm SU}(3)$ and
$R=F$ in \cite{b5su3}.

In addition to being of interest and value in their own right, our new
scheme-independent calculations, performed to the highest order yet achieved,
are useful in several ways.  First, we will compare our results for
$\gamma_{\bar\psi\psi,IR}$ and $\beta'_{IR}$ for various $G$ and $R$ with the
values that we obtained at comparable order with the conventional $n$-loop
approach in \cite{bvh}-\cite{lnn}.  Our new results have the merit of being
scheme-independent at each order in $\Delta_f$, in contrast to scheme-dependent
series expansions of $\gamma_{\bar\psi\psi,IR}$ and $\beta'_{IR}$ in powers of
the IR coupling. Second, there is, at present, an intensive program to study
this IR behavior on the lattice \cite{lgtreviews}. Thus, it is of considerable
interest to compare our scheme-independent results for
$\gamma_{\bar\psi\psi,IR}$ for various theories with values measured in lattice
simulations of these theories.  We have done this in \cite{gsi,dex,dexs} (as
well as in our work on conventional $n$-loop calculations \cite{bvh,flir}), and
we will expand upon this comparison here.  Third, we believe that our
scheme-independent expansions for these physical quantities are of interest in
the context of the great current resurgence of research activity on conformal
field theories (CFT).  Much of this current activity makes use of
operator-product expansions and the associated bootstrap approach
\cite{cft_bootstrap}. Our method of scheme-independent series expansions for
physical quantities at an IR fixed point is complementary to this bootstrap
approach in yielding information about a conformal field theory.

Our calculations rely on $\alpha_{IR}$ being an exact zero of the beta function
and thus an exact IR fixed point of the renormalization group, and this
property holds in the non-Abelian Couloumb phase (conformal window). In this
phase, the chiral symmetry associated with the massless fermions is preserved
in the presence of the gauge interaction.  However, there has also been
interest in vectorial asymptotically free gauge theories that exhibit
quasi-conformal behavior associated with an approximate IRFP in the phase with
broken chiral symmetry, which could feature a substantial value of an effective
$\gamma_{\bar\psi\psi,IR} \sim O(1)$ \cite{wtc}.  Our scheme-independent
calculations are also relevant to this area of research in two ways: (i) if
$N_f \lsim N_{f,cr}$, then the effective values of quantities such as
$\gamma_{\bar\psi\psi,IR}$ may be close to the values calculated via the
$\Delta_f$ expansion from within the NACP; (ii) combining our calculations of
$\gamma_{\bar\psi\psi,IR}$ with an upper bound on this anomalous dimension from
conformal invariance and an assumption that this bound is saturated as $N_f
\searrow N_{f,cr}$ yields an estimate of the value of $N_{f,cr}$.  This is
useful, since the value of $N_{f,cr}$ for a given $G$ and $R$ is not known 
exactly at present and is the subject of current investigation, including
lattice studies, as discussed further below.

Although most of our paper deals with new scheme-independent results for
physical quantities, one of the ouputs of our calculations is a new type of
series expansion for a scheme-dependent quantity, namely $\alpha_{IR}$.  The
conventional procedure for calculating the IR zero of a beta function at the
$n$-loop order, which we have applied in earlier work to four-loop order for
arbitrary $G$ and $R$ \cite{bvh}-\cite{lnn} (see also \cite{ps}) is to examine
the $n$-loop beta function, which has the form of $\alpha^2$ times a polynomial
of degree $n-1$ in $\alpha$, and then determine the $n$-loop value
$\alpha_{IR,n\ell}$ as the (real, positive) root of this polynomial closest to
the origin.  However, in \cite{flir}, we investigated the five-loop beta
function for $G={\rm SU}(3)$ and $R=F$, as calculated in the standard
$\overline{\rm MS}$ scheme, and found that, over a substantial range of values
of $N_f$ in the non-Abelian Coulomb phase, it does not have any positive real
root.  We were able to circumvent this problem in \cite{flir} by the use of
Pad\'e approximants, but nevertheless, it is a complication for this
conventional approach to calculating $\alpha_{IR}$.  The new calculation of
$\alpha_{IR}$ as an expansion in powers of $\Delta_f$ up to $O(\Delta_f^4)$ for
general $G$ and $R$ that we present here has the advantage that it always
yields a physical value, in contrast to the situation with the $n$-loop beta
function.

The paper is organized as follows. Some relevant background and methods are
discussed in Section \ref{methods_section}. We present our calculation of
$\kappa_4$ in the scheme-independent expansion of $\gamma_{\bar\psi\psi,IR}$
for general $G$ and $R$ in Section \ref{kappa_section}, together with
evaluations for $G={\rm SU}(N_c)$ and $R=F, \ adj, \ S_2$, and $A_2$.  These
are compared with values from $n$-loop calculations and with lattice
measurements.  In this section we also present results for case $R=F$ in the
limit $N_c \to \infty, \ N_f \to \infty$, with $N_f/N_c$ fixed, which we call
the LNN limit.  In Section \ref{betaprime_section} we present our calculation
of the coefficient $d_5$ in the scheme-independent expansion of $\beta'_{IR}$
for general $G$ and $R$, with evaluations for the above-mentioned specific
representations.  Section \ref{xiir_section} gives an analysis of the five-loop
rescaled beta function in the LNN limit and a determination of the interval
over which it exhibits a physical IR zero. Section \ref{alfir_Deltasection} is
devoted to the calculation of the coefficients in an expansion of $\alpha_{IR}$
in powers of $\Delta_f$ up to $O(\Delta_f^4)$.  Our conclusions are given in
Section \ref{conclusion_section}, and some auxiliary formulas are listed in an
appendix.

% =========================================================================

% section II 
\section{Background and Methods} 
\label{methods_section}

In this section we review some background and methods relevant for our
calculations.  The series expansion of $\beta$ in powers of $\alpha$ is
\beq
\beta = -2\alpha \sum_{\ell=1}^\infty b_\ell \,
\Big (\frac{\alpha}{4\pi} \Big )^\ell \ . 
\label{beta}
\eeq
where $b_\ell$ is the $\ell$-loop coefficient.  For a general operator
${\cal O}$, we denote the full scaling dimension as $D_{\cal O}$ and its
free-field value as $D_{{\cal O},free}$.  The anomalous dimension of this
operator, denoted $\gamma_{\cal O}$, is defined via the relation
\cite{gammaconvention}
\beq
D_{\cal O} = D_{{\cal O},free} - \gamma_{\cal O} \ . 
\label{anomdim}
\eeq
An operator of particular interest is the (gauge-invariant) fermion bilinear,
$\bar\psi\psi$.  The expansion of the anomalous dimension of this operator, 
$\gamma_{\bar\psi\psi}$, in powers of $\alpha$ is 
\beq
\gamma_{\bar\psi\psi} = \sum_{\ell=1}^\infty
c_\ell \Big ( \frac{\alpha}{4\pi} \Big )^\ell \ ,
\label{gamma}
\eeq
where $c_\ell$ is the $\ell$-loop coefficient.  As noted above, the
coefficients $b_1$, $b_2$, and $c_1$ are scheme-independent, while the $b_\ell$
with $\ell \ge 3$ and the $c_\ell$ with $\ell \ge 2$ are scheme-dependent
\cite{gross75}.  For a general gauge group $G$ and fermion representation $R$,
the coefficients $b_1$ and $b_2$ were calculated in \cite{b1} and \cite{b2},
and $b_3$ and $b_4$ were calculated in \cite{b3} and \cite{b4} (and checked in
\cite{b4p}) in the commonly used $\overline{\rm MS}$ scheme \cite{msbar}.  For
$G={\rm SU}(3)$ and $R=F$, $b_5$ was calculated in \cite{b5su3} and recently,
an impressive calculation of $b_5$ for general gauge group $G$ and fermion
representation $R$ was presented in \cite{b5}, again in the $\overline{\rm MS}$
scheme.  We also make use of the $c_\ell$ up to loop order $\ell=4$, calculated
in \cite{c4}.  Although we use these coefficients as calculated in the
$\overline{\rm MS}$ scheme below, we emphasize that the main results of this
paper are calculations of the quantities $\kappa_4$ and $d_5$ which, like all
of the $\kappa_j$ and $d_j$, are scheme-independent.  We denote the 
$n$-loop $\beta$, $\beta'$, and $\gamma_{\bar\psi\psi}$ as 
$\beta_{n\ell}$, $\beta'_{n\ell}$, and $\gamma_{\bar\psi\psi,n\ell}$.  
As discussed above, we denote the IR zero of $\beta_{n\ell}$ as
$\alpha_{IR,n\ell}$, and the corresponding evaluations of $\beta'_{n\ell}$ and 
$\gamma_{\bar\psi\psi,n\ell}$ at $\alpha_{IR,n\ell}$ as 
$\beta'_{IR,n\ell}$ and $\gamma_{\bar\psi\psi,IR,n\ell}$. The symbols 
$\alpha_{IR}$, $\gamma_{\bar\psi\psi,IR}$, and $\beta'_{IR}$ refer to the 
exact values of these quantities.

For a given $G$ and $R$, as $N_f$ increases, $b_1$ decreases through positive
values and vanishes with sign reversal at $N_f=N_u$, with
\beq
N_u = \frac{11C_A}{4T_f} \ ,
\label{nfb1z}
\eeq
where $C_A$ and $T_f$ are
group invariants \cite{casimir,nfintegral}.  Hence, the asymptotic freedom
condition yields the upper bound $N_f < N_u$.

There is a range of $N_f < N_u$ where $b_2 < 0$, so the
two-loop beta function has an IR zero, at the value
\beq
\alpha_{IR,2\ell}=-\frac{4\pi b_1}{b_2} \ . 
\label{alfir_2loop}
\eeq
The $n$-loop beta function has a double UV zero at $\alpha=0$ and $n-1$ zeros
away from the origin.  Among the latter zeros of the beta function, the
smallest (real, positive) zero, if there is such a zero, is the physical IR
zero, $\alpha_{IR,n\ell}$, of $\beta_{n\ell}$.  As $N_f$ decreases below 
$N_u$, $b_2$ passes through zero to positive values as $N_f$ decreases 
through 
\beq
N_\ell = \frac{17C_A^2}{2T_f(5C_A+3C_f)} \ .
\label{nfb2z}
\eeq
Hence, with $N_f$ formally extended
from nonnegative integers to nonnegative real numbers \cite{nfintegral}, 
$\beta_{2\ell}$ has an IR zero (IRZ) for $N_f$ in the interval
\beq
I_{IRZ}: \quad N_\ell < N_f < N_u \  .
\label{nfinterval}
\eeq
Thus, $N_\ell$ is the lower ($\ell$) end of this interval \cite{phasenote}

As $N_f$ decreases in this interval, $\alpha_{IR,2\ell}$ increases. 
Therefore, in order to investigate the IR zero of the beta function 
for $N_f$ toward the middle and
lower part of $I_{IRZ}$ with reasonable accuracy, one requires higher-loop
calculations. These were performed in \cite{gkgg,elias}, 
\cite{bvh}-\cite{lnn}, \cite{ps},\cite{flir} for
$\alpha_{IR,n\ell}$ and for the anomalous dimension of the fermion bilinear
operator (see also \cite{bfs,bfss}). Since the $b_\ell$ with $\ell \ge 3$ are
scheme-dependent, it is necessary to determine the degree of
sensitivity of the value obtained for $\alpha_{IR,n\ell}$ for $n \ge 3$ to the
scheme used for the calculation. This was done in 
\cite{sch}-\cite{gracey2015}. 

The nonanomalous global flavor symmetry of the theory is
\beq
G_{fl} = {\rm SU}(N_f)_L \otimes {\rm SU}(N_f)_R \otimes
{\rm U}(1)_V \ .
\label{gfl}
\eeq
This $G_{fl}$ symmetry is preserved in the (deconfined) non-Abelian Coulomb
phase. As in \cite{gtr,flir,gsi,dex,dexs}, we focus on this phase in the
present work, since both the expansion in a small $\alpha_{IR}$ and the
scheme-independent expansion in powers of $\Delta_f$ start from the upper end
of the interval $I_{IRZ}$ in this phase. In contrast, in the phase with
confinement and spontaneous chiral symmetry breaking, the gauge interaction
produces a bilinear fermion condensate, $\langle \bar\psi \psi\rangle$, 
and this breaks $G_{fl}$ to ${\rm SU}(N_f)_V \otimes {\rm U}(1)_V$, where 
${\rm SU}(N_f)_V$ is the diagonal subgroup of 
${\rm SU}(N_f)_L \otimes {\rm SU}(N_f)_R$. 

We will consider the flavor-nonsinglet ($fns$) and 
flavor-singlet ($fs$) bilinear fermion operators
$\sum_{j,k=1}^{N_f} \bar\psi_j (T_a)_{jk} \psi_k$ and
$\sum_{j=1}^{N_f} \bar\psi_j \psi_j$, where 
here $T_a$ with $a=1,...,N_f^2-1$ is an generator of the global flavor 
group SU($N_f$).  We will usually suppress the explicit flavor indices 
and thus write these operators as 
$\bar\psi T_a \psi$ and $\bar\psi \psi$.  These have the same anomalous 
dimension (e.g., \cite{gracey_gammatensor}), which we denote simply as the
anomalous dimension for the flavor-singlet operator, $\gamma_{\bar\psi\psi}$.  
In vectorial gauge theories of the type considered here, these fermion bilinear
operators are gauge-invariant, and hence the anomalous dimension 
$\gamma_{\bar\psi\psi}$ and its IR value, $\gamma_{\bar\psi\psi,IR}$, are 
physical.  (In contrast, in a chiral gauge theory, fermion bilinears are
generically not gauge-invariant, and hence neither are their anomalous
dimensions.)  

Since $\alpha_{IR}$ vanishes (linearly) with 
$\Delta_f$ as $\Delta_f \to 0$, we can express it as a series expansion
in this variable, $\Delta_f$.  We thus write 
\beq
\alpha_{IR} \equiv 4\pi a_{IR} = 4\pi \sum_{j=1}^\infty a_j \Delta_f^j \ . 
\label{alfir_Deltaseries}
\eeq
The calculation of the $a_j$ requires, as input, the $b_\ell$ with $1 \le \ell
\le j+1$ \cite{gtr,dex}.

A basic question concerns the part of the interval $I_{IRZ}$ in which the 
series expansions for $\gamma_{\bar\psi\psi,IR}$
and $\beta'_{IR}$ in Eqs. (\ref{gamma_ir_Deltaseries}) and 
(\ref{betaprime_ir_Deltaseries}) are reliable. We analyzed this
question in \cite{gtr,gsi,dex,dexs} and concluded that these expansions 
for $\gamma_{IR}$ and $\beta'_{IR}$ should be reasonably reliable
throughout much of the interval $I_{IRZ}$ and 
non-Abelian Coulomb phase. We will use our higher-order
calculations in this paper to extend this analysis here.  We recall that 
the properties of the theory change qualitatively as $N_f$
decreases through the value $N_{f,cr}$ and spontaneous chiral symmetry breaking
occurs, with the fermions gaining dynamical masses.  The (chirally symmetric)
non-Abelian Coulomb phase with $N_{f,cr} < N_f < N_u$ is clearly
qualitatively different from the confined phase with spontaneous chiral
symmetry breaking at smaller $N_f$ below $N_{f,cr}$.  Therefore, one does not,
in general, expect the small-$\Delta_f$ series expansion to hold below
$N_{f,cr}$.  Estimating the range of applicability of this expansion is thus
connected with estimating the value of $N_{f,cr}$.  For 
general $G$ and $R$, as $N_f$, formally continued from the nonnegative
integers to the nonnegative real numbers, decreases from the upper end of the
interval $I_{IRZ}$ at $N_u$ to the lower end of this interval at
$N_f=N_\ell$, $\Delta_f$ increases from 0 to the maximal value
\beqs
& & (\Delta_f)_{max} = N_u - N_\ell \cr\cr
  & = & \frac{3C_A(7C_A+11C_f)}{4T_f(5C_A+3C_f)} \quad {\rm for} \ N_f \in
  I_{IRZ} .
\label{Deltaf_max_irz}
\eeqs

Recall that for a function $f(z)$ that is analytic about $z=0$ and has a 
Taylor series expansion 
\beq
f(z) = \sum_{j=1}^\infty f_j z^j \ , 
\label{fz}
\eeq
the radius of convergence of this series, $z_c$, can be determined by the ratio
test 
\beq
z_c = \lim_{j \to \infty} \frac{|f_{j-1}|}{|f_j|} \ . 
\label{zc}
\eeq
Of course, we cannot apply the full ratio test here, since we have only
calculated the $\kappa_j$ and $d_j$ to finite order.  However, we can get a
rough measure of the range of applicability of the series expansions in
$\Delta_f$ (and also $\Delta_r$ in the LNN limit \cite{lnn} discussed below) by
computing the ratios $\kappa_{j-1}/\kappa_j$ and $d_{j-1}/d_j$ for the values
of $j$ for which we have calculated these coefficients.

The series expansion (\ref{gamma_ir_Deltaseries}) for $\gamma_{IR}$ starts at
$\Delta_f=0$, i.e., at the upper end of the non-Abelian Coulomb
phase, and extends downward through this phase.  Given that the theory at
$\alpha_{IR}$ in this phase is conformal, there is an upper bound from
conformal invariance, namely \cite{gammabound} 
\beq
\gamma_{\bar\psi\psi,IR} \le 2 \ . 
\label{gamma_upperbound}
\eeq
We have used this in our earlier work \cite{bvh,gtr,flir,gsi,dex,dexs} and we
will apply it with our higher-order calculations here.  As discussed in
\cite{bvh}, in the phase with spontaneous chiral symmetry breaking (S$\chi$SB),
there is a similar upper bound, $\gamma_{\bar\psi\psi,IR} < 2$. This follows
from the requirement that if $m(k)$ is the momentum-dependent running dynamical
mass generated in association with the S$\chi$SB, then $\lim_{k \to
  \infty}m(k)=0$ (see Eqs. (4.1)-(4.2) of \cite{bvh}).  Thus, if the
approximate calculation of the anomalous dimension of a given quantity at a
fixed value of $\Delta_f$, computed up to order $\Delta_f^p$, yields a value
greater than 2, then we can infer that the perturbative calculation is not
applicable at this value of $\Delta_f$ or equivalently, $N_f$.

In particular, this can give information on the extent of the non-Abelian
Coulomb phase and the value of $N_{f,cr}$.  The application of this bound is
particularly powerful in the context of our present scheme-independent
calculations because we find that the $\kappa_j$ in
Eq. (\ref{gamma_ir_Deltaseries}) are positive for all of the representations
considered here, and hence, for a given $p$, $\gamma_{IR,\Delta_f^p}$ is a
monotonically increasing function of $\Delta_f$ or equivalently it increases
monotonically as $N_f$ decreases from its upper limit, $N_u$. If one assumes
that $\gamma_{IR}$ saturates its upper bound, (\ref{gamma_upperbound}) and if a
calculation of $\gamma_{IR}$ is reliable in the regime where it is approaching
2 from below, then one can, in principle, determine the value of $N_{f,cr}$,
where $\gamma_{IR}$ reaches this upper bound after approaching it from
below. In this context, it should be mentioned that in a supersymmetric
(vectorial) gauge theory (SGT) with $N_f$ pairs of massless chiral superfields
transforming according the representations $R$ and $\bar R$ of a gauge group
$G$, the exact expression for $\gamma_{IR}$ is known \cite{nsvz,seiberg}, and
(i) it increases monotonically with decreasing $N_f$ in the NACP; and (ii) it
saturates its upper bound (which, in the SGT case is $\gamma_{IR,SGT} \le 1$)
at the lower end of the non-Abelian Coulomb phase.  Specifically, in this
supersymmetri gauge theory, the upper and lower ends of the NACP occur at
\cite{nfintegral}
\beq
N_{u,SGT}= \frac{3C_A}{2T_f} \ , 
\label{nfb1z_sgt}
\eeq
and
\beq
N_{\ell,SGT} = \frac{3C_A}{4T_f} = \frac{N_u}{2} \ , 
\label{nfb2z_sgt}
\eeq
and 
\beqs
\gamma_{\bar\psi\psi,IR,SGT} & = & \frac{3C_A-2T_fN_f}{2T_fN_f}
= \frac{N_u}{N_f} - 1 \cr\cr
& = & 
\frac{\frac{2T_f}{3C_A}\Delta_f}{1-\frac{2T_f}{3C_A}\Delta_f } \ . 
\label{gamma_ir_sgt}
\eeqs
Thus, $\gamma_{\bar\psi\psi,IR,SGT}$ increases from 0 to 1 as
$N_f$ decreases from $N_{u,SGT}$ to $N_{\ell,SGT}$. 
However, it is not known if this saturation occurs
in the non-supersymmetric case.  In practice, we are only able to apply this
test in an approximate manner because for a given $G$ and $R$, as $N_f$
decreases toward the lower part of $I_{IRZ}$, the ratio test already shows that
higher-order terms in the $\Delta_f$ expansion are becoming increasingly
non-negligible, so that the truncation of the infinite series
(\ref{gamma_ir_Deltaseries}) to maximal power $p=4$ involves an increasingly
great uncertainty, as does an extrapolation to $p=\infty$.  

For some perspective, we note that in order to asses the accuracy of the
$\Delta_f$ expansion, the coefficients $\kappa_{j,SGT}$ were calculated for
$j=1, \ 2$ in \cite{gtr} and were found to be in perfect agreement with the
corresponding Taylor series expansion of the exact expression
(\ref{gamma_ir_sgt}).  This check was carried to one higher order in
\cite{dexs} for the case $G={\rm SU}(N_c)$ and $R=F$ with a calculation of
$\gamma_{IR,SGT,\Delta_f^3}$, and again, perfect agreement was found with the
exact result.  This agreement explicitly demonstrated the scheme independence
of the $\kappa_{j,SGT}$, since the calculations were carried out using inputs
computed in the $\overline{DR}$ scheme, while (\ref{gamma_ir_sgt}) was derived
in the NSVZ scheme \cite{nsvz}.  Furthermore, as a consequence of
electric-magnetic duality \cite{seiberg}, as $N_f \searrow N_{\ell,SGT}$ in the
non-Abelian Coulomb phase, the physics is described by a magnetic theory with
coupling strength going to zero, or equivalently, by an electric theory with
divergent $\alpha_{IR}$. Therefore, this perfect agreement, order-by-order,
between the $\kappa_{j,SGT}$ and the expansion of the exact expression
(\ref{gamma_ir_sgt}) for
$\gamma_{IR,SGT}$ in powers of $\Delta_f$, showed that the $\Delta_f$ expansion
in this supersymmetric gauge theory is able to treat situations with strong, as
well as weak, coupling.  This could not be done with conventional perturbative
series expansions in powers of $\alpha$ \cite{bfs,bfss}.

% ======================================================================

% section III 
\section{Calculation of $\gamma_{\bar\psi\psi,IR}$ to $O(\Delta_f^4)$ }
\label{kappa_section}

\subsection{General $G$ and $R$} 
\label{kappa_general}

The coefficients $\kappa_j$ in the scheme-independent expansion of 
$\gamma_{\bar\psi\psi,IR}$ in powers of $\Delta_f$, Eq. 
(\ref{gamma_ir_Deltaseries}), contain important information about the theory. 
For a general asymptotically free vectorial gauge theory with gauge group $G$
and $N_f$ massless fermions in a representation $R$, the coefficients
$\kappa_j$ were given in \cite{gtr} up to order $j=3$, yielding the 
expansion of $\gamma_{\bar\psi\psi,IR}$ to order $\Delta_f^3$.  
It is convenient to define
\beq
D = 7C_A+11C_f \ , 
\label{d}
\eeq
since this factor occurs repeatedly in denominators of various expressions. 
For reference, we list the $\kappa_j$ for $1 \le j \le 3$ below:
\beq
\kappa_1 = \frac{8C_fT_f}{C_AD} \ , 
\label{kappa1}
\eeq
\beq
\kappa_2 = \frac{4C_fT_f^2(5C_A+88C_f)(7C_A+4C_f)}{3C_A^2 D^3} \ , 
\label{kappa2}
\eeq
and 
\begin{widetext}
\beqs
\kappa_3 &=& \frac{4C_fT_f}{3^4 C_A^4 D^5} \Bigg [ 
3C_AT_f^2 \bigg ( -18473C_A^4 + 144004 C_A^3C_f 
+650896C_A^2C_f^2 +356928C_AC_f^3+569184C_f^4 \bigg ) \cr\cr
&-&2560T_f^2D\frac{d_A^{abcd}d_A^{abcd}}{d_A}
+45056C_AT_fD\frac{d_R^{abcd}d_A^{abcd}}{d_A}
-170368C_A^2D\frac{d_R^{abcd}d_R^{abcd}}{d_A} \cr\cr
&+& 33 \cdot 2^{10}D\bigg (2T_f^2 \frac{d_A^{abcd}d_A^{abcd}}{d_A}
-13 C_AT_f \frac{d_R^{abcd}d_A^{abcd}}{d_A}
+11 C_A^2 \frac{d_R^{abcd}d_R^{abcd}}{d_A} \bigg )\zeta_3 \ \Bigg ] \ . 
\label{kappa3}
\eeqs
Here, $\zeta_s = \sum_{n=1}^\infty n^{-s}$ 
is the Riemann zeta function, the quantities $C_A$, $C_f$, and $T_f$ are group
invariants, the contractions $d_A^{abcd}d_A^{abcd}$,
$d_R^{abcd}d_A^{abcd}$, $d_R^{abcd}d_R^{abcd}$ are additional group-theoretic
quantities given in \cite{b4}, and $d_A$ is the dimension of the adjoint
representation of $G$.  In \cite{gtr,dex}, the expression for $\kappa_3$ was
given with terms written in order of descending powers of $C_A$.  
It is also useful to express this coefficient $\kappa_3$ 
in an equivalent form that renders certain factors of $D$ explicit and 
shows the simple factorization of terms multiplying 
$\zeta_3$, and we have done this in Eq. (\ref{kappa3}).

Our new result here for $\kappa_4$ for a general gauge group $G$ and fermion 
representation $R$ is
\beqs
&&\kappa_4 = \frac{T_f^2}{3^5 C_A^5D^7} \Bigg [ 
C_AC_fT_f^2 \bigg ( 19515671C_A^6-131455044C_A^5C_f+1289299872C_A^4C_f^2 
+2660221312C_A^3C_f^3 \cr\cr
&+&1058481072C_A^2C_f^4+6953709312C_AC_f^5 +1275715584C_f^6 \bigg ) 
+2^{10}C_fT_f^2D \bigg ( 5789C_A^2-4168C_AC_f-6820C_f^2\bigg ) 
\frac{d_A^{abcd}d_A^{abcd}}{d_A} \cr\cr
&-& 2^{10}C_AC_fT_fD\bigg ( 41671C_A^2-125477C_AC_f-53240C_f^2\bigg ) 
\frac{d_R^{abcd}d_A^{abcd}}{d_A} \cr\cr
&-&2^8 \cdot 11^2 C_A^2C_fD(2569C_A^2+18604C_AC_f-7964C_f^2 \bigg )
\frac{d_R^{abcd}d_R^{abcd}}{d_A} \cr\cr
&-&2^{14} \cdot 3C_AT_f^2D^3 \frac{d_R^{abcd}d_A^{abcd}}{d_R}
+2^{13} \cdot 33 C_A^2T_fD^3 \frac{d_R^{abcd}d_R^{abcd}}{d_R}\cr\cr
&+&2^8D \bigg [ -3C_AC_fT_f^2D\bigg (
4991C_A^4-17606C_A^3C_f+33240C_A^2C_f^2-30672C_AC_f^3+9504C_f^4 \bigg ) \cr\cr
&-& 2^4 C_fT_f^2 \frac{d_A^{abcd}d_A^{abcd}}{d_A}
\Big ( 17206C_A^2-60511C_AC_f-45012C_f^2 \Big )
+40C_AC_fT_f\frac{d_R^{abcd}d_A^{abcd}}{d_A}
\Big (35168C_A^2-154253C_AC_f-88572C_f^2\Big ) \cr\cr
&-&88C_A^2C_f \frac{d_R^{abcd}d_R^{abcd}}{d_A}
\Big (973C_A^2-93412C_AC_f-56628C_f^2\Big )
+1440C_AT_f^2D^2\frac{d_R^{abcd}d_A^{abcd}}{d_R}
-7920C_A^2T_fD^2\frac{d_R^{abcd}d_R^{abcd}}{d_R} \ \bigg]\zeta_3 \cr\cr
&+&\frac{4505600C_AC_fD^2}{d_A} \bigg [ -4T_f^2 d_A^{abcd}d_A^{abcd}
+2T_f d_R^{abcd}d_A^{abcd}(10C_A+3C_f) 
+11 C_A d_R^{abcd}d_R^{abcd}(C_A-3C_f) \ \bigg ]\zeta_5 \ \Bigg ] \ . 
\label{kappa4}
\eeqs
\end{widetext}
Here, $d_R$ is the dimension of the fermion representation $R$.  As before, we
have indicated the simple factors in the prefactor and, for sufficiently simple
cases, also factorizations of numbers in numerator terms.  We will follow the
same format for indicating numerical factorizations below.  We proceed to
evaluate this general expression for the gauge group $G={\rm SU}(N_c)$ and
several specific fermion representations $R$, namely the fundamental, adjoint,
and symmetric and antisymmetric rank-2 tensor. As stated in the introduction,
we will use the abbreviations $F$, $adj$, $S_2$, and $A_2$ to refer to these
representations. It is also worthwhile to evaluate our general formulas for
other gauge groups and their representations, including orthogonal, symplectic,
and exceptional groups.  We will report these evaluations for other groups and
their representations elsewhere. There has, indeed, been interest in conformal
phases for theories with these other gauge groups \cite{othergroups}.

The coefficients $\kappa_1$ and $\kappa_2$ are manifestly positive for all $G$
and $R$. For $G={\rm SU}(N_c)$ with all physical $N_c$, and for representations
$R = F, \ adj, \ S_2$, we have found that $\kappa_3$ and $\kappa_4$ are also
positive \cite{gtr}-\cite{dexs}.  As one of the results in the present paper,
we generalize this further to include $R=A_2$.  That is, for all physical $N_c$
and for all of these representations, we find that $\kappa_j > 0$ for $j=3,\ 4$
as well as the manifestly positive cases $j=1, \ 2$.  Thus, extending our
previous discussion in \cite{gtr}-\cite{dexs}, the property that, for all of
these representations $R$, $\kappa_j > 0$ for $1 \le j \le 4$ and for all $N_c$
implies two important monotonicity results: (i) for these $R$, and with a fixed
$p$ in the interval $1 \le p \le 4$, $\gamma_{\bar\psi\psi,IR,\Delta_f^p}$ is a
monotonically increasing function of $\Delta_f$, i.e., it increases
monotonically with decreasing $N_f$; and (ii) for these $R$,
and with a fixed $N_f \in I_{IRZ}$, $\gamma_{\bar\psi\psi,IR,\Delta_f^p}$ is a
monotonically increasing function of $p$ in the range $1 \le p \le 4$.  In
addition to the manifestly positive $\kappa_1$ and $\kappa_2$, a plausible
conjecture is that, for these $R$, $\kappa_j > 0$ for all $j \ge 3$.  Assuming
that this conjecture is valid, then three consequences are that for these
representations $R$, (iii) for fixed $N_f$, 
$\gamma_{\bar\psi\psi,IR,\Delta_f^p}$ is a monotonically increasing function of
$p$ for all $p$; (iv) $\gamma_{\bar\psi\psi,IR,\Delta_f^p}$ is a
monotonically increasing function of $\Delta_f$, i.e. it increases with
decreasing $N_f$, for all $p$; and hence (v) 
(assuming that the infinite series (\ref{gamma_ir_Deltaseries}) converges), 
the quantity 
$\gamma_{\bar\psi\psi,IR}$ defined by this infinite series, and equivalent to 
$\lim_{p \to \infty} \gamma_{\bar\psi\psi,IR,\Delta_f^p}$, is a monotonically 
increasing function of $\Delta_f$, i.e., it increases monotonically with
decreasing $N_f$. 

% =====================================================================

\subsection{$\gamma_{\bar\psi\psi,IR,\Delta_f^4}$ for 
$G={\rm SU}(N_c)$ and $R=F$}
\label{kappa_fund_section}

An important special case is $G={\rm SU}(N_c)$ with $R$ being the fundamental
representation.  For this case, the general expression for the interval 
$I_{IRZ}$, Eq. (\ref{nfinterval}), is \cite{nfintegral}
\beq
I_{IRZ}: \quad \frac{34N_c^3}{13N_c^2-3} < N_f < \frac{11N_c}{2} \quad 
{\rm for} \ R=F \ . 
\label{interval_fund}
\eeq
The factor $D$ in Eq. (\ref{d}) has the explicit form
\beq
D = \frac{25N_c^2-11}{2N_c} \quad {\rm for} \ R = \ {\rm fund.}
\label{dfund}
\eeq
The general results for $\kappa_p$ with 
$1 \le p \le 3$ in (\ref{kappa1})-(\ref{kappa3}) from \cite{gtr} take the
following forms given in \cite{dex}:
\beq
\kappa_{1,F} = \frac{4(N_c^2-1)}{N_c(25N_c^2-11)}
\label{kappa1f}
\eeq
\beq
\kappa_{2,F} = \frac{4(N_c^2-1)(9N_c^2-2)(49N_c^2-44)}
{3N_c^2(25N_c^2-11)^3} 
\label{kappa2f}
\eeq
and
\begin{widetext}
\beqs
\kappa_{3,F} &=& 
\frac{8(N_c^2-1)}{3^3N_c^3(25N_c^2-11)^5} 
\bigg [ \Big ( 274243N_c^8-455426N_c^6-114080N_c^4+47344N_c^2+35574 \Big ) 
\cr\cr
&-& 4224N_c^2(4N_c^2-11)(25N_c^2-11)\zeta_3 \bigg ] \ . 
\label{kappa3f}
\eeqs
For $\kappa_{4,F}$, we have \cite{dexs} 
\beqs
\kappa_{4,F} &=& 
\frac{4(N_c^2-1)}{3^4N_c^4(25N_c^2-11)^7} 
\bigg [ \Big (263345440N_c^{12} - 673169750N_c^{10} + 256923326N_c^8 \cr\cr
&-& 290027700N_c^6 + 557945201N_c^4 - 208345544N_c^2 + 6644352 \Big ) \cr\cr
&+& 384(25N_c^2-11)\Big ( 4400N_c^{10}-123201N_c^8+480349N_c^6-486126N_c^4
+84051N_c^2+1089 \Big )\zeta_3 \cr\cr
&+& 211200N_c^2(25N_c^2-11)^2(N_c^6+3N_c^4-16N_c^2+22)\zeta_5 \ \bigg ] \ . 
\label{kappa4f}
\eeqs
\end{widetext}
We have checked that when we substitute the value $N_c=3$ in our expression for
$\kappa_{4,F}$ in Eq. (\ref{kappa4f}), the
result agrees with our previous calculation of $\kappa_{4,F}$ for this case in
Eq. (9) of Ref. \cite{gsi}.

\begin{figure}
  \begin{center}
    \includegraphics[height=6cm]{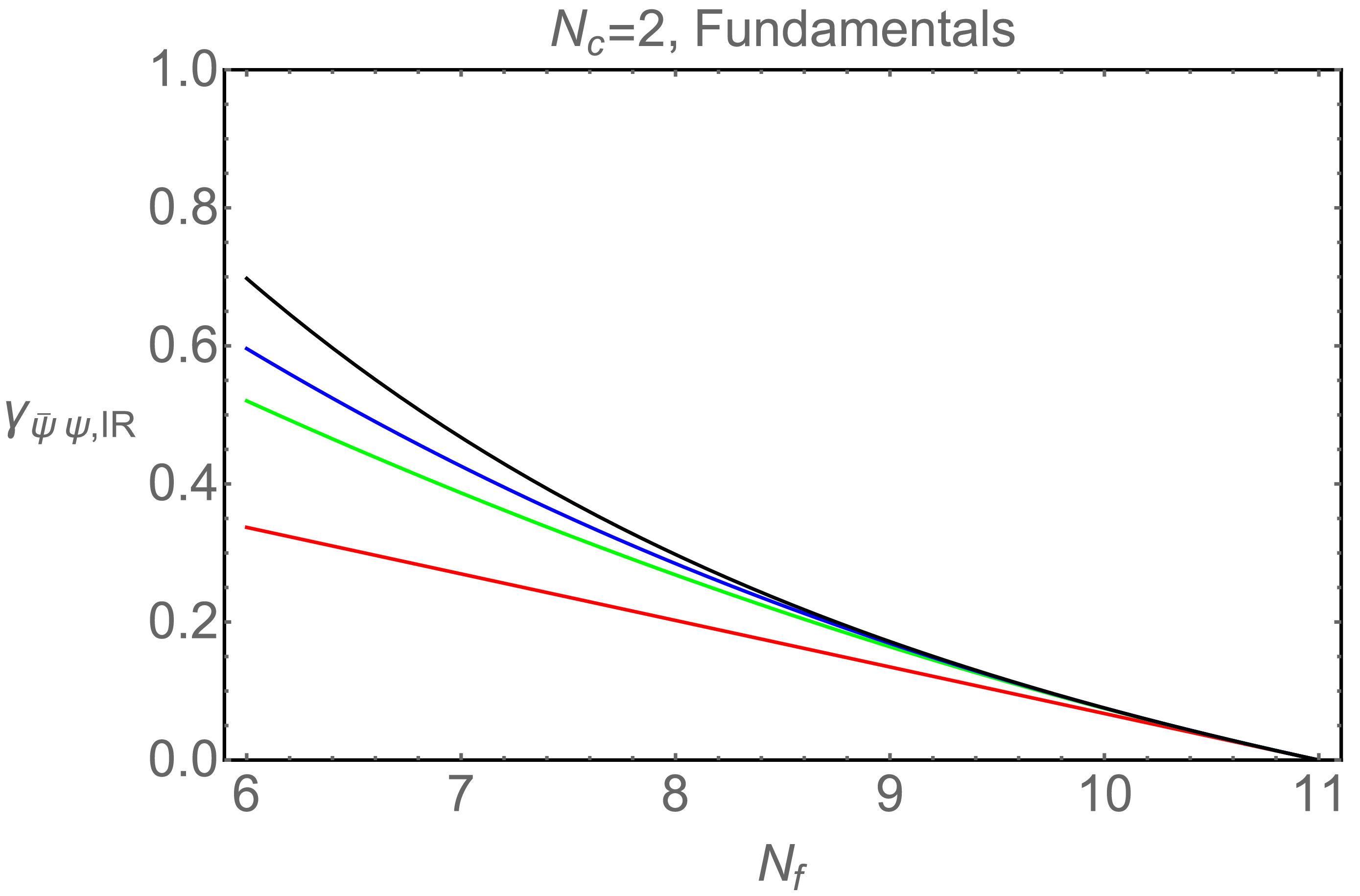}
  \end{center}
\caption{Plot of $\gamma_{\bar\psi\psi,IR,F,\Delta_f^p}$ (labelled as 
$\gamma_{\bar\psi\psi,IR}$ on the vertical axis in this and subsequent graphs) 
for $N_c=2$, i.e., $G={\rm SU}(2)$, and
$1 \le p \le 4$ as a function of $N_f \in I_{IRZ}$. From bottom to top, 
the curves (with colors online) refer to 
$\gamma_{\bar\psi\psi,IR,F,\Delta_f}$ (red),                           
$\gamma_{\bar\psi\psi,IR,F,\Delta_f^2}$ (green),
$\gamma_{\bar\psi\psi,IR,F,\Delta_f^3}$ (blue), and
$\gamma_{\bar\psi\psi,IR,F,\Delta_f^4}$ (black).}
\label{gammaNc2fund_plot}
\end{figure}

\begin{figure}
  \begin{center}
    \includegraphics[height=6cm]{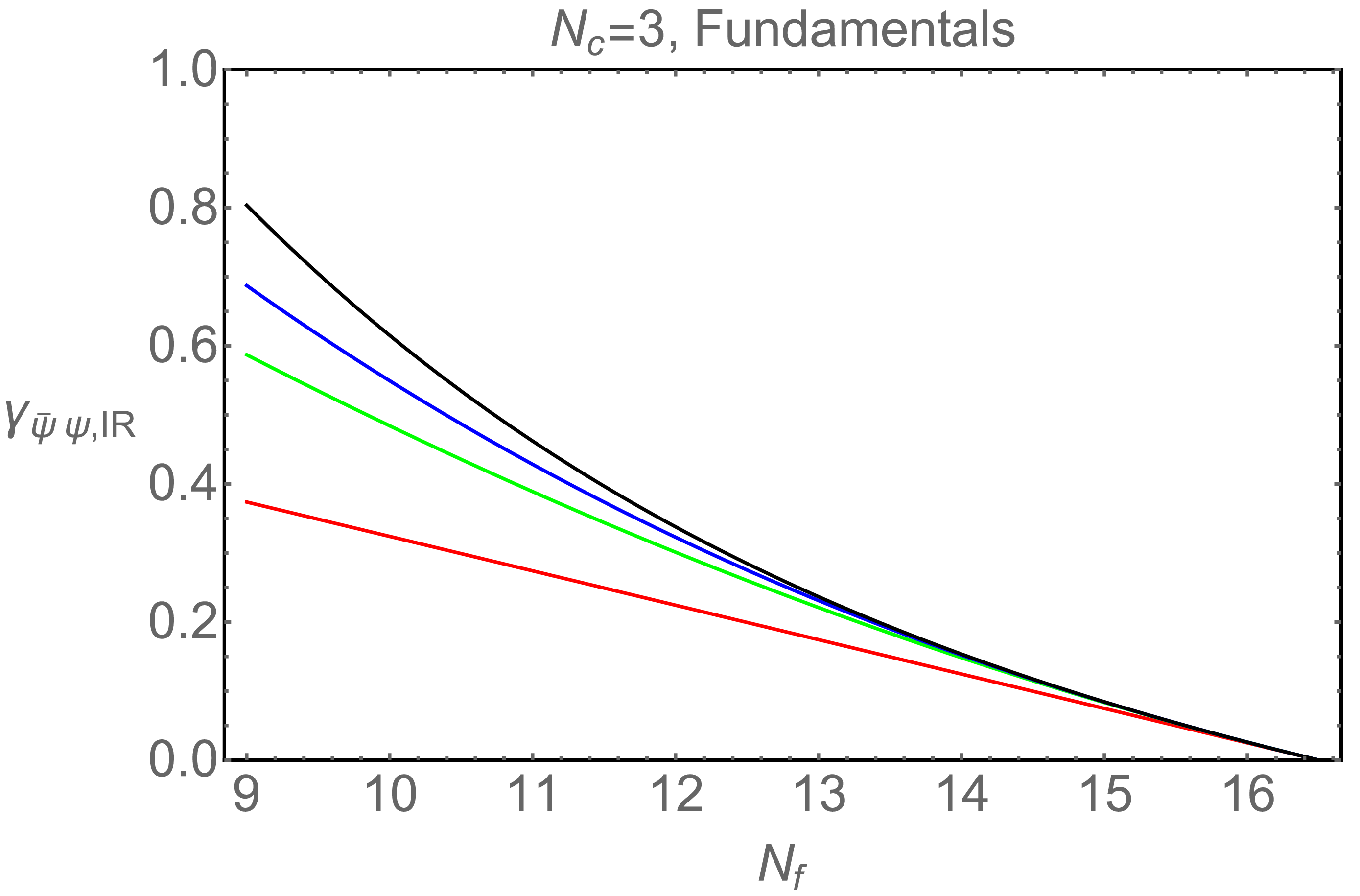}
  \end{center}
\caption{Plot of $\gamma_{\bar\psi\psi,IR,F,\Delta_f^p}$ for $N_c=3$ and
$1 \le p \le 4$ as a function of $N_f \in I_{IRZ}$.  From bottom to top, 
the curves (with colors online) refer to 
$\gamma_{\bar\psi\psi,IR,F,\Delta_f}$ (red),                           
$\gamma_{\bar\psi\psi,IR,F,\Delta_f^2}$ (green),
$\gamma_{\bar\psi\psi,IR,F,\Delta_f^3}$ (blue), and
$\gamma_{\bar\psi\psi,IR,F,\Delta_f^4}$ (black).}
\label{gammaNc3fund_plot}
\end{figure}

\begin{figure}
  \begin{center}
    \includegraphics[height=6cm]{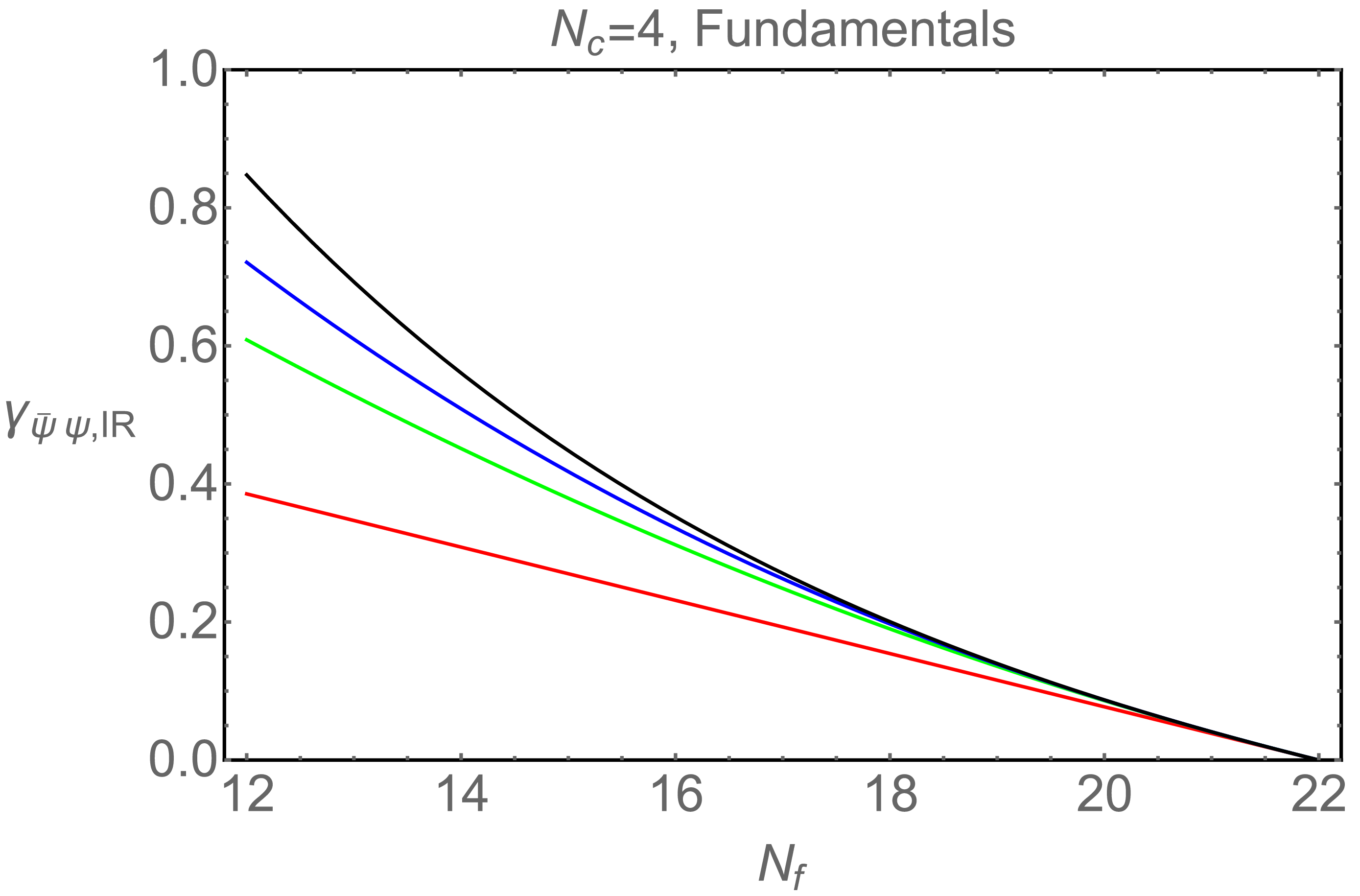}
  \end{center}
\caption{Plot of $\gamma_{\bar\psi\psi,IR,F,\Delta_f^p}$ for $N_c=4$ and
$1 \le p \le 4$ as a function of $N_f \in I_{IRZ}$.  From bottom to top, 
the curves (with colors online) refer to 
$\gamma_{\bar\psi\psi,IR,F,\Delta_f}$ (red),                           
$\gamma_{\bar\psi\psi,IR,F,\Delta_f^2}$ (green),
$\gamma_{\bar\psi\psi,IR,F,\Delta_f^3}$ (blue), and
$\gamma_{\bar\psi\psi,IR,F,\Delta_f^4}$ (black).}
\label{gammaNc4fund_plot}
\end{figure}

The explicit numerical expressions for the scheme-independent series expansions
of $\gamma_{\bar\psi\psi,IR}$ to order $\Delta_f^4$ for $R=F$ and $N_c=2, \ 3,
\ 4$ are as follows:
\begin{widetext}
\beqs
{\rm SU}(2): \ \gamma_{\bar\psi\psi,IR,F,\Delta_f^4} & = & 
\Delta_f \Big [ 0.067416 
+ (0.73308 \times 10^{-2}) \Delta_f 
+ (0.60531 \times 10^{-3}) \Delta_f^2 
+ (1.62662 \times 10^{-4}) \Delta_f^3 \ \Big ] 
\cr\cr
& & 
\label{gamma_Delta_p4_su2}
\eeqs
\beqs
{\rm SU}(3): \ \gamma_{\bar\psi\psi,IR,F,\Delta_f^4} & = & 
\Delta_f \Big [ 0.049844
+ (0.37928 \times 10^{-2}) \Delta_f 
+ (0.23747 \times 10^{-3}) \Delta_f^2 
+ (0.36789 \times 10^{-4}) \Delta_f^3 \ \Big ] 
\cr\cr
& & 
\label{gamma_Delta_p4_su3}
\eeqs
and
\beqs
{\rm SU}(4): \ \gamma_{\bar\psi\psi,IR,F,\Delta_f^4} & = & 
\Delta_f \Big [ 0.038560
+ (0.22314 \times 10^{-2}) \Delta_f 
+ (0.11230 \times 10^{-3}) \Delta_f^2 
+ (0.126505 \times 10^{-4}) \Delta_f^3 \ \Big ] \ . 
\cr\cr
& & 
\label{gamma_Delta_p4_su4}
\eeqs
\label{gamma_psibarpsi_series}
\end{widetext} 
In these equations, 
\beq
\Delta_f = \frac{11N_c}{2} - N_f \quad {\rm for} \ R=F \ . 
\label{Deltaf_fund}
\eeq
Plots of $\gamma_{\bar\psi\psi,IR,F,\Delta_f^p}$ for $N_c=2$ and $N_c=3$ and 
$1 \le p \le 4$ were given in \cite{dexs}.  These showed the two monotonicity
properties mentioned above.  For an extended comparison, we show the plots of 
$\gamma_{\bar\psi\psi,IR,F,\Delta_f^p}$ for $2 \le N_c \le 4$ and
$1 \le p \le 4$ in Figs. \ref{gammaNc2fund_plot}-\ref{gammaNc4fund_plot}. 

In Table \ref{gamma_values} we list the values of
$\gamma_{\bar\psi\psi,IR,F,\Delta_f^p}$ for $1 \le p \le 4$ for the SU(2),
SU(3), and SU(4) theories, with $N_f$ in the respective interval $I_{IRZ}$ for
each.  For comparison, we also include the values of
$\gamma_{\bar\psi\psi,IR,n\ell}$ obtained with our earlier $n$-loop
calculations in \cite{bvh}, using series expansions in powers of $\alpha$
evaluated at $\alpha=\alpha_{IR,n\ell}$ for $1 \le n \le 4$ with $b_3$ and
$b_4$ and $c_n$, $2 \le n \le 4$ calculated in the $\overline{\rm MS}$
scheme. (See Table VI in \cite{bvh} for a list of numerical values of values of
$\gamma_{\bar\psi\psi,IR,n\ell}$.)  As discussed above, if, for a given $N_c$
and $N_f$, a calculated value of $\gamma_{\bar\psi\psi,IR}$ violates the upper
bound $\gamma_{\bar\psi\psi,IR} \le 2$ in (\ref{gamma_upperbound}), this is
unphysical (marked with a symbol ``u'' in Table \ref{gamma_values}) and
indicates that the perturbative calculation is unreliable and hence not
applicable for this $N_f$.  In the case of the $n$-loop values
$\gamma_{IR,n\ell}$, if this occurs at the two-loop level, it also leads to
caution concerning $\gamma_{IR,n\ell}$ for $n=3, \ 4$, and this is similarly
indicated with a ``u''. The computations of $\gamma_{IR,n\ell}$ in
\cite{bvh,ps} made use of the $b_n$ and $c_n$ up to the $n=4$ loop level, where
the scheme-dependent $b_3$, $b_4$, and $c_n$ with $2 \le n \le 4$ had been
calculated in the widely used $\overline{\rm MS}$ scheme \cite{b3,b4,b4p,c4}.
As we pointed out in \cite{flir}, the five-loop beta function in the
$\overline{\rm MS}$ scheme does not exhibit a physical IR zero over a
substantial lower part of $I_{IRZ}$. We discuss this further below.  For
compact notation, we will often leave the subscript $\bar\psi\psi$ implicit on
these and other quantities and thus write $\gamma_{\bar\psi\psi,IR} \equiv
\gamma_{IR}$, $\gamma_{\bar\psi\psi,IR,n\ell} \equiv \gamma_{IR,n\ell}$, etc.
From Eqs. (\ref{nfb1z}) and (\ref{nfb2z}) it follows that the respective lower
and upper ends of the intervals $I_{IRZ}$ for these theories are 
$(N_u, \, N_\ell) = (5.55, \, 11)$, (8.05, \, 16.5), and (10.61, \, 22) for 
SU(2), SU(3), and SU(4),
and hence the physical intervals $I_{IRZ}$ are $6 \le N_f \le 10$ for SU(2), $9
\le N_f \le 16$ for SU(3), and $11 \le N_f \le 21$ for SU(4).

Since the calculation of $\kappa_j$ and the resultant $\gamma_{IR,\Delta_f^j}$
uses information from the $(j+1)$-loop beta function from (\ref{beta}) and the
$j$-loop expansion of $\gamma_{\bar\psi\psi}$ in (\ref{gamma}), it is natural
to compare the (SI) $\gamma_{IR,\Delta_f^p}$ with the (SD) $\gamma_{IR,p'\ell}$
for $p'=p$ and $p'=p+1$.  In the upper and middle part of the interval
$I_{IRZ}$ for a given $N_c$, we find that $\gamma_{IR,\Delta_f^4}$ is slightly
larger than $\gamma_{IR,4\ell}$, with the difference increasing as $N_f$
decreases below $N_u$, i.e., as $\Delta_f$ increases.

It is important to assess the range of applicability and reliability of these
results from the $\Delta_f$ expansion. We did this in \cite{gtr,gsi,dex} and
extend our analysis here, using our new result for $\kappa_4$.  Following our
discussion above on the ratio test for the determination of the radius of
convergence of a Taylor series, the ratios of successive coefficients, 
$\kappa_{j-1}/\kappa_j$, give an approximate measure of the range of
applicability of the $\Delta_f$ expansion for $\gamma_{IR}$.  For a given $G$
and $R$, this range may be compared with the
maximum size of $\Delta_f$ in the interval $I_{IRZ}$ where the 
scheme-independent two-loop beta function $\beta_{2\ell}$ has an IR zero.  
For the present case of $G={\rm SU}(N_c)$ and $R=F$, the general
formula (\ref{Deltaf_max_irz}) takes the form
\beq
R=F: \quad (\Delta_f)_{max} = \frac{3N_c(25N_c^2-11)}{2(13N_c^2-3)} \ . 
\label{Deltaf_max_ir_fund}
\eeq
This has the respective values
\beq
(\Delta_f)_{max}=5.45, \ 8.45, \ 11.39 \quad {\rm for} \ 
N_c=2, \ 3, \ 4 \ .  
\label{Deltamax_values_fund}
\eeq
We begin by reviewing the SU(3) theory, for which 
\beqs
{\rm SU}(3): \quad && \frac{\kappa_{F,1}}{\kappa_{,F,2}} = 13.14, \quad 
\frac{\kappa_{F,2}}{\kappa_{F,3}} = 15.97, \cr\cr
&& \frac{\kappa_{F,3}}{\kappa_{F,4}} = 6.455 \ . 
\label{kapparatios_su3_fund}
\eeqs
As discussed in  \cite{gtr,gsi,dex}, these results suggest that for the SU(3)
theory with $R=F$, the $\Delta_f$ expansion calculated to this order should be
reasonably reliable over a substantial part, including the upper and middle 
portions, of the interval $I_{IRZ}$ and the non-Abelian Coulomb phase.  

Using our new results, we now extend this analysis to the SU(2) and SU(4)
theories (and will give a further analysis in the LNN limit of
Eq. (\ref{lnn})).  We find 
\beqs
{\rm SU}(2): \quad && \frac{\kappa_{F,1}}{\kappa_{,F,2}} = 9.20, \quad
\frac{\kappa_{F,2}}{\kappa_{F,3}} = 12.11, \cr\cr
&& \frac{\kappa_{F,3}}{\kappa_{F,4}} = 3.72  
\label{kapparatios_su2_fund}
\eeqs
and
\beqs
{\rm SU}(4): \quad && \frac{\kappa_{F,1}}{\kappa_{,F,2}} = 17.28,  \quad
\frac{\kappa_{F,2}}{\kappa_{F,3}} = 19.87, \cr\cr
&& \frac{\kappa_{F,3}}{\kappa_{F,4}} = 8.88 \ .
\label{kapparatios_su4_fund}
\eeqs
Since $(\Delta_f)_{max}$ has the respective values 5.45 and 11.39 for the 
SU(2) and SU(4) theories, we are led to the same conclusion for these theories
that we reached for the SU(3) theory, namely that the $\Delta_f$ expansion
should be reasonably reliable over a substantial portion of the respective
intervals $I_{IRZ}$.  

As discussed above, another way to assess the range of applicability of the
$\Delta_f$ expansion is to check to see whether the resultant values of
$\gamma_{IR,\Delta_f^p}$ obey the upper bound $\gamma_{IR} \le 2$ in
(\ref{gamma_upperbound}). As is evident from Table \ref{gamma_values}, all of
our values of $\gamma_{IR,\Delta_f^p}$ listed there obey this bound.  This
again shows the advantages of the scheme-independent $\Delta_f$ expansion as a
way of calculating $\gamma_{IR}$ to a given order, as compared with the
conventional $n$-loop calculation of $\gamma_{IR,n\ell}$.  As is also evident
from Table \ref{gamma_values} for each of the cases listed there, namely
$N_c=2, \ 3, \ 4$, one finds unphysically large values of $\gamma_{IR,n\ell}$
for values of $N_f$ in the lower portions of the respective intervals
$I_{IRZ}$.  In \cite{bvh} and later works we explained this as a consequence of
the fact that, for a given $G$ and $R$, as $N_f$ decreases toward $N_\ell$ in
the interval $I_{IRZ}$, the coupling $\alpha_{IR}$ increases from weak toward
strong coupling. Thus, toward the lower end of the respective intervals
$I_{IRZ}$, the IR coupling $\alpha_{IR,n\ell}$ become too large for the
perturbative $n$-loop calculations of $\gamma_{IR,n\ell}$ to be applicable. In
contrast, the $\Delta_f$ expansion can be applied over a considerably greater
portion of the interval $I_{IRZ}$ to yield results for $\gamma_{IR,\Delta_f^p}$
that obey the upper bound (\ref{gamma_upperbound}).  We will show this further
below for the LNN limit (\ref{lnn}).  This also demonstrates that the
$\Delta_f$ expansion for $\gamma_{IR}$ is able to be used in situations with
substantially stronger IR coupling than is the case with the conventional 
expansion in powers of this coupling yielding the $n$-loop value
 $\gamma_{IR,n\ell}$. 

 We proceed to compare our values in Table \ref{gamma_values} with lattice
 measurements. The SU(3) theory with $R=F$ and $N_f=12$ has been the subject of
 many lattice measurements. In \cite{gsi}, we compared our results for this
 theory with lattice measurements, so we only briefly review that discussion
 here. We recall that there is not, at present, a consensus among all lattice
 groups as to whether this theory is in an IR-conformal phase or is in a
 chirally broken phase \cite{lgtreviews}.  There is a considerable spread of
 values of $\gamma_{IR}$ in published papers, including the values (where
 uncertainties in the last digits are indicated in parentheses) $\gamma_{IR}
 \sim 0.414(16)$ \cite{lsd}, $\gamma_{IR} \simeq 0.35$ \cite{degrand},
 $\gamma_{IR} \simeq 0.4$ \cite{latkmi}, $\gamma_{IR}=0.27(3)$ \cite{ah1},
 $\gamma_{IR} \simeq 0.25$ \cite{ah2} (see also \cite{ah3}),
 $\gamma_{IR}=0.235(46)$ \cite{lmnp}, and $0.2 \lsim \gamma_{IR} \lsim 0.4$
 \cite{kuti}.  We refer the reader to \cite{lgtreviews} and 
 \cite{lsd}-\cite{kuti} for
 discussions of estimates of overall uncertaintites in these measurements.
 Our value $\gamma_{IR,\Delta_f^4}=0.338$ and our extrapolated value for
 $\lim_{p \to \infty} \gamma_{IR,\Delta_f^p} = \gamma_{IR}$, namely
 $\gamma_{IR} = 0.40$, are consistent with this range of lattice measurements
 and are somewhat higher than our five-loop value $\gamma_{IR,5\ell}=0.255$
 from the conventional $\alpha$ series that we obtained in \cite{flir}. It is
 hoped that further work by lattice groups will lead to a consensus concerning
 whether this theory is IR conformal or not and concerning the value of
 $\gamma_{IR}$. 

 The SU(3) theory with $N_f=10$ has been investigated on the lattice in
 \cite{lsdnf10}, with the result $\gamma_{IR} \sim 1$.  While our
 highest-order $n$-loop values, namely our four-loop result,
 $\gamma_{IR,4\ell}=0.156$ \cite{bvh}, and our five-loop result,
 $\gamma_{IR,5\ell}=0.211$ obtained using Pad\'e methods \cite{flir}, are
 smaller than this lattice value, our extrapolated scheme-independent value,
 $\gamma_{IR} = 0.95 \pm 0.06$ \cite{gsi}, is consistent with it.

 There have also been a number of lattice studies of the SU(3) theory with
 $N_f=8$ \cite{latkminf8,lsdnf8,appelquist_eft}, which have yielded the
 estimate $\gamma_{IR} \simeq 1$.  As is evident from
 Fig. \ref{gammaNc3fund_plot}, if we were to continue the curve for
 $\gamma_{IR,\Delta_f^4}$ plotted there downward further to $N_f=8$, the
 resultant value would be compatible with $\gamma_{IR} \sim 1$.  We note that
 this theory may well be in the chirally broken phase, and there is not yet a
 clear consensus as to whether it is in this phase or possibly near the lower
 end of the IR-conformal non-Abelian Coulomb phase.  In this context, one may
 recall that if, for a given $G$ and $R$, $N_f < N_{f,cr}$, so that there is
 spontaneous chiral symmetry breaking, then the IR zero of the beta function is
 only approximate, since the theory flows away from this value as the fermions
 gain dynamical mass and are integrated out, leaving a pure gluonic low-energy
 effective field theory. For such a theory, the quantity extracted from either
 continuum or lattice analyses as $\gamma_{IR}$ is only an effective anomalous
 dimension that describes the renormalization-group behavior as the theory is
 flowing near to the approximate zero of the beta function.  A general comment
 is that the determination of $N_{f,cr}$ relies upon effective methods to
 analyze the lattice data \cite{lgtreviews}; progress on this continues
 \cite{lsd}-\cite{fleming_eft}.

 Theories with an SU(2) gauge group and $N_f=8$ have been of interest in the
 context of certain ideas for physics beyond the Standard Model (SM)
 \cite{ckm}, in which the number of Dirac fermions is $N_f=N_{wk}(N_c+1)=8$,
 where $N_{wk}=2$, corresponding to the SU(2) factor group in the SM and
 $N_c=3$ colors. There have been several lattice of this SU(2) theory with
 $N_f=8$, including \cite{su2nf8,lgtreviews,tuominen2017}.  These are
 consistent with this theory being IR-conformal, and the recent study
 \cite{tuominen2017} has reported the measurement $\gamma_{IR}=0.15 \pm
 0.02$. For comparison, as listed in Table \ref{gamma_values}, our previous
 higher $n$-loop values were $\gamma_{IR,3\ell}=0.272$ and
 $\gamma_{IR,4\ell}=0.204$ \cite{bvh}, and our current highest-order
 scheme-independent value is $\gamma_{IR,\Delta_f^4}=0.298$.  These are
 somewhat higher than this lattice result.

There have also been a number of lattice studies of the SU(2) theory
with $N_f=6$ \cite{su2nf6,yamada_su2nf6,tuominen_su2nf6,lgtreviews}. 
From this work, it is not yet clear if
this theory is IR-conformal or chirally broken.  Ref. 
\cite{yamada_su2nf6} obtained the range $0.26 < \gamma_{IR} < 0.74$, while
Ref. \cite{tuominen_su2nf6} found $\gamma_{IR} \simeq 0.275$.  Our higher-order
scheme-independent values, as listed in Table \ref{gamma_values}, in
particular, $\gamma_{IR,\Delta_f^4}=0.698$, are in agreement with the range 
given in \cite{yamada_su2nf6} and are somewhat higher than the value from
\cite{tuominen_su2nf6}. 

% =====================================================================

\subsection{LNN Limit for $G={\rm SU}(N_c)$ and $R=F$}
\label{gamma_lnn_section}

For $G={\rm SU}(N_c)$ and $R=F$, it is of interest to consider the limit
\beqs
& & LNN: \quad N_c \to \infty \ , \quad N_f \to \infty \cr\cr
& & {\rm with} \ r \equiv \frac{N_f}{N_c} \ {\rm fixed \ and \ finite}  \cr\cr
& & {\rm and} \ \ \xi(\mu) \equiv \alpha(\mu) N_c \ {\rm is \ a \ 
finite \ function \ of} \ \mu \ . 
\cr\cr
& &
\label{lnn}
\eeqs
We will use the symbol $\lim_{LNN}$ for this limit, where ``LNN'' stands
for ``large $N_c$ and $N_f$'' with the constraints in Eq. (\ref{lnn})
imposed.  This is also called the 't Hooft-Veneziano limit.  Anticipating our
later discussion of theories with fermions in two-index representations
(adjoint and symmetric and antisymmetric rank-2 tensor), we will use the symbol
$\lim_{LN}$, where ``LN'' stands for ``large $N_c$'', to denote the original 't
Hooft limit 
\beqs
&& LN: \quad N_c \to \infty \cr\cr
&& {\rm with} \ \xi(\mu) \equiv \alpha(\mu) N_c \ {\rm a \ 
finite \ function \ of} \ \mu \cr\cr
&&
\label{ln}
\eeqs
and $N_f$ fixed and finite.  

Continuing our discussion of the LNN limit, as relevant to theories with 
fermions in the fundamental represention, we define the following quantities in
this limit: 
\beq
\xi = 4\pi x = \lim_{LNN} \alpha N_c \ , 
\label{xlnn}
\eeq
\beq
r_u = \lim_{LNN} \frac{N_u}{N_c} \ , 
\label{rb1zdef}
\eeq
and
\beq
r_\ell = \lim_{LNN} \frac{N_\ell}{N_c} \ , 
\label{rb2zdef}
\eeq
with values 
\beq
r_\ell = \frac{11}{2} =5.5
\label{rb1z}
\eeq
and
\beq
r_\ell = \frac{34}{13}=2.615  \ . 
\label{rb2z}
\eeq
(to the indicated floating-point accuracy).  
With $I_{IRZ}: \ N_\ell < N_f < N_u$, it follows that the corresponding 
interval in the ratio $r$ is 
\beq
I_{IRZ,r}: \quad \frac{34}{13} < r < \frac{11}{2}, \ i.e., 
\ 2.615 < r < 5.5 
\label{intervalr}
\eeq
The critical value of $r$ such that for $r > r_{cr}$, the
LNN theory is IR-conformal and for $r< r_{cr}$, it exhibits spontaneous chiral
symmetry breaking is denoted $r_{cr}$ and is defined as 
\beq
r_{cr} = \lim_{LNN} \frac{N_{f,cr}}{N_c} \ . 
\label{rcr}
\eeq
We define the scaled scheme-independent expansion parameter for the LNN limit

\beq
\Delta_r \equiv \frac{\Delta_f}{N_c} = r_u-r = \frac{11}{2}-r \ . 
\label{deltar}
\eeq
As $r$ decreases from $r_u$ to $r_\ell$ in the interval $I_{IRZ,r}$,
$\Delta_r$ increases from 0 to a maximal value 
\beqs
& & (\Delta_r)_{max} = r_u-r_\ell = \frac{75}{26} = 2.8846 \quad
{\rm for} \ r \in I_{IRZ,r} \ . \cr\cr
& & 
\label{Deltar_max_irz}
\eeqs

We define rescaled coefficients $\hat \kappa_{j,F}$
\beq
\hat \kappa_{j,F} \equiv \lim_{N_c \to \infty} N_c^j \, \kappa_{j,F}
\label{kappahatn}
\eeq
that are finite in this LNN limit. The anomalous dimension
$\gamma_{IR}$ is also finite in this limit and is given by
\beq
R=F: \quad \lim_{LNN} \gamma_{IR} = \sum_{j=1}^\infty \kappa_{j,F} \Delta_f^j 
= \sum_{j=1}^\infty \hat \kappa_{j,F} \Delta_r^j \ . 
\label{gamma_ir_lnn}
\eeq

From the results for $\kappa_j$, $j=1, \ 2, \ 3$ in \cite{gtr} or the special
cases given above for $G={\rm SU}(N_c)$ and $R=F$ in
Eqs. (\ref{kappa1f})-(\ref{kappa3f}), we have 
\beq
\hat\kappa_{1,F} = \frac{2^2}{5^2} = 0.1600 \ , 
\label{kappahat1}
\eeq
\beq
\hat\kappa_{2,F} = \frac{588}{5^6} = 0.037632 \ , 
\label{kappahat2}
\eeq
and
\beq
\hat\kappa_{3,F} = 
\frac{2193944}{3^3 \cdot 5^{10}} = 0.83207 \times 10^{-2} \ , 
\label{kappahat3}
\eeq
where, as above, we indicate the factorizations of the denominators. 
(The numerators do not, in general, have such simple factorizations; for
example, in $\kappa_{3,F}$, $2193944=2^3 \cdot 274243$.) 
From our new expression for $\kappa_4$, we calculate
\beqs
\hat\kappa_{4,F} & = & \frac{210676352}{3^4 \cdot 5^{13}} 
+ \frac{90112}{3^3 \cdot 5^{10}} \zeta_3 
+ \frac{11264}{3^3 \cdot 5^8} \zeta_5 \cr\cr
& = & 0.36489 \times 10^{-2} \ .
\label{kappahat4}
\eeqs
Hence, numerically, to order $O(\Delta_r^4)$, 
\beqs
R=F: && \ \  \gamma_{IR,LNN,\Delta_r^4} =  
\Delta_r \Big [ 0.160000 + 0.037632 \Delta_r \cr\cr
&+&0.0083207 \Delta_r^2 + 0.003649 \Delta_r^3 \ \Big ] \ . 
\label{gamma_ir_lnn_Delta3}
\eeqs

Using these results for $\gamma_{IR,F,\Delta_r^p}$ with $1 \le p \le 4$ for 
$R=F$ in the LNN limit, we can now carry out a polynomial
extrapolation to $p=\infty$.  To do this, we fit an expression for
$\gamma_{IR,F,\Delta_r^p}$ with some subset of the $p$ terms to a
polynomial in $1/p$. We denote the resultant value generically as 
$\gamma_{IR,F,s}$, where here $s$ denotes the subset of the $p$ terms used for
the extrapolation.  We shall use, as a necessary condition for
$\gamma_{IR,F,s}$ to be reliable, the requirement that it not differ
too much from the highest-order value, $\gamma_{IR,F,\Delta_r^4}$.
Quantitatively, we require that for the given subset $s$, 
$\gamma_{IR,F,s}/\gamma_{IR,F,\Delta_r^4} < 1.5$. We find 
that this condition is satisfied if $r \in I_{IRZ,r}$ is $r \gsim 3.5$, but
that it is not satisfied as $r$ decreases below this value toward the lower end
of the interval $I_{IRZ,r}$ at $r_\ell=2.615$.  As an example, at $r=4.0$, 
depending on the subset of terms used for the extrapolation, we obtain 
$\gamma_{IR,F,s}/\gamma_{IR,F,\Delta_r^4} \simeq 1.2$, while at $r=3.6$, this
ratio increases to $\simeq 1.4$.  We remark that the value $r=4.0$
corresponds to $N_f=12$ for the SU(3) theory and $N_f=8$ for the SU(2) theory.

Previously, in \cite{gsi} we performed this analysis for the special case
$G={\rm SU}(3)$ and $R=F$ and, for that work, we studied how the extrapolated
value depends on the subset of terms that one includes for the fit.  We perform
the corresponding analysis here for this LNN case.  We study three sets of
terms:
\beq
{\rm set}_{34}: \quad \{ 
  \gamma_{IR,F,\Delta_r^3}, \ 
  \gamma_{IR,F,\Delta_r^4} \}
\label{gamma_set34}
\eeq
\beq
{\rm set}_{234}: \quad 
\{ \gamma_{IR,F,\Delta_r^2}, \ 
   \gamma_{IR,F,\Delta_r^3}, \ 
   \gamma_{IR,F,\Delta_r^4} \}
\label{gamma_set234}
\eeq
\beq
{\rm set}_{1234}: \quad 
\{ \gamma_{IR,F,\Delta_r}, \ 
   \gamma_{IR,F,\Delta_r^2}, \ 
   \gamma_{IR,F,\Delta_r^3}, \ 
   \gamma_{IR,F,\Delta_r^4} \}
\label{gamma_set1234}
\eeq
There are countervailing advantages of these sets of terms.  The two-term set
(\ref{gamma_set34}) has the advantage of using the two highest-order terms,
while the three-term and four-term sets have the advantage of using
progressively more terms in the fit.  The fits to the sets
(\ref{gamma_set34})-(\ref{gamma_set1234}) yield polynomials in the variable
$p^{-1}$ of the respective forms
\beq
{\rm set}_{34} \ \Rightarrow \ 
\gamma_{IR,F,ex34,p} = s_{34,0} + s_{34,1} p^{-1}
\label{gamma_pinv_ex34}
\eeq
\beqs
{\rm set}_{234} \ \Rightarrow \ 
 \gamma_{IR,F,ex234,p} & = & s_{234,0} + s_{234,1} p^{-1} + s_{234,2}p^{-2} 
\cr\cr
& & 
\label{gamma_pinv_ex234}
\eeqs
and
\beqs
{\rm set}_{1234} \ & \Rightarrow & \ 
\gamma_{IR,F,ex1234,p} = s_{1234,0} + s_{1234,1} p^{-1} \cr\cr
 & + & s_{1234,2}p^{-2} + s_{1234,3}p^{-3} \ . 
\label{gamma_pinv_ex1234}
\eeqs
The extrapolated values in the limit $p \to \infty$ given by these 
fits are, respectively, 
as
\beq
\lim_{p \to \infty} \gamma_{IR,F,ex34,p} = s_{34,0} \equiv \gamma_{IR,F,ex34}
\label{gamma_ex34}
\eeq
\beq
\lim_{p \to \infty}\gamma_{IR,F,ex234,p}=s_{234,0} \equiv \gamma_{IR,F,ex234}
\label{gamma_ex234}
\eeq
and
\beq
\lim_{p \to \infty} \gamma_{IR,F,ex1234,p} = s_{1234,0} \ . 
\equiv \gamma_{IR,F,ex1234}
\label{gamma_ex1234}
\eeq
We have calculated these quantities analytically.  Below, we list the
corresponding expressions with coefficients given to the indicated 
floating-point precision: 
\beqs
\gamma_{IR,F,ex34} & = & 16.758754-11.042531r+2.8240528r^2 \cr\cr
&-& 0.32942724r^3 + 0.014595750r^4
\label{gamma_ex34_num}
\eeqs
\beqs
\gamma_{IR,F,ex234} & = &27.346053-19.2457889r+5.1985972r^2 \cr\cr
&-&0.63389228 r^3 + 0.0291915006r^4
\label{gamma_ex234_num}
\eeqs
and
\beqs
\gamma_{IR,F,ex1234} & = &33.901799-24.4060664r+6.71925275r^2 \cr\cr
&-&0.832708600 r^3 + 0.038922001 r^4 \ . 
\label{gamma_ex1234_num}
\eeqs
Note that there are strong cancellations between individual terms for relevant
values of $r \in I_{IRZ,r}$.  Some examples will show the range of resultant
values of extrapolations for these different choices of sets of terms used in
the fits.  As anticipated, for values of $r$ in the upper part of the interval
$I_{IRZ,r}$, all of the different types of extrapolation give quite similar 
results.  For example,
\beqs
r=5.0 \ &\Longrightarrow& \ 
\gamma_{IR,F,ex,34}=0.0914, \quad 
\gamma_{IR,F,ex234}=0.0902, \cr\cr 
& & \gamma_{IR,F,ex1234}=0.0905 \ . 
\label{r5_extrap}
\eeqs
As $r$ decreases in the interval $I_{IRZ,r}$, the differences between the
extrapolations using the different sets of terms increase slightly, e.g., for a
value roughly in the middle of this interval, namely $r=4.0$, we find 
\beqs
r=4.0 \ &\Longrightarrow& \ 
\gamma_{IR,F,ex34}=0.427, \quad 
\gamma_{IR,F,ex234}=0.444, \cr\cr
& & \gamma_{IR,F,ex1234}=0.456 \ . 
\label{r4_extrap}
\eeqs
Toward the lower part of the interval $I_{IRZ,r}$, these differences increase
further, but also, as discussed above, for a given $r$, all of the different
types of extrapolations involve greater uncertainties, since each of the
extrapolated values differs more from the value of highest-order explicitly
calculated quantity, $\gamma_{IR,\Delta_r^4}$.  For example, for $r=3.0$, 
\beqs
r=3.0 \ &\Longrightarrow& \ 
\gamma_{IR,F,ex34}=1.335, \quad 
\gamma_{IR,F,ex234}=1.645, \cr\cr
& & \gamma_{IR,F,ex1234}=1.826 \ . 
\label{r3_extrap}
\eeqs
The ratios of these values divided by the highest-order explicitly calculated
value, $\gamma_{IR,F,\Delta_r^4}$, are
\beqs
r=3.0 \ &\Longrightarrow& \ 
\frac{\gamma_{IR,F,ex34  }}{\gamma_{IR,F,\Delta_r^4}}=1.47, \quad 
\frac{\gamma_{IR,F,ex234 }}{\gamma_{IR,F,\Delta_r^4}}=1.82 \cr\cr
&& \frac{\gamma_{IR,F,ex1234}}{\gamma_{IR,F,\Delta_r^4}}=2.01 \ . 
\label{r3_extrap_ratios}
\eeqs
Given our fiducial requirement that the ratio of the extrapolated value for 
$p \to \infty$ divided by the highest-order explicitly calculated value, 
should not be greater than 1.5 for the extrapolation to be considered
reasonably reliable, it follows that we would not consider the latter two
extrapolations in Eq. (\ref{r3_extrap}) to be sufficiently reliable to meet
this requirement.  

It is interesting to compare these scheme-independent calculations of
$\gamma_{IR,F,\Delta_r^p}$ to order $1 \le p \le 4$ with the results from the
conventional $n$-loop calculations as truncated expansions in
$\alpha_{IR,F,n\ell}$, denoted $\gamma_{IR,F,n\ell}$ from Table V of \cite{lnn}
up to $n=4$ loop order.  We list our scheme-independent values together with
these $n$-loop values in Table \ref{gamma_values_lnn}.  For each value of $r$,
we also include the extrapolated value, $\gamma_{IR,F,ex234}$ for the $p \to
\infty$ limit, and the ratio $\gamma_{IR,F,ex234}/\gamma_{IR,\Delta_r^4}$. We
do not include the results from the $n=5$ loop conventional calculation,
because of the absence of a physical IR zero in the five-loop beta function for
$2.615 < r < 4.323$ in $I_{IRZ,r}$.  Although the extrapolated values
$\gamma_{IR,F,ex234}$ for $r$ values below $r=3.5$ are included, we caution
that these do not satisfy our fiducial criterion for sufficient reliability of
extrapolation, since they differ by too much from our highest-order calculated
values, $\gamma_{IR,\Delta_r^4}$. For this reason, although we can roughly
apply the method discussed in Section \ref{methods_section} to use the
extrapolated value of $\gamma_{IR}$ to estimate the lower end, $r_{cr}$, of the
IR-conformal non-Abelian Coulomb phase (defined in Eq. (\ref{rc})), this
involves a substantial degree of uncertainty.  Bearing this caveat in mind,
the resulting estimate would be that $r_{cr} \sim 2.7$. 
If one were to pull back from the LNN limit and multiply this value of $r_{cr}$
by a specific finite value of $N_c$ to get an estimate of the corresponding
$N_{f,cr}$, then, for example, for $N_c=3$, i.e., $G={\rm SU}(3)$, this would
yield $N_{f,cr} \sim 8$.  This estimate is consistent with the estimate $8
\lsim N_{f,cr} \lsim 9$ that we derived from our calculation of
$\gamma_{IR,F,\Delta_f^4}$ for this theory and extrapolation to obtain $\lim_{p
  \to \infty} \gamma_{IR,F,\Delta_f^p}$ in \cite{gsi}.  Clearly, the lower that
one goes in $N_c$ away from the LNN limit, the greater is the error in
performing this conversion from a specific $r$ value in the LNN limit to a
corresponding ratio $N_f/N_c$ with finite $N_f$ and $N_c$, so we do not perform
this conversion for $N_c=2$.

\begin{figure}
  \begin{center}
    \includegraphics[height=6cm]{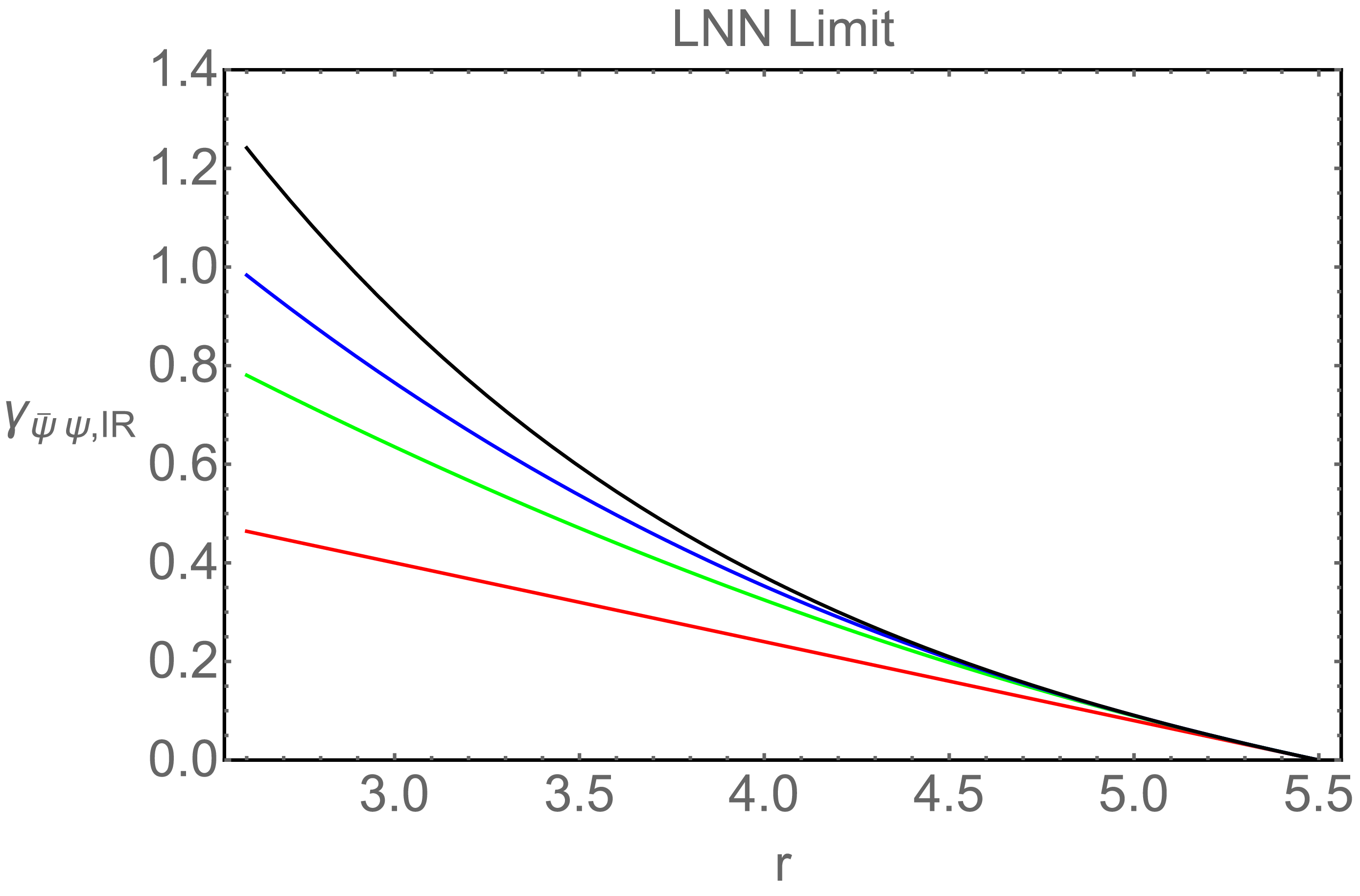}
  \end{center}
\caption{Plot of $\gamma_{IR,F,\Delta_r^p}$ for $1 \le p \le 4$ as a
  function of $r \in I_{IRZ,r}$ in the LNN limit (\ref{lnn}).  
 From bottom to top, the curves (with colors online)
 refer to  $\gamma_{IR,F,\Delta_r}$ (red),
           $\gamma_{IR,F,\Delta_r^2}$ (green)
           $\gamma_{IR,F,\Delta_r^3}$ (blue)
           $\gamma_{IR,F,\Delta_r^4}$ (black).}
\label{gamma_p4_lnn_plot}
\end{figure}

In Fig. \ref{gamma_p4_lnn_plot} we plot $\gamma_{IR,F,\Delta_r^p}$, i.e., the
value of $\gamma_{IR}$ for $R=F$, calculated to order $\Delta_r^p$ with $1 \le
p \le 4$, in the scheme-independent expansion, as a function of $r \in
I_{IRZ,r}$. As a consequence of the positivity of the $\hat \kappa_{p,F}$ in
Eqs. (\ref{kappahat1})-(\ref{kappahat3}), for a fixed $r$,
$\gamma_{IR,F,\Delta_r^p}$ is a monotonically increasing function of the order
of calculation, $p$. As $r$ decreases toward the lower end of the interval
$I_{IRZ,r}$ at $r=r_\ell=2.615$, the value of $\gamma_{IR}$ calculated
to the highest order in this LNN limit, namely $O(\Delta_r^4)$, is slightly
greater than 1.

As we did for specific SU($N_c$) theories above, here we proceed to investigate
the range of applicability of the scheme-independent series expansion for
$\gamma_{IR}$ in the LNN limit.  As is evident from Table
\ref{gamma_values_lnn}, all of our values of $\gamma_{IR,F,\Delta_r^p}$ for $1
\le p \le 4$ satisfy the bound $\gamma_{IR} \le 2$.  This is also true for all
of our extrapolated values, $\gamma_{IR,F,ex234}$, except for the lowest value
of $r$ listed, namely $r=2.8$, for which $\gamma_{IR,F,ex234}=2.09$, slightly
above this bound.  Thus, these results in the LNN limit again demonstrate the
advantage of the scheme-independent expansions, since they enable us to
calculate self-consistent values of $\gamma_{IR,F,\Delta_r}$ over a greater
range of the interval $I_{IRZ,r}$ than is the case with the conventional
$n$-loop calculations.  To show the latter in detail, we have explicitly listed
the values of $\gamma_{IR,F,3\ell}$ and $\gamma_{IR,F,4\ell}$ for values of $r$
where $\gamma_{IR,F,2\ell}$ was unphysically large.  

To investigate the range of applicability of the scheme-independent expansions
further, it is worthwhile to obtain an estimate of this range from ratios of
successive coefficients.  From the coefficients $\hat\kappa_{j,F}$ that we have
calculated with $1 \le n \le 3$, we compute the ratios
\beq
\frac{\hat\kappa_{1,F}}{\hat\kappa_{2,F}} = 4.252 
\label{kappa1hat_over_kappa2hat}
\eeq
\beq
\frac{\hat\kappa_{2,F}}{\hat\kappa_{3,F}} = 4.523 
\label{kappa2hat_over_kappa3hat}
\eeq
and
\beq
\frac{\hat\kappa_{3,F}}{\hat\kappa_{4,F}} = 2.280 \ . 
\label{kappa3hat_over_kappa4hat}
\eeq
Recalling that the maximal value of $\Delta_r$ in the interval $I_{IRZ,r}$ is
2.885 (Eq. (\ref{Deltar_max_irz}), these ratios are consistent with the
inference that the small-$\Delta_r$ series expansion may be reasonably accurate
throughout most of this interval $I_{IRZ,r}$.  

% ======================================================================

\subsection{$\gamma_{\bar\psi\psi,IR,\Delta_f^4}$ for 
$G={\rm SU}(N_c)$ and $R=adj$}
\label{kappa_adj_section}

Here we present our results for the $\kappa_j$ coefficients and thus
$\gamma_{\bar\psi\psi,IR,\Delta_f^j}$ with $1 \le j \le 4$ for $G={\rm
  SU}(N_c)$ and $N_f$ fermions in the adjoint representation, $R=adj$. We will
usually denote these as $\kappa_{j,adj}$ and
$\gamma_{\bar\psi\psi,IR,adj,\Delta_f^j}$ but sometimes, when no confusion will
result, we will omit this $adj$ subscript for brevity of notation.

In this theory, Eqs. (\ref{nfb2z}) and (\ref{nfb2z})
yield, for the upper and lower ends of the interval $I_{IRZ}$, the values 
\beq
N_{u,adj} = \frac{11}{4} =2.75 
\label{nfb1z_adj}
\eeq
and
\beq
N_{\ell,adj}=\frac{17}{16} = 1.0625 \ , 
\label{nfb2z_adj}
\eeq
so this interval includes only one integral value of $N_f$, namely $N_f=2$.  We
note that since the adjoint representation is self-conjugate, a theory with
$N_f$ Dirac fermions with $R=adj$ is equivalent to a theory with $N_{f,Maj} =
2N_f$ Majorana fermions.  Hence, here, one may also allow the half-integral
values $N_f=3/2, \ 5/2$ corresponding to $N_{f,Maj}=3, \ 5$.  We have
\beq
R=adj: \quad \Delta_f = N_u - N_f = \frac{11}{4} - N_f \ . 
\label{Delta_adj}
\eeq
For this case, the factor $D$ in Eq. (\ref{d}) is simply $D=18$. 
In \cite{dex} we gave the coefficients $\kappa_{j,adj}$ for 
$1 \le n \le 3$.  These are as follows:
\beq
\kappa_{1,adj}= \bigg ( \frac{2}{3} \bigg )^2 = 0.44444 \ , 
\label{kappa1_adj}
\eeq
\beq
\kappa_{2,adj}= \frac{341}{2 \cdot 3^6} = 0.23388 \ ,  
\label{kappa2_adj}
\eeq
and
\beqs
\kappa_{3,adj} & = & 
\frac{61873}{2^3 \cdot 3^{10}} - \frac{592}{3^8 N_c^2} \cr\cr
& = & 0.130978 - 0.090230N_c^{-2}  \ , 
\label{kappa3_adj}
\eeqs
where, as before, we indicate the simple factorizations of the denominators.
The coefficient $\kappa_{4,adj}$ is
\beqs
\kappa_{4,adj} & = & \frac{53389393}{2^7 \cdot 3^{14}} +
\frac{368}{3^{10}}\zeta_3 \cr\cr
& + & 
\bigg ( -\frac{2170}{3^{10}} + \frac{33952}{3^{11}}\zeta_3 \bigg ) N_c^{-2}
\cr\cr
& = & 0.0946976 + 0.193637N_c^{-2} \ . 
\label{kappa4_adj}
\eeqs
The coefficients $\kappa_{1,adj}$ and $\kappa_{2,adj}$ are manifestly positive,
and we find that for all physical $N_c$, the coefficients $\kappa_{3,adj}$ and
$\kappa_{4,adj}$ are also positive.  Although $\kappa_{1,adj}$ and
$\kappa_{2,adj}$ are independent of $N_c$, the coefficients $\kappa_{j,adj}$
for $j=3, \ 4$ do depend on $N_c$.  We find that $\kappa_{3,adj}$ and
$\kappa_{4,adj}$ are, respectively, monotonically increasing and monotonically
decreasing functions of $N_c$. The $N_c \to \infty$ limits of $\kappa_{3,adj}$
and $\kappa_{4,adj}$ are given by the respective first terms in
Eqs. (\ref{kappa3_adj}) and (\ref{kappa4_adj}).

Thus, to order $\Delta_f^4$, we have 
\begin{widetext}
\beq
\gamma_{\bar\psi\psi,IR,adj,\Delta_f^4} = \Delta_f 
\bigg [ 0.44444 + 0.23388 \Delta_f 
+ (0.13098 - 0.090230N_c^{-2})\Delta_f^2 
+ (0.094698 + 0.19364N_c^{-2})\Delta_f^3 \ \bigg ] \ .
\label{gamma_ir_p4}
\eeq
\end{widetext}

\begin{figure}
  \begin{center}
    \includegraphics[height=6cm]{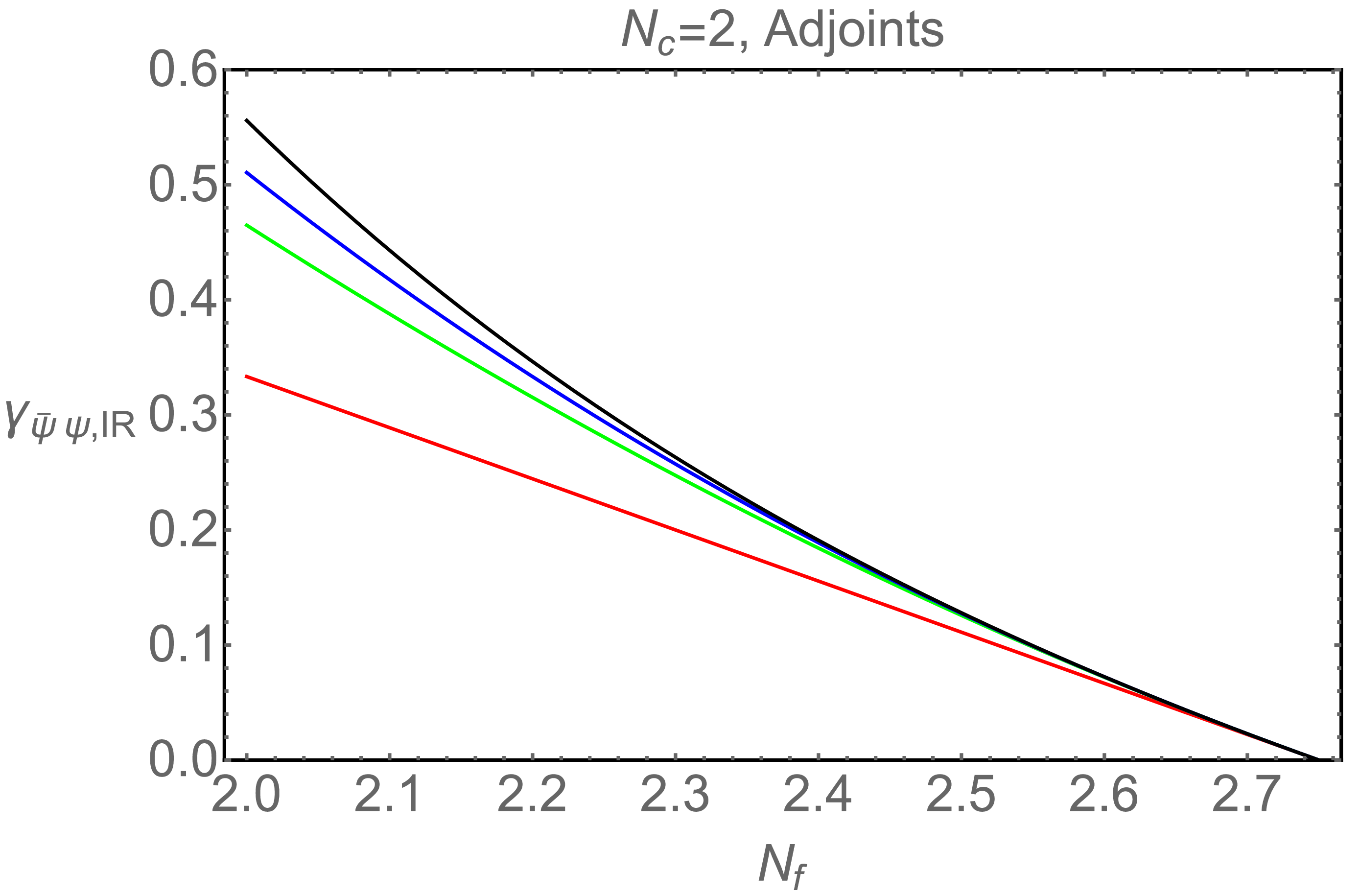}
  \end{center}
\caption{Plot of $\gamma_{\bar\psi\psi,IR,adj,\Delta_f^p}$ for $G={\rm SU}(2)$ 
and $1 \le p \le 4$ as a function of $N_f \in I_{IRZ}$ for $R=adj$ and 
$N_f=2$. From bottom to top, the curves (with colors online) refer to 
$\gamma_{IR,adj,\Delta_f}$ (red),                           
$\gamma_{IR,adj,\Delta_f^2}$ (green),
$\gamma_{IR,adj,\Delta_f^3}$ (blue), and
$\gamma_{IR,adj,\Delta_f^4}$ (black).}
\label{gammaNc2adj_plot}
\end{figure}

In Fig. \ref{gammaNc2adj_plot} we show 
$\gamma_{\bar\psi\psi,IR,adj,\Delta_f^p}$ with $1 \le p \le 4$ for the
SU(2) theory, as a function of $N_f$, formally generalized from the nonnegative
integers to the real numbers.  In Table \ref{gamma_ir_adj_values} we list
values of $\gamma_{\bar\psi\psi,IR,adj,\Delta_f^p}$ with $1 \le p \le 4$ 
for $N_f=2$ and $N_c=2$ and $N_c=3$. For comparison, we also include our 
$n$-loop values $\gamma_{\bar\psi\psi,IR,adj,n\ell}$ calculated in the
conventional manner via power series in the coupling (in the 
$\overline{\rm MS}$ scheme), from Table VIII of \cite{bvh}. 

Among SU($N_c$) theories with fermions in the adjoint representation, the SU(2)
theory with $N_f=2$ (Dirac) fermions has been of particular interest
\cite{sannino_su2adj}.  In the following, for notational brevity, the subscript
$adj$ is understood implicitly.  For this theory, as listed in Table
\ref{gamma_ir_adj_values} we obtain the values $\gamma_{IR,\Delta_f^2}=0.465$,
$\gamma_{IR,\Delta_f^3}=0.511$, and $\gamma_{IR,\Delta_f^4}=0.556$, which are
close to our earlier higher-order $n$-loop calculations in \cite{bvh}, namely
$\gamma_{IR,3\ell}=0.543$ and $\gamma_{IR,4\ell}=0.500$.  It is of interest to
compare these values with the results of lattice studies. There have been a
number of such studies, and these are consistent with the conclusion that this
theory is conformal in the infrared
\cite{catterall2010}-\cite{montvay},\cite{lgtreviews}.  These studies have
yielded a rather large range of measured values for $\gamma_{IR}$, including
the following (where the published estimated uncertainties in the last digits
are indicated in parentheses): $\gamma_{IR}=0.49(13)$ \cite{catterall2010},
$\gamma_{IR}=0.22(6)$ \cite{deldebbio2010}, $\gamma_{IR}=0.31(6)$
\cite{degrand2011}, $\gamma_{IR}=0.17(5)$ \cite{lsd2011}, $\gamma_{IR}=0.37(2)$
\cite{deldebbio2016}, $\gamma_{IR}=0.20(3)$ \cite{tuominen2016}, and
$\gamma_{IR}=0.50(26)$ \cite{giedt2016}. (See these references and
\cite{montvay} for additional discussion of estimates of overall
uncertainties.)  Our scheme-independent calculation of $\gamma_{IR}$ to
$O(\Delta_f^4)$ and our earlier $n$-loop calculations of $\gamma_{IR,n\ell}$ up
to $n=4$ loops are clearly consistent with the larger among these lattice
values.  Before carrying out a comparison of our results with the full set of
lattice values, it will be necessary to narrow the current wide range of
lattice measurements.  

It is of interest to investigate the $N_c \to \infty$ limit for an
SU($N_c$) gauge theory with fermions in the adjoint representation.   Since in
this case, the upper and lower ends of the interval $I_{IRZ}$, given by 
$N_u=11/4$ in Eq. (\ref{nfb1z_adj}) and $N_\ell=17/16$ in Eq. (\ref{nfb2z}) 
are independent of $N_c$, it follows that $\Delta_f$ is also independent of
$N_c$.  Hence, for $R=adj$, 
\beq
\lim_{LN} \gamma_{IR} = \sum_{j=1}^\infty \hat\kappa_{j,adj} \Delta_f^j
\label{gamma_ir_adj_ln}
\eeq
where 
\beq
\hat \kappa_{j,adj} = \lim_{LN} \kappa_{j,adj} \ . 
\label{kappahat_adj}
\eeq
The values of $\hat\kappa_{j,adj}$ are evident from the full expressions 
for $\kappa_{j,adj}$ that we have given above in Eqs. 
(\ref{kappa1_adj})-(\ref{kappa4_adj}); for example, 
$\hat\kappa_{3,adj}=61873/(2^3 \cdot 3^{10})$.

% =====================================================================

\subsection{$\gamma_{\bar\psi\psi,IR,\Delta_f^4}$ for 
$G={\rm SU}(N_c)$ and $R=S_2, \ A_2$}
\label{kappa_tensor_section}

Here we present our results for the $\kappa_j$ coefficients and thus
$\gamma_{\bar\psi\psi,IR,\Delta_f^j}$ with $1 \le j \le 4$ for $G={\rm
  SU}(N_c)$ and $N_f$ fermions in the symmetric and antisymmetric rank-2 tensor
representations of SU($N_c$), $S_2$ and $A_2$. Since many formulas for these
two cases are simply related to each other by sign reversals in certain terms,
it is convenient to treat these cases together.  As before \cite{bvh}, we shall
use the symbol $T_2$ (rank-2 tensor) to refer to these cases together. (Do not
confuse this use of $T$ with our use of the symbol $T$ in Section VII of
Ref. \cite{dex} for the anomalous dimension of the operators $\bar\psi
\sigma_{\mu\nu} \psi$ and operators $\bar\psi T_a \sigma_{\mu\nu} \psi$, where
it referred to the antisymmetric Dirac tensor $\sigma_{\mu\nu} =
(i/2)[\gamma_\mu, \gamma_\nu]$.) 

The values of $N_u$ and $N_\ell$ for $R=T_2$ are \cite{bvh}
\beq
N_{u,T_2} = \frac{11N_c}{2(N_c \pm 2)} 
\label{nfb1z_t2}
\eeq
and
\beq
N_{\ell,T_2} = \frac{17N_c^3}{(N_c \pm 2)(8N_c^2 \pm 3N_c - 6)} \ , 
\label{nfb2z_t2}
\eeq
so that
\beq
R=T_2: \quad \Delta_f = \frac{11N_c}{2(N_c \pm 2)} - N_f \ . 
\label{Deltaf_tensor}
\eeq
The factor $D$ in Eq. (\ref{d}) takes the explicit form
\beq
R=T_2: \quad D = \frac{18N_c^2 \pm 11N_c-22}{N_c} \equiv \frac{F_\pm}{N_c}
\label{dtensor}
\eeq
whence 
\beq
F_\pm = 18N_c^2 \pm 11N_c-22 \ . 
\label{fplusminus}
\eeq
Both $F_+$ and $F_-$ are positive-definite for the physical range $N_c \ge 2$.
At the lower end of the interval $I_{IRZ}$, $\Delta_f$ takes on the maximum 
value
\beq
R=T_2: \quad (\Delta_f)_{max} = \frac{3N_c F_\pm}
{2(N_c \pm 2)(8N_c^2 \pm 3N_c-6)} \ . 
\label{Deltaf_ir_max_tensor}
\eeq

If $N_c=2$, then $S_2$ is the same as the adjoint representation, so we focus
on $N_c \ge 3$ here. For this $R=S_2$ theory, the illustrative values $N_c=3$
and $N_c=4$ yield the respective intervals $I_{IRZ}$ $1.22 < N_f < 3.30$ and
$1.35 < N_f < 3.67$.  Hence, the physical integral values of $N_f$ in these
respective intervals $I_{IRZ}$ are $N_f=2, \ 3$ for both $N_c=3$ and
$N_c=4$. Furthermore, the $A_2$ representation is the singlet if $N_c=2$ and is
the same as the conjugate fundamental, $\bar F$ if $N_c=3$, so in the case of
$A_2$, we restrict to $N_c \ge 3$ and focus mainly on $N_c \ge 4$.  In the
SU(4) theory with $R=A_2$, the interval $I_{IRZ}$ is $4.945 < N_f < 11$,
including the integral values $5 \le N_f \le 10$.

Here, using our general results (\ref{kappa1})-(\ref{kappa4}), we give 
explicit expressions for the $\kappa_j$ with $1 \le j \le 4$ for the case 
$G={\rm SU}(N_c)$ and fermion representation $R=T_2$. 
From the general expressions for $\kappa_j$ with $1 \le j \le
4$, Eqs. (\ref{kappa1})-(\ref{kappa4}), we calculate the following. In each
expression, the $+$ and $-$ signs refer to the $S_2$ and $A_2$ special cases of
$T_2$, respectively:
\beq
\kappa_{1,T_2}=\frac{4(N_c \mp 1)(N_c \pm 2)^2}{N_cF_\pm} 
\label{kappa1_tensor}
\eeq
\begin{widetext}
\beq
\kappa_{2,T_2} = \frac{(N_c \mp 1)(N_c \pm 2)^3
(11N_c^2 \pm 4N_c-8)(93N_c^2 \pm 88N_c-176)}{3N_c^2F_\pm^3} 
\label{kappa2_tensor}
\eeq
\beqs
\kappa_{3,T_2} &=& 
\frac{(N_c \mp 1)(N_c \pm 2)^3}{2 \cdot 3^3 N_c^3 F_\pm^5} 
\bigg [ \Big ( 1670571N_c^9 \pm 7671402N_c^8 + 2181584N_c^7
\mp 25294256N_c^6 \cr\cr
& - & 13413856N_c^5 \pm 17539136N_c^4 + 16707328N_c^3
\mp 3046912N_c^2 -27320832N_c \pm 18213888 \Big ) \cr\cr
& \pm & 8448N_c^2(N_c \mp 2)F_\pm (3N_c^3 \pm 28N_c^2 \mp 176)\zeta_3 \
\bigg ] 
\label{kappa3_tensor}
\eeqs
and
\beqs
\kappa_{4,T_2} &=& 
\frac{(N_c \mp 1)(N_c \pm 2)^4}{2^4 \cdot 3^4 N_c^4 F_\pm^7}
\bigg [ \Big ( 4324540833 N_c^{13} \pm 26924228982N_c^{12}
+30086550336N_c^{11} \mp 106026091536N_c^{10} \cr\cr
&-&224952825968N_c^9
\pm 105492861344N_c^8 +600583055488N_c^7 \pm 45292329216N_c^6
-1067559840512N_c^5
\cr\cr
&\pm & 68261028352N_c^4 +982655860736N_c^3
\mp 385868775424N_c^2-136076328960N_c\pm 54430531584 \Big ) \cr\cr
&+&
2^9 F_\pm \Big ( 33534N_c^{11}\pm 702000N_c^{10} + 4448403N_c^9
\mp 2216812N_c^8 - 38600660N_c^7 \pm 22594304N_c^6 \cr\cr
&+& 124680384N_c^5
\mp 82679040N_c^4 - 90554112N_c^3 \pm 64551168N_c^2 - 6690816N_c
\pm 3345408 \Big )\zeta_3 \cr\cr
&\mp & 563200 N_c^2(N_c \mp 2)F_\pm^2 \Big (15N_c^5 \pm 158N_c^4 + 
240N_c^3 \mp 912N_c^2 - 1056N_c \pm 2112 \Big )\zeta_5 \ \bigg ] \ . 
\label{kappa4_tensor}
\eeqs
\end{widetext}
We comment on some factors in these $\kappa_{j,T_2}$ expressions.  The property
that the $\kappa_{j,A_2}$ coefficients contain an overall factor of $(N_c-2)$
(possibly raised to a power higher than 1), and hence vanish for $N_c=2$, is a
consequence of the fact that for $N_c=2$, the $A_2$ representation is a
singlet, so for SU(2), fermions in the $A_2=$ singlet representation have no
gauge interactions and hence no anomalous dimensions.  Clearly, this property
holds in general; i.e., the coefficients $\kappa_{j,A_2}$ for all $j$ contain
an overall factor of $(N_c-2)$ (as well as possible additional factors of
$(N_c-2)$).

As noted above, if $N_c=2$, then the $S_2$ representation is the same as the
adjoint representation, so the coefficients must satisfy the equality
$\kappa_{j,S_2}=\kappa_{j,adj}$ for this SU(2) case, and we have checked that
they do.  Note that this equality requires (i) that the term proportional to
$\zeta_3$ in $\kappa_{3,S_2}$ must be absent if $N_c=2$, since $\kappa_{3,adj}$
does not contain any $\zeta_3$ term, and, indeed, this is accomplished by the
factor $(N_c-2)$ multiplying the $\zeta_3$ term in $\kappa_{3,S_2}$; and (ii)
the term proportional to $\zeta_5$ in $\kappa_{4,S_2}$ must be absent if
$N_c=2$, since $\kappa_{4,adj}$ does not contain any $\zeta_5$ term, and this
is accomplished by the factor $(N_c-2)$ multiplying this $\zeta_5$ term in
$\kappa_{4,S_2}$. Similarly, as we observed above, if $N_c=3$, then the $A_2$
representation is the same as the conjugate fundamental representation, $\bar
F$, so the coefficients must satisfy the equality $\kappa_{j,A_2}=\kappa_{j,F}$
for this SU(3) case, and we have checked that they do.

The resultant $\Delta_f$ expansions for 
$\gamma_{\bar\psi\psi,IR,S_2,\Delta_f^4}$ with $2 \le N_c \le 4$ are 
\begin{widetext}
\beq
{\rm SU}(2): \quad 
\gamma_{\bar\psi\psi,IR,S_2,\Delta_f^4} = 
\Delta_f \bigg [ 0.44444 + 0.23388\Delta_f 
+0.10842\Delta_f^2 +0.14311\Delta_f^3 \ \bigg ] 
\label{gamma_ir_sym_su2}
\eeq
\beq
{\rm SU}(3): \quad 
\gamma_{\bar\psi\psi,IR,S_2,\Delta_f^4} = 
\Delta_f \bigg [ 0.38536 + 0.17038\Delta_f 
+0.078062\Delta_f^2 +0.060081\Delta_f^3 \ \bigg ] 
\label{gamma_ir_sym_su3}
\eeq
and
\beq
{\rm SU}(4): \quad 
\gamma_{\bar\psi\psi,IR,S_2,\Delta_f^4} = 
\Delta_f \bigg [ 0.34839 + 0.13875\Delta_f 
+0.059680\Delta_f^2 +0.38102\Delta_f^3 \ \bigg ] \ . 
\label{gamma_ir_sym_su4}
\eeq

For $R=A_2$, we give illustrative results for the 
$\Delta_f$ expansion of $\gamma_{\bar\psi\psi,IR}$ for $N_c=4, \ 5$:
\beq
{\rm SU}(4): \quad 
\gamma_{\bar\psi\psi,IR,A_2,\Delta_f^4} = 
\Delta_f \bigg [ 0.090090 + 
(1.1114 \times 10^{-2})\Delta_f + 
(1.6013 \times 10^{-3})\Delta_f^2 +
(2.9668 \times 10^{-4})\Delta_f^3 \ \bigg ]
\label{gamma_ir_asym_su4}
\eeq
and
\beq
{\rm SU}(5): \quad 
\gamma_{\bar\psi\psi,IR,A_2,\Delta_f^4} = \Delta_f \bigg [ 0.11582 + 
(1.7570 \times 10^{-2})\Delta_f + 
(2.9243 \times 10^{-3})\Delta_f^2 +
(0.59791\times 10^{-3})\Delta_f^3 \ \bigg ] \ . 
\label{gamma_ir_asym_su5}
\eeq
\end{widetext}

\begin{figure}
  \begin{center}
    \includegraphics[height=6cm]{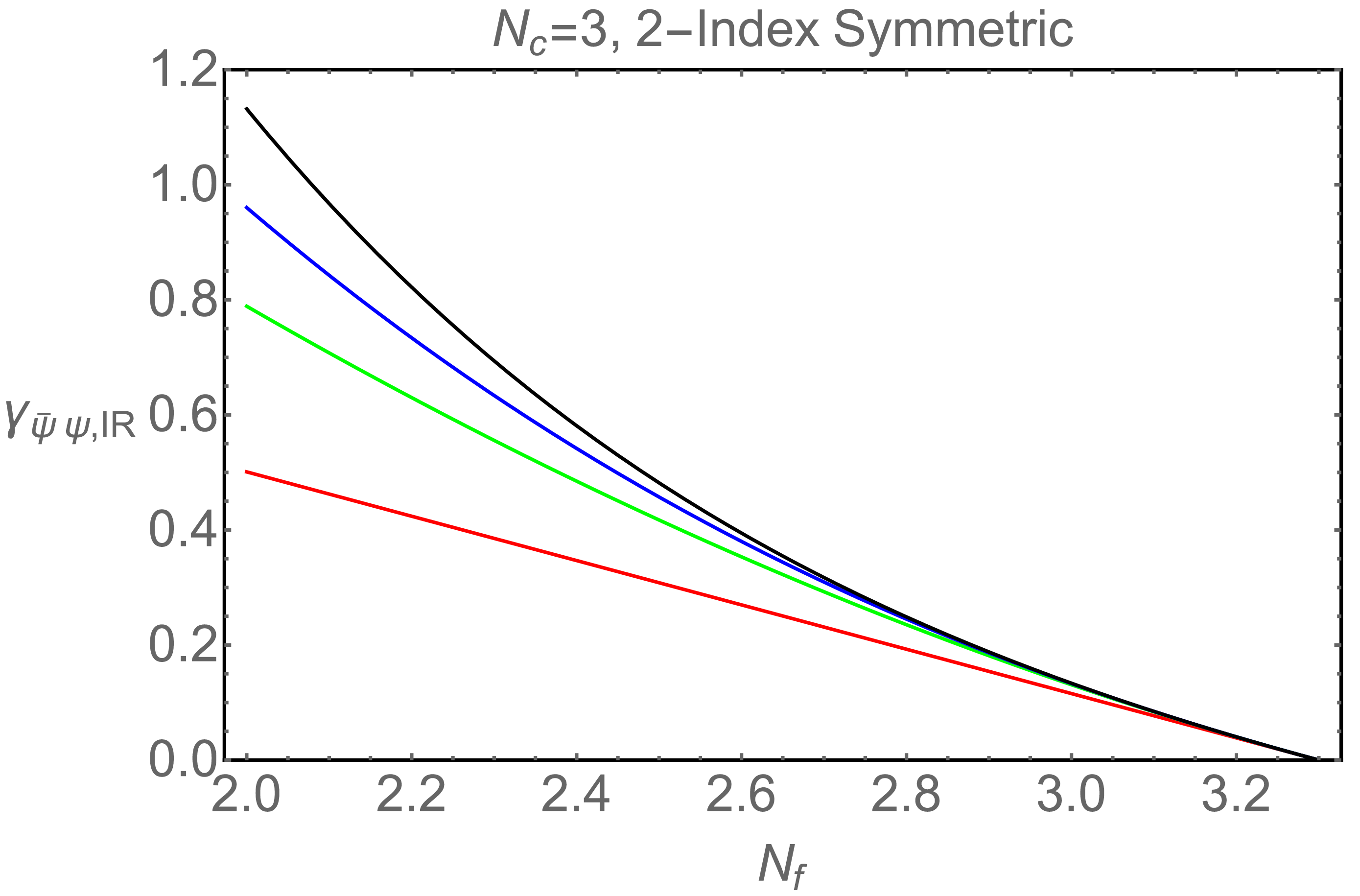}
  \end{center}
\caption{Plot of $\gamma_{\bar\psi\psi,IR,S_2,\Delta_f^p}$ for $N_c=3$
and $1 \le p \le 4$ as a function of $N_f$. Here, $S_2$ denotes
the symmetric rank-2 tensor representation. From bottom to top, 
the curves (with colors online) refer to 
$\gamma_{\bar\psi\psi,IR,S_2,\Delta_f}$ (red),                           
$\gamma_{\bar\psi\psi,IR,S_2,\Delta_f^2}$ (green),
$\gamma_{\bar\psi\psi,IR,S_2,\Delta_f^3}$ (blue), and
$\gamma_{\bar\psi\psi,IR,S_2,\Delta_f^4}$ (black).}
\label{gammaNc3sym_plot}
\end{figure}

In Fig. \ref{gammaNc3sym_plot} we present a plot of
$\gamma_{\bar\psi\psi,S_2,IR,\Delta_f^p}$ for $G={\rm SU}(3)$, $R=S_2$, and $1
\le p \le 4$, as a function of $N_f$. We list values of the
$\gamma_{IR,S_2,\Delta_f^p}$ with $1 \le p \le 4$ for the SU(3) and SU(4)
theories with $R=S_2$ in Table \ref{gamma_ir_sym_values}.  In both of these
theories, the interval $I_{IRZ}$ includes the two integer values $N_f=2, \ 3$.
For comparison, we also include the values $\gamma_{IR,S_2,n\ell}$ for $2 \le n
\le 4$ calculated via the conventional power series expansion to $n$-loop order
and evaluated at $\alpha=\alpha_{IR,n\ell}$ from Table XI in our previous work,
Ref. \cite{bvh}.  As is evident from this table, for a given $N_c$ and $N_f$,
there is reasonable agreement between the $n=4$ loop values
$\gamma_{IR,S_2,\Delta_f^4}$ and $\gamma_{IR,S_2,4\ell}$. For example, for
SU(3) and $N_f=2$, $\gamma_{IR,S_2,4\ell}=1.12$ while
$\gamma_{IR,S_2,\Delta_f^4}=1.13$.

We next compare our calculation of $\gamma_{\bar\psi\psi,IR,S_2,\Delta_f^p}$ to
order $p=4$ with lattice measurements.  A theory of particular interest is the
SU(3) gauge theory with $N_f=2$ flavors of fermions in the $S_2$
representation, and lattice studies of this theory include 
\cite{degrand_sextet} and \cite{kuti_sextet} (see also 
\cite{lgtreviews}).   As indicated in Table 
\ref{gamma_ir_sym_values}, our higher-order scheme-independent results are 
$\gamma_{IR,\Delta_f^3}=0.960$, and $\gamma_{IR,\Delta_f^4}=1.132$.
By comparison, our $n$-loop results from \cite{bvh} for this theory 
are $\gamma_{IR,3\ell}=0.500$ and $\gamma_{IR,4\ell}=0.470$.  The lattice study
\cite{degrand_sextet} concluded that this theory is IR-conformal and
obtained $\gamma_{IR} < 0.45$ \cite{degrand_sextet}, while Ref. 
\cite{kuti_sextet} concluded that it is not IR-conformal and got an 
effective $\gamma_{IR} \sim 1$ \cite{kuti_sextet}.  One hopes that further
work by lattice groups will lead to a consensus concerning whether this theory
is IR conformal or not and concerning the value of $\gamma_{IR}$.

Regarding the range of applicability of the $\Delta_f$ expansion for these
cases, we compute the following ratios of successive coefficients for the 
$G={\rm SU}(3)$, $R=S_2$ case:
\beq
\frac{\kappa_{1,S_2}}{\kappa_{2,S_2}} = 2.26176
\label{kappa_ratio12_sym_su3}
\eeq
\beq
\frac{\kappa_{2,S_2}}{\kappa_{3,S_2}} = 2.1826
\label{kappa_ratio23_sym_su3}
\eeq
and
\beq
\frac{\kappa_{3,S_2}}{\kappa_{4,S_2}} = 1.2993 \ . 
\label{kappa_ratio34_sym_su3}
\eeq
The first two ratios, (\ref{kappa_ratio12_sym_su3}) and 
(\ref{kappa_ratio23_sym_su3}), are slightly larger than 
$(\Delta_f)_{max,S_2}=519/250=2.076$ in $I_{IRZ}$ for this theory.  However,
the third ratio is about 40 \% less than this maximal value of 
$\Delta_{f,S_2}$.  This suggests that because of slow convergence, one must use
the $\Delta_f$ expansion with caution in the lower part of the interval
$I_{IRZ}$ in this theory. 

We list values of the $\gamma_{IR,A_2,\Delta_f^p}$ with $1 \le p \le 4$ for the
SU(4) theory with $R=A_2$ and $N_f \in I_{IRZ}$ for this
theory in Table \ref{gamma_ir_asym_values}. Again, for 
comparison, we include the values $\gamma_{IR,A_2,n\ell}$ for
$2 \le n \le 4$ calculated via the conventional power series expansion to
$n$-loop order and evaluated at $\alpha=\alpha_{IR,n\ell}$ from Table XII in 
our previous work \cite{bvh}.  As expected, the agreement between the two
methods of calculation is best at the upper end of the interval $I_{IRZ}$,
where the IRFP occurs at weak coupling.  For example, for $N_f=9$, 
$\gamma_{IR,A_2,\Delta_f^4}=0.242$, while $\gamma_{IR,4\ell}=0.232$.

It is of interest to consider the $N_c \to \infty$ (LN) limit of Eq. (\ref{ln})
for these theories with $R=S_2$ and $A_2$.  In this LN limit, the upper ends of
the interval $I_{IRZ}$ for the $S_2$ and $A_2$ representations approach the
same limit, and similarly for the lower ends:
\beq
\lim_{LN} N_{u,T_2} = \frac{11}{2} = 5.5
\label{nfb1z_tensor_nc_inf}
\eeq
\beq
\lim_{LN} N_{\ell,T_2} = \frac{17}{8} = 2.125 \ . 
\label{nfb2z_tensor_nc_inf}
\eeq
Hence, in this $N_c \to \infty$ limit, the interval $I_{IRZ}$ is formally 
$2.125 < N_f < 5.5$, including the physical integer values $3 \le N_f \le
5$. Similarly, in this limit, the variable $\Delta_f$ is given by 
$\Delta_f=(11/2)-N_f$ and reaches a maximum value, at 
$N_f=N_{\ell,T_2}$, of 
\beq
\lim_{LN} (\Delta_f)_{max,T_2} = \frac{27}{8} = 3.375 \ . 
\label{Deltaf_ir_max_tensor_nc_inf}
\eeq
This the $N_c \to \infty$ limit of (\ref{Deltaf_ir_max_tensor}). 

As with the adjoint representation, we define 
\beq
\hat \kappa_{j,T_2} = \lim_{LN} \kappa_{j,T_2} \ . 
\label{kappahat_tensor}
\eeq
We find that 
\beq
\hat\kappa_{j,S_2} = \hat\kappa_{j,A_2} \ . 
\label{kappahat_s2a2}
\eeq
From our general expressions for $\kappa_{j,T_2}$ with $1 \le j \le 4$, we
calculate 
\beq
\hat\kappa_{1,T_2} = \frac{2}{3^2} = 0.2222 
\label{kappa1hat_tensor}
\eeq
\beq
\hat\kappa_{2,T_2} = \frac{341}{2^3 \cdot 3^6} = 
0.0584705
\label{kappahat2_tensor}
\eeq
\beq
\hat\kappa_{3,T_2} = \frac{61873}{2^6 \cdot 3^{10}} =
0.016372
\label{kappa3hat_tensor}
\eeq
and
\beq
\hat\kappa_{4,T_2} = \frac{53389393}{2^{11} \cdot 3^{14}} 
+ \frac{23\zeta_3}{3^{10}} = 0.59186 \times 10^{-2} \ . 
\label{kappa4hat_tensor}
\eeq
Hence, 
\beq
\lim_{LN} \gamma_{IR,S_2,\Delta_f^p} = 
\lim_{LN} \gamma_{IR,A_2,\Delta_f^p} 
\label{gamma_s2as_Ncinf}
\eeq
and, in the limit $p \to \infty$, 
\beq
\lim_{LN} \gamma_{IR,S_2} = \lim_{LN} \gamma_{IR,A_2} \ . 
\label{gamma_ir_s2a2}
\eeq
Thus, for both $R=S_2$ and $R=A_2$, 
\beqs
&& \lim_{LN} \gamma_{\bar\psi\psi,IR,T_2,\Delta_f^4} = 
\Delta_f \bigg [ 0.22222 + 0.0584705\Delta_f \cr\cr
&+& 0.016372\Delta_f^2 + 0.0059186\Delta_f^3 \ \bigg ] \ . 
\label{gamma_ir_tensor_Ncinf}
\eeqs

We observe that for all of the cases we have calculated, namely $1 \le j \le
4$, 
\beq
\hat\kappa_{j,T_2} = 2^{-j} \hat \kappa_{j,adj} \ . 
\label{kappahat_tensor_adj}
\eeq
One can understand this relation from the structure of the relevant group
invariants, including the fact that the trace invariant $T(R)$ satisfies
\beq
\lim_{N_c \to \infty} \frac{T_{T_2}}{T_{adj}} = \frac{1}{2} \ . 
\label{traceratio}
\eeq
We thus infer more generally that the relation (\ref{kappahat_tensor_adj})
holds for all $j$. 
In Table \ref{gamma_ir_tensor_ncinf_values} we list the resultant common 
values of $\gamma_{IR,T_2,\Delta_f^p}$ for $1 \le p \le 4$ and $N_f \in 
I_{IRZ}$ in the LN limit.  As noted above, in this LN limit, this interval
consists of the integral values $N_f = 3, \ 4, \ 5$. 

Concerning the range of applicability of the $\Delta_f$ expansion in this
LN limit, we compute the ratios
\beq
\frac{\hat\kappa_{1,T_2}}{\hat\kappa_{2,T_2}} = 
\frac{1296}{341} = 3.8006
\label{kappa_ratio12_tensor_nc_inf}
\eeq
\beq
\frac{\hat\kappa_{2,T_2}}{\hat\kappa_{3,T_2}} = 
\frac{220968}{61873} = 3.5713
\label{kappa_ratio23_tensor_nc_inf}
\eeq
and
\beqs
\frac{\hat\kappa_{3,T_2}}{\hat\kappa_{4,T_2}} &=&
\frac{160374816}{53389393+3815424\zeta_3} \cr\cr
&=& 2.76624 \ . 
\label{kappa_ratio34_tensor_nc_inf}
\eeqs
The first two ratios, (\ref{kappa_ratio12_tensor_nc_inf}) and
(\ref{kappa_ratio23_tensor_nc_inf}), are slightly greater than the maximum
value $(\Delta_f)_{max,T_2}=3.375$, but the third ratio,
(\ref{kappa_ratio34_tensor_nc_inf}), is smaller than this maximum value,
suggesting that in this limit, for these tensor representations, because of
slow convergence, one must use caution in applying the $\Delta_f$ expansion in
the lower part of the interval $I_{IRZ}$.  This is similar to what we found for
the $S_2$ representation in the SU(3) theory.

% ======================================================================

% section IV 
\section{Calculation of $\beta'_{IR}$ to $O(\Delta_f^5)$}
\label{betaprime_section}

\subsection{General $G$ and $R$}
\label{betaprime_general}

The derivative $\beta'_{IR}$ is an important physical quantity characterizing
the conformal field theory at $\alpha_{IR}$.  We denote the gauge field of the
theory as $A_\mu^a$ (where $a$ is a group index), the field strength-tensor as 
$F_{\mu\nu}^a = \partial_\mu A_\nu^a - \partial_\nu A_\mu^a + 
g c_{abc} A_\mu^b A_\nu^c$ (where $c_{abc}$ is the structure constant of the
Lie algebra of $G$) and the rescaled field-strength tensor as 
$F_{\mu\nu,r}^a = gF_{\mu\nu}^a$, so that the gauge field kinetic term in the
Lagrangian is ${\cal L}_g = -[1/(4g^2)]F_{\mu\nu,r}^a F_r^{a \ \mu\nu}$.  The
trace anomaly states that the trace of the energy-momentum tensor $T^\mu_\nu$
satisfies the relation \cite{traceanomaly}
\beq
T^\mu_\mu = \frac{\beta}{16\pi \alpha^2} F_{\mu\nu,r}^a F_r^{a \ \mu\nu} \ . 
\label{tracerelation}
\eeq
Therefore, the full scaling dimension of the operator 
$F_{r,\mu\nu} F_r^{a \ \mu\nu}$, which we denote as $D_{F^2}$, satisfies
\cite{tracerel}
\beq
D_{F^2} = 4 + \beta' - \frac{2\beta}{\alpha} \ , 
\label{betaprimerelation}
\eeq
where we use the shorthand notation 
$F^2 \equiv F_{r,\mu\nu}^a F_r^{a \ \mu\nu}$.
We denote the anomalous dimension of $F^2$, $\gamma_{F^2}$ via the equation
\cite{gammaconvention}
\beq
D_{F^2} = D_{F^2,free}-\gamma_{_{F^2}} = 4 - \gamma_{_{F^2}} 
\label{fsqrel}
\eeq
and its evaluation
at $\alpha=\alpha_{IR}$ as $\gamma_{_{F^2,IR}}$. From Eq. 
(\ref{betaprimerelation}), it
follows that at a zero of the beta function away from the origin, in
particular, at $\alpha_{IR}$, the derivative $\beta'_{IR}$ is equivalent to 
the anomalous dimension of the operator $F^a_{r,\mu\nu}F_r^{a \ \mu\nu}$:
\beq
\beta'_{IR} = -\gamma_{_{F^2,IR}} \ .
\label{betaprime_fsquared_anomdim}
\eeq

In \cite{dex} we calculated the expansion coefficients $d_j$ of $\beta'_{IR}$
in Eq. (\ref{betaprime_ir_Deltaseries}) to order $\Delta_f^4$ for general $G$
and $R$, and to order $\Delta_f^5$ for the special case $G={\rm SU}(3)$ and
fermion representation $R=F$, the fundamental.  Here we calculate the next
higher-order coefficient, namely $d_5$, for general $G$ and $R$.  For this
purpose, we make use of the recent computation of the five-loop beta function
coefficient, $b_5$, in \cite{b5}.  The computation in \cite{b5} was performed
in the $\overline{\rm MS}$ scheme, so that we can combine it with the
scheme-independent $b_1$ and $b_2$ \cite{b1,b2} and the results for $b_3$ and
$b_4$ that have also been calculated in the $\overline{\rm MS}$ scheme
\cite{b3,b4}.  However, we again stress that since the $d_n$ coefficients are
scheme-independent, it does not matter which scheme one uses to calculate them.
We first recall our previous results from Ref. \cite{dex}:
\beq
d_1 = 0 \ , 
\label{d1}
\eeq
\beq
d_2 = \frac{2^5 T_f^2}{3^2 C_A D} \ , 
\label{d2}
\eeq
\beq
d_3 = \frac{2^7 T_f^3(5C_A+3C_f)}{3^3 C_A^2 D^2} \ , 
\label{d3}
\eeq
and
\begin{widetext}
\beqs    
d_4 &=& -\frac{2^3T_f^2}{3^6C_A^4 D^5} \, \Bigg [
-3C_AT_f^2 \bigg ( 137445 C_A^4 + 103600C_A^3C_f+72616C_A^2C_f^2 
+951808C_AC_f^3 - 63888C_f^4 \bigg ) \cr\cr
&-& 5120T_f^2D \frac{d_A^{abcd}d_A^{abcd}}{d_A}
+90112C_AT_fD\frac{d_R^{abcd}d_A^{abcd}}{d_A}
-340736C_A^2D\frac{d_R^{abcd}d_R^{abcd}}{d_A} \cr\cr
&+& 8448D \bigg [ C_A^2T_f^2\Big ( 21C_A^2+12C_AC_f-33C_f^2\Big ) 
+ 16T_f^2 \frac{d_A^{abcd}d_A^{abcd}}{d_A}
- 104 C_AT_f \frac{d_R^{abcd}d_A^{abcd}}{d_A}
+ 88C_A^2\frac{d_R^{abcd}d_R^{abcd}}{d_A} \bigg ] \zeta_3 \ \Bigg ] \ . \cr\cr
& & 
\label{d4}
\eeqs
In Ref. \cite{dex} we presented the expression for $d_4$ with terms written 
in order of descending powers of $C_A$.  
It is also useful to express this coefficient $d_4$ 
in an equivalent form that renders certain factors of $D$ explicit and 
shows the simple factorization of terms multiplying 
$\zeta_3$, and we have done this in Eq. (\ref{d4}).  

Here we present our calculation of $d_5$ for arbitrary $G$ and $R$: 
\beqs
d_5 &=& \frac{2^4T_f^3}{3^7C_A^5 D^7} \, \Bigg [ 
-C_AT_f^2 \bigg ( 39450145C_A^6+235108272C_A^5C_f + 1043817726C_A^4C_f^2
+765293216C_A^3C_f^3 \cr\cr
&-& 737283360C_A^2C_f^4+730646400C_AC_f^5 -356750592C_f^6 \bigg ) 
-2^9T_f^2D\frac{d_A^{abcd}d_A^{abcd}}{d_A}
(6139C_A^2+2192C_AC_f-3300C_f^2)
\cr\cr
&+& 2^9C_AT_fD\frac{d_R^{abcd}d_A^{abcd}}{d_A}
(43127C_A^2-28325C_AC_f-2904C_f^2)
+15488C_A^2D\frac{d_R^{abcd}d_R^{abcd}}{d_A}
(2975C_A^2+8308C_AC_f-12804C_f^2) \cr\cr
&+& 2^7D\bigg [ 3C_AT_f^2 D \bigg ( 6272C_A^4-49823C_A^3C_f 
+40656C_A^2C_f^2 +13200C_AC_f^3 + 2112C_f^4 \bigg ) \cr\cr
&+& 2^4T_f^2\frac{d_A^{abcd}d_A^{abcd}}{d_A}
(19516C_A^2 - 18535C_AC_f-21780C_f^2)
-2^3 C_AT_f\frac{d_R^{abcd}d_A^{abcd}}{d_A}
(182938C_A^2-297649C_AC_f-197472C_f^2) \cr\cr
&-&88C_A^2 \frac{d_R^{abcd}d_R^{abcd}}{d_A}
(245C_A^2 + 62524C_AC_f + 42108 C_f^2) \bigg ]\zeta_3 \cr\cr
&+& 2^{10} \cdot 55C_AD^2 \bigg [ 9C_AT_f^2D(C_A+2C_f)(C_A-C_f) 
+160T_f^2 \frac{d_A^{abcd}d_A^{abcd}}{d_A} \cr\cr
&-&80T_f(10C_A+3C_f)\frac{d_R^{abcd}d_A^{abcd}}{d_A}
-440C_A(C_A-3C_f)\frac{d_R^{abcd}d_R^{abcd}}{d_A} \bigg ]\zeta_5 \
 \Bigg ] \ .
\label{d5}
\eeqs
\end{widetext} 
We proceed to evaluate these coefficients $d_j$ up to $j=5$, and hence the
derivative $\beta'_{IR}$ up to $O(\Delta_f^5)$ below for $G={\rm SU}(N_c)$ and
several specific representations. The coefficients $d_2$ and
$d_3$ are manifestly positive for arbitrary $G$ and $R$.  These signs are
indicated in Table \ref{dj_signs}.  We discuss the signs of $d_4$ and $d_5$ for
various representations below. 

% ========================================================================

\subsection{$\beta'_{IR,\Delta_f^4}$ for $G={\rm SU}(N_c)$ and $R=F$}
\label{betaprime_fundamental_section}

Here we present the evaluation of our general result (\ref{d5}) for the case
$G={\rm SU}(N_c)$ and $R=F$.  For reference, we first recall our results from
\cite{dex} for $d_j$ with $2 \le j \le 4$ (and also recall that $d_1=0$ for all
$G$ and $R$):
\beq
d_{2,F} = \frac{2^4}{3^2(25N_c^2-11)} \ , 
\label{d2_fund}
\eeq
\bigskip
\beq
d_{3,F} = \frac{2^5(13N_c^2-3)}{3^3N_c(25N_c^2-11)^2} \ , 
\label{d3_fund}
\eeq
and
\begin{widetext} 
\beqs
d_{4,F} & = & -\frac{2^4}{3^5N_c^2(25N_c^2-11)^5} \bigg [
N_c^8\Big (-366782+660000\zeta_3 \Big )+
N_c^6\Big (865400-765600\zeta_3 \Big ) \cr\cr
&+& N_c^4\Big (-1599316+2241888\zeta_3 \Big ) 
 +  N_c^2\Big (571516-894432\zeta_3 \Big ) + 3993 \ \bigg ] \ . 
\label{d4_fund}
\eeqs
This coefficient can be written equivalently in a form that shows the simple
factorization of the terms multiplying $\zeta_3$:
\beqs
d_{4,F} & = & -\frac{2^4}{3^5N_c^2(25N_c^2-11)^5} \bigg [
\Big (-366782N_c^8 + 865400N_c^6 - 1599316N_c^4 +571516N_c^2 + 3993 \Big ) 
\cr\cr
&+& 1056N_c^2(25N_c^2-11)(25N_c^4-18N_c^2+77)\zeta_3 \ \bigg ] \ . 
\label{d4_fundfac}
\eeqs

In \cite{dexs} we presented the expression for $d_{5,F}$ with terms ordered as
descending powers of $N_c$. As with $d_{4,F}$, it is also useful to display
this coefficient in an equivalent form that shows the simple factorizations of
the terms multiplying $\zeta_3$ and $\zeta_5$: 
\beqs
d_{5,F} & = & \frac{2^5}{3^6N_c^3(25N_c^2-11)^7} \bigg [
\Big ( -298194551N_c^{12} + 414681770N_c^{10} + 80227411N_c^8 \cr\cr
&+& 210598856N_c^6
-442678324N_c^4+129261880N_c^2+ 3716152 \Big ) \cr\cr
&-&96(25N_c^2-11)\Big (176375N_c^{10} -564526N_c^8 +1489367N_c^6 
-1470392N_c^4 + 290620N_c^2 + 968 \Big )\zeta_3 \cr\cr
&+& 21120N_c^2(25N_c^2-11)^2 
\Big (40N_c^6-27N_c^4+124N_c^2-209\Big )\zeta_5 \ \bigg ] \ . 
\label{d5_fund}
\eeqs
\end{widetext}
We have checked that when we set $N_c=3$ in our general result for 
$d_{5,F}$ in Eq. (\ref{d5_fund}), the result agrees with our earlier
calculation of $d_{5,F}$ in Eq. (5.20) of Ref. \cite{dex}.  

As observed above, the coefficients $d_2$ and $d_3$ are manifestly positive for
any $G$ and $R$.  We find that $d_{4,F}$ and $d_{5,F}$ are negative-definite
for $G={\rm SU}(N_c)$ and all physical values of $N_c \ge 2$.  These results
are summarized in Table \ref{dj_signs}.

We list below the explicit numerical expressions for $\beta'_{IR}$ to order
$\Delta_f^5$, denoted $\beta'_{IR,{\rm SU}(N_c),F,\Delta_f^5}$, for 
the gauge groups SU($N_c$) with $N_c=2, \ 3, \ 4$, with fermions in the
fundamental representation, to the indicated floating-point precision:
\begin{widetext}
\beqs
{\rm SU}(2): \quad \beta'_{IR,F,\Delta_f^5} & =& \Delta_f^2 \Big [ 
  (1.99750 \times 10^{-2} 
+ (3.66583 \times 10^{-3})\Delta_f 
- (3.57303 \times 10^{-4})\Delta_f^2
- (2.64908 \times 10^{-5})\Delta_f^3 \ \Big ] \cr\cr
& &  
\label{betaprime_sunf_p5_su2}
\eeqs
\beqs
{\rm SU}(3): \quad \beta'_{IR,F,\Delta_f^5} & =& \Delta_f^2 \Big [ 
  (0.83074 \times 10^{-2})
+ (0.98343 \times 10^{-3}\Delta_f
- (0.46342 \times 10^{-4})\Delta_f^2
- (0.56435 \times 10^{-5})\Delta_f^3 \ \Big ] \cr\cr
& & 
\label{betaprime_sunf_p5_su3}
\eeqs
and
\beqs
{\rm SU}(4): \quad \beta'_{IR,F,\Delta_f^5} & =& \Delta_f^2 \Big [ 
  (0.45701 \times 10^{-2})
+ (0.40140 \times 10^{-3}\Delta_f
- (0.12938 \times 10^{-4})\Delta_f^2
- (0.15498 \times 10^{-5})\Delta_f^3 \ \Big ] \ . \cr\cr
& & 
\label{betaprime_sunf_p5_su4}
\eeqs
\end{widetext}

In Table \ref{betaprime_values} we list the (scheme-independent) values that we
calculate for $\beta'_{IR,F,\Delta_f^p}$ with $2 \le p \le 4$ for the
illustrative gauge groups $G={\rm SU}(2)$, SU(3), and SU(4), as functions of
$N_f$ in the respective intervals $I_{IRZ}$ given in
Eq. (\ref{nfinterval}). For comparison, we list the $n$-loop values of
$\beta'_{IR,F,n\ell}$ with the $2 \le n \le 4$ from \cite{bc,dex}, where
$\beta'_{IR,F,3\ell}$ and $\beta'_{IR,F,4\ell}$ are computed in the
$\overline{\rm MS}$ scheme.  Although, for completeness, we list values of
$\beta'_{IR,F,2\ell}$ with $N_f$ extending down to the lower end of the
respective intervals $I_{IRZ}$ for each value of $N_c$, we caution that in a
number of cases, including $N_f=6$ for SU(2), $N_f=9$ for SU(3), and $10 \le
N_f \le 12$ for SU(4), the corresponding values of $\alpha_{IR,2\ell}$
(discussed further below) are too large for the perturbative $n$-loop
calculations to be applicable.  Moreover, since for a considerable
range of values of $N_f \in I_{IRZ}$ for each $N_c$, the five-loop beta
function $\beta_{5\ell}$ calculated via the conventional power series expansion
has no physical IR zero, we restrict the resultant $\beta'_{IR,F,n\ell}$
evalulations to $1 \le n \le 4$ loops.

\begin{figure}
  \begin{center}
    \includegraphics[height=6cm]{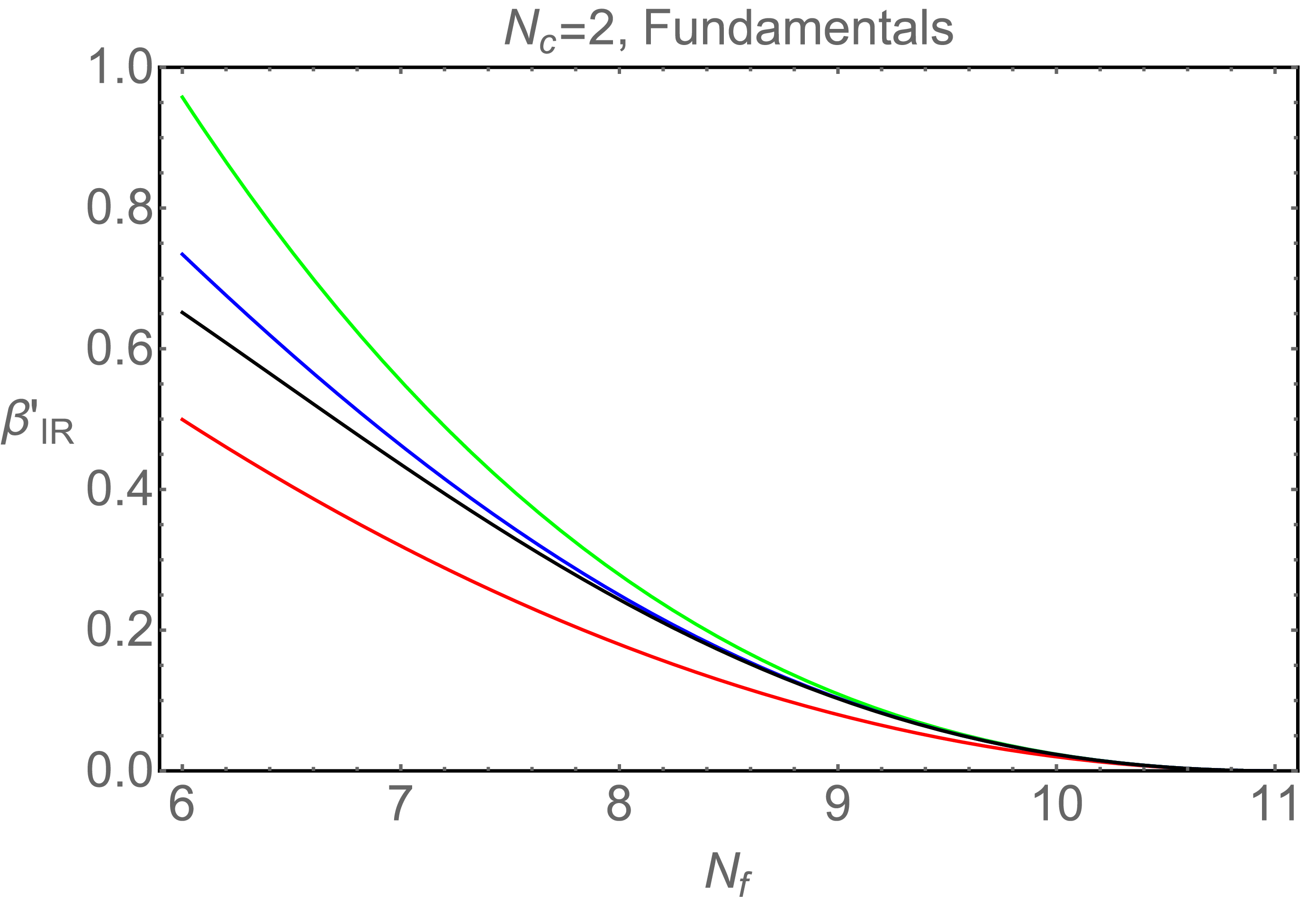}
  \end{center}
\caption{Plot of $\beta'_{IR,F,\Delta_f^p}$ (labelled as 
$\beta'_{IR}$ on the vertical axis) for $N_c=2$ and
$2 \le p \le 5$ as a function of $N_f \in I_{IRZ}$.  
From bottom to top, the curves (with colors online) refer to 
$\beta'_{IR,F,\Delta_f^2}$ (red),                           
$\beta'_{IR,F,\Delta_f^3}$ (green),
$\beta'_{IR,F,\Delta_f^4}$ (blue), and
$\beta'_{IR,F,\Delta_f^5}$ (black).}
\label{betaprime_Nc2_fund_plot}
\end{figure}

\begin{figure}
  \begin{center}
    \includegraphics[height=6cm]{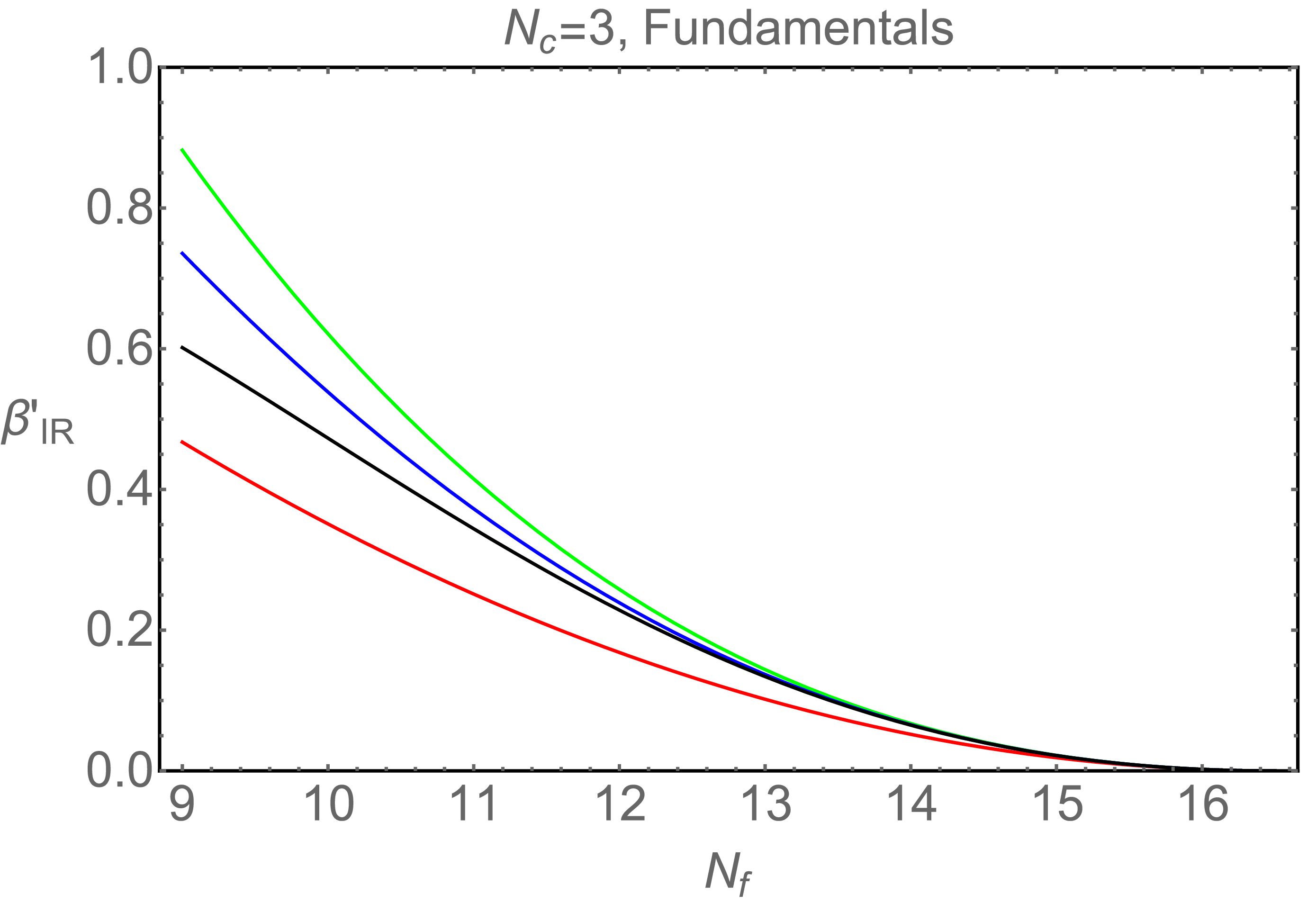}
  \end{center}
\caption{Plot of $\beta'_{IR,F,\Delta_f^p}$ for $N_c=3$ and
$2 \le p \le 5$ as a function of $N_f \in I_{IRZ}$.  From bottom to top, 
the curves (with colors online) refer to 
$\beta'_{IR,F,\Delta_f^2}$ (red),                           
$\beta'_{IR,F,\Delta_f^3}$ (green),
$\beta'_{IR,F,\Delta_f^4}$ (blue), and
$\beta'_{IR,F,\Delta_f^5}$ (black).}
\label{betaprime_Nc3_fund_plot}
\end{figure}

\begin{figure}
  \begin{center}
    \includegraphics[height=6cm]{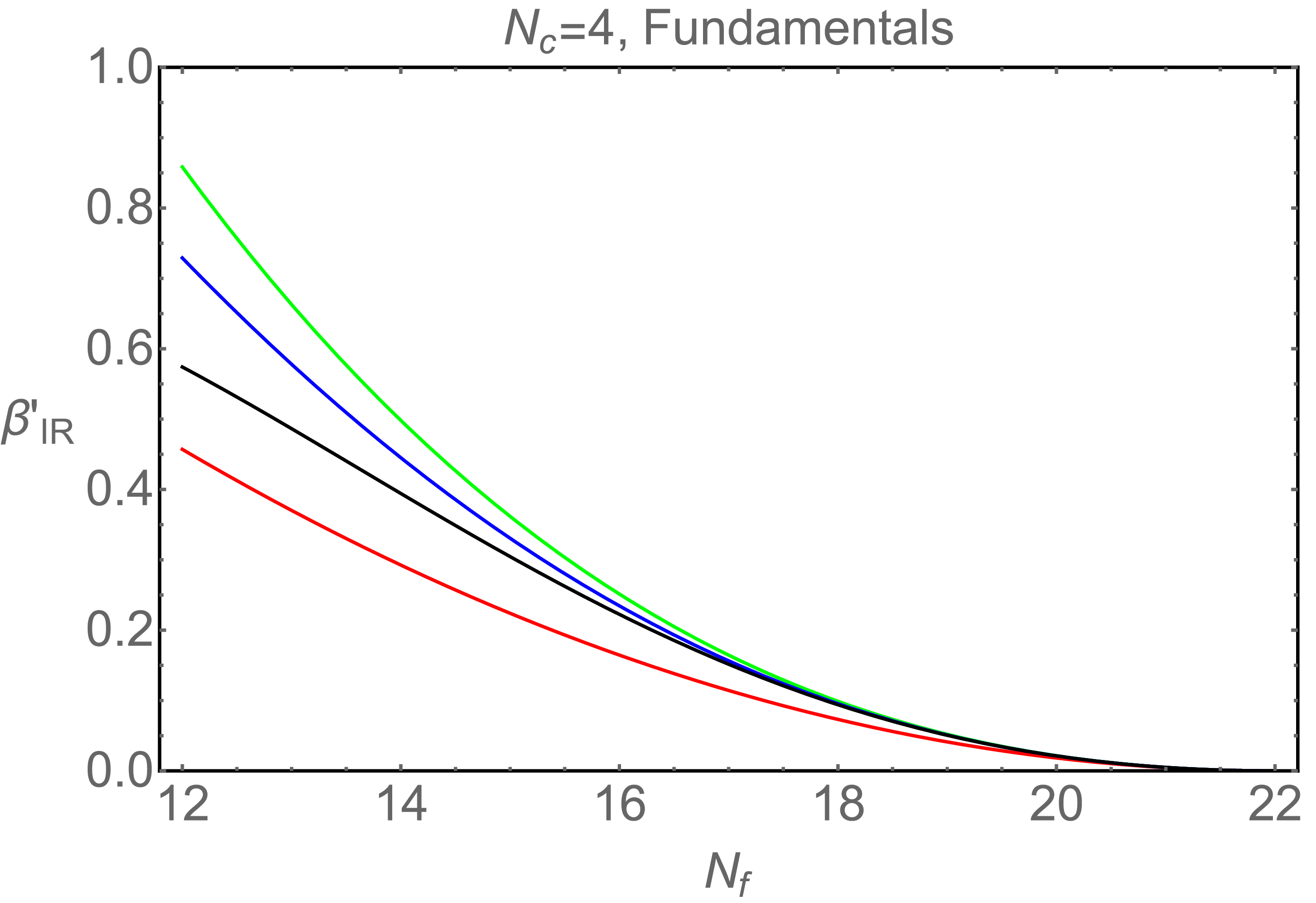}
  \end{center}
\caption{Plot of $\beta'_{IR,F,\Delta_f^p}$ for $N_c=4$ and
$2 \le p \le 5$ as a function of $N_f \in I_{IRZ}$.  From bottom to top, 
the curves (with colors online) refer to 
$\beta'_{IR,F,\Delta_f^2}$ (red),                           
$\beta'_{IR,F,\Delta_f^3}$ (green),
$\beta'_{IR,F,\Delta_f^4}$ (blue), and
$\beta'_{IR,F,\Delta_f^5}$ (black).}
\label{betaprime_Nc4_fund_plot}
\end{figure}

In Figs. \ref{betaprime_Nc2_fund_plot}-\ref{betaprime_Nc4_fund_plot} 
we plot the values of $\beta'_{IR}$, calculated to order
$\Delta_f^p$ with $2 \le p \le 5$, for $R=F$ for the gauge groups 
SU(2), SU(3), and SU(4). In the general
calculations of $\gamma_{IR}$ as a series in powers of $\Delta_f$ to maximal
power $p=3$ (i.e., order $\Delta_f^3$) in \cite{gtr} and, for $G={\rm SU}(3)$
and $R=F$, to maximal power $p=4$ in \cite{gsi}, it was found that, for a fixed
value of $N_f$, or equivalently, $\Delta_f$, in the interval $I_{IRZ}$, these
anomalous dimensions increased monotonically as a function of $p$. This feature
motivated our extrapolation to $p=\infty$ in \cite{gtr} to obtain estimates for
the exact $\gamma_{IR}$.  In contrast, here we find that, for a fixed value of
$N_f$, or equivalently, $\Delta_f$, in $I_{IRZ}$, as a consequence of the fact
that different coefficients $d_n$ do not all have the same sign,
$\beta'_{IR,\Delta_f^p}$ is not a monotonic function of $p$. Because of this
non-monotonicity, we do not attempt to extrapolate our series to $p=\infty$.

A lattice measurement of $\beta'_{IR}$ has been reported in
\cite{hasenfratz_betaprime} for the SU(3) theory with $R=F$ and $N_f=12$,
namely $\beta'_{IR}=0.26(2)$. The earlier higher-order values calculated in
\cite{bc} via $n$-loop expansions in the coupling are
$\beta'_{IR,3\ell}=0.2955$ and $\beta'_{IR,4\ell}=0.282$, which agree with this
lattice measurement.  As indicated in Table \ref{betaprime_values}, our
higher-order scheme-independent values for this theory are
$\beta'_{IR,\Delta_f^3}=0.258$, $\beta'_{IR,\Delta_f^4}=0.239$, and
$\beta'_{IR,\Delta_f^5}=0.228$. Given the possible contributions of
higher-order terms in the $\Delta_f$ expansion, we consider that our
scheme-independent calculation of $\beta'_{IR}$ to this order is also
consistent with the lattice measurement from Ref. \cite{hasenfratz_betaprime}.

To get a rough estimate of the range of accuracy and applicability of the
series expansion for $\beta'_{IR}$ for this $R=F$ case, we can compute ratios
of coefficients, as discussed before. For the illustrative case of SU(3), we
have
\beq
\frac{d_{2,F}}{d_{3,F}} = 8.447 \quad {\rm for \ SU(3)}, 
\label{d2_over_d3_fund}
\eeq
\beq
\frac{d_{3,F}}{|d_{4,F}|} = 21.221 \quad {\rm for \ SU(3)}, 
\label{d3_over_d4_fund}
\eeq
and
\beq
\frac{|d_{4,F}|}{|d_{5,F}|} = 8.2115  \quad {\rm for \ SU(3)} \ . 
\label{d4_over_d5_fund}
\eeq
Since $N_u=16.5$ and $N_\ell=153/19=8.053$ in this SU(3) theory, the
maximal value of $\Delta_f$ in the interval $I_{IRZ}$, as given by
(\ref{Deltaf_max_ir_fund}), is 
\beq
(\Delta_f)_{\rm max} = \frac{321}{38} = 8.447 \quad {\rm for} \ {\rm SU}(3), \ 
N_f \in I_{IRZ} \ .  
\label{Delta_max_su3_irz}
\eeq
Therefore, these ratios suggest that the small-$\Delta_f$ expansion may be
reasonably reliable in most of this interval, $I_{IRZ}$ and the associated
non-Abelian Coulomb phase.

% =======================================================================

\subsection{$\beta'_{IR,\Delta_f^5}$ in LNN Limit} 
\label{betaprime_lnn_section}

The appropriately rescaled beta function that is finite in the LNN limit 
is
\beq
\beta_\xi = \frac{d\xi}{dt} = \lim_{LNN} N_c \beta \ , 
\label{betaxi}
\eeq
where $\xi = 4\pi x = \lim_{LNN} \alpha N_c$ was defined in Eq. (\ref{lnn}). 
This has the series expansion 
\beq
\beta_\xi \equiv \frac{d\xi}{dt}
= -8\pi x \sum_{\ell=1}^\infty \hat b_\ell x^\ell 
=   -2\xi \sum_{\ell=1}^\infty \tilde b_\ell \xi^\ell 
\label{betaxiseries}
\eeq
where 
\beq
\hat b_\ell = \lim_{LNN} \frac{b_\ell}{N_c^\ell} \ . 
\label{bellhat}
\eeq
and $\tilde b_\ell = \hat b_\ell/(4\pi)^\ell$. 
The $\hat b_\ell$ with $1 \le \ell \le 4$ were analyzed in \cite{bc,lnn} and
are listed for the reader's convenience in the Appendix.

From the recent calculation of $b_5$ in \cite{b5}, for general $G$ and $R$, 
in the $\overline{\rm MS}$ scheme \cite{b5}, we calculate 
\beqs
\hat b_5 & = & \frac{8268479}{3888} + \frac{38851}{162}\zeta_3 
- \frac{121}{6}\zeta_4 - 330\zeta_5 \cr\cr
& + & \biggr ( -\frac{11204369}{5184} - \frac{231619}{648}\zeta_3 
+ \frac{77}{6}\zeta_4 + \frac{4090}{9}\zeta_5 \biggr ) \, r \cr\cr
& + & \biggr ( \frac{3952801}{7776} + \frac{33125}{108}\zeta_3 
- \frac{241}{6}\zeta_4 - \frac{1630}{9}\zeta_5 \biggr ) \, r^2 \cr\cr
& + & \biggr ( -\frac{5173}{432} - \frac{1937}{81}\zeta_3 + 7\zeta_4 
+ \frac{20}{3}\zeta_5 \biggr ) \, r^3 \cr\cr
& + & \biggr ( \frac{61}{486} - \frac{52}{81}\zeta_3 \biggr ) \, r^4 \cr\cr
& = & 2050.932 - 2105.880 r + 645.7474 r^2 \cr\cr
& - & 26.2309 r^3 - 0.64618 r^4 \ . 
\label{b5hat}
\eeqs
(In this expression although $\zeta_4$ could be expressed explicitly as 
$\zeta_4=\pi^4/90$, we leave it in abstract form to be parallel with the
$\zeta_3$ and $\zeta_5$ terms.)  We find that this coefficient $\hat b_5$ is
positive throughout the entire asymptotically free interval $0 \le r < 5.5$.
(Considered formally as a function of $r \in {\mathbb R}$, $\hat b_5$ is
negative for $r < -58.609$, positive for $-58.609 < r < 14.336$, and negative
for $r > 14.336$, where the numbers are quoted to the given floating-point
accuracy.)

Since the derivative $d\beta_\xi/d\xi$ satisfies the relation 
\beq
\frac{d \beta_\xi}{d\xi} = \frac{d\beta}{d\alpha} \equiv \beta' \ , 
\label{dbetarelation}
\eeq
it follows that $\beta'$ is finite in the
LNN limit (\ref{lnn}).  In terms of the variable $x$ 
defined in Eq. (\ref{xlnn}), we have
\beq
\beta' = -2\sum_{\ell=1}^\infty (\ell+1) \hat b_\ell \, x^\ell \ . 
\label{betaprime_lnn}
\eeq

Because $\beta'_{IR}$ is scheme-independent and is finite in the LNN limit, one
is motivated to calculate the LNN limit of the scheme-independent expansion
(\ref{betaprime_ir_Deltaseries}).  For this purpose, in addition to the
rescaled quantities $\Delta_r$ defined in Eq. (\ref{deltar}), we define the
rescaled coefficient
\beq
\hat d_{j,F} = \lim_{LNN} N_c^j \, d_{j,F} \ , 
\label{dnhat}
\eeq
which is finite.  Then each term
\beq
\lim_{LNN} d_{j,F} \Delta_f^j= (N_c^j d_{j,F})
\Big ( \frac{\Delta_f}{N_c} \Big )^j = \hat d_{j,F} \Delta_r^j 
\label{finiteproduct}
\eeq
is finite in this limit. 
Thus, writing $\lim_{LNN} \beta'_{IR}$ as $\beta'_{IR,LNN}$ for this 
$R=F$ case, we have 
\beqs
& & \beta'_{IR,LNN} = \sum_{j=1}^\infty d_{j,F} \Delta_f^j = 
\sum_{j=1}^\infty \hat d_{j,F} \Delta_r^j \ . \cr\cr
& &  
\label{betaprime_ir_lnn}
\eeqs
We denote the value of $\beta'_{IR,LNN}$ obtained from 
this series calculated to order $O(\Delta_f^p)$ as 
$\beta'_{IR,LNN,\Delta_f^p}$. 

From Eqs. (\ref{d1})-(\ref{d4}), we find that the approach to the LNN limits
for $\hat d_{j,F}$ involves correction terms that vanish like $1/N_c^2$.  This
is the same property that was found in \cite{bc,lnn} and, in the same way, it
means that the approach to the LNN limit for finite $N_c$ and $N_f$ with fixed
$r=N_f/N_c$ is rather rapid, as discussed in \cite{lnn}.  In \cite{dex} we gave
the $\hat d_{j,F}$ for $1 \le n \le 4$; in addition to $\hat d_1=0$ (which
holds for any $G$ and $R$), these are
\beq
\hat d_{2,F} = \frac{2^4}{3^2 \cdot 5^2} = 0.0711111 \ , 
\label{d2hat_lnn}
\eeq
\beq
\hat d_{3,F} = \frac{416}{3^3 \cdot 5^4} = 2.465185 \times 10^{-2} \ , 
\label{d3hat_lnn}
\eeq
and
\beq
\hat d_{4,F} = \frac{5868512}{3^5 \cdot 5^{10}}
-\frac{5632}{3^4 \cdot 5^6} \zeta_3 = -(2.876137 \times 10^{-3}) \ . 
\label{d4hat_lnn}
\eeq
Here we give the next higher coefficient: 
\beqs
\hat d_{5,F} &=& -\frac{9542225632}{3^6 \cdot 5^{14}} 
- \frac{1444864}{3^5 \cdot 5^9}\zeta_3 
+ \frac{360448}{3^5 \cdot 5^8}\zeta_5 \cr\cr
& = & -(1.866490 \times 10^{-3}) \ .
\label{d5hat_lnn}
\eeqs
In these equations we have indicated the simple factorizations of the
denominators that were already evident in the general analytic expressions
(\ref{d1})-(\ref{d4}).  Although the numerical coefficients in the numerators
of terms in Eq. (\ref{d5hat_lnn}) do not, in general, have simple
factorizations, they do contain various powers of 2; for example, in $\hat
d_{5,F}$, $1444864 = 2^{10} \cdot 17 \cdot 83$, etc.
Thus, numerically, to order $\Delta_r^5$, for the LNN
limit of this theory with $R=F$, we have
\beqs
& & \beta'_{IR,LNN} = \Delta_r^2\Big [7.1111 \times 10^{-2} + 
(2.4652 \times 10^{-2})\Delta_r \cr\cr
&-& (2.8761 \times 10^{-3})\Delta_r^2 - (1.8665 \times 10^{-3})\Delta_r^3 
\cr\cr
&+& O(\Delta_r^4) \Big ] \ ,
\label{betaprime_numerical}
\eeqs
where the coefficients are given to the indicated floating-point precision.
We may again calculate ratios of successive magnitudes of these coefficients to
get a rough estimate of the range over which the small-$\Delta_r$ expansion is
reliable in this LNN limit.  We find
\beq
\frac{\hat d_{2,F}}{\hat d_{3,F}} = 2.885 \ , 
\label{d2hat_over_d3hat}
\eeq
\beq
\frac{\hat d_{3,F}}{|\hat d_{4,F}|} = 8.571 \ , 
\label{d3hat_over_d4hat}
\eeq
and
\beq
\frac{|\hat d_{4,F}|}{|\hat d_{5,F}|} =  1.541\ . 
\label{d4hat_over_d5hat}
\eeq
For $r \in I_{IRZ,r}$, the maximal value of $\Delta_r$ is
$(\Delta_r)_{\rm max} = 75/26 = 2.885$.  The first two ratios, 
(\ref{d2hat_over_d3hat}) and (\ref{d3hat_over_d4hat}) suggest that the
$\Delta_r$ expansion for $\beta'_{IR}$ may be reasonably reliable over a 
reasonable fraction of the interval $I_{IRZ,r}$.  From the third ratio, 
(\ref{d4hat_over_d5hat}), we infer that the expansion is expected to be more
accurate in the upper portion of the interval $I_{IRZ,r}$ than the lower
portion. 

In Ref. \cite{dex} we presented a comparison of these scheme-independent
calculations of $\beta'_{IR,LNN}$ calculated up to the $\Delta_r^4$ order with
the results of conventional $n$-loop calculations, denoted
$\beta'_{IR,n\ell,LNN}$, computed up to the $n=4$ loop order for which the
$b_n$ were known at that time.  We refer the reader to \cite{dex} for details
of this discussion. Here we shall extend this comparison to the $\Delta_r^5$
order. In Table \ref{betaprime_values_lnn} we list the numerical values of
these conventional $n$-loop calculations up to $n=4$, in comparison with our
scheme-independent results calculated to $O(\Delta_r^p)$ with $p$ up to 5.
(The conventional 4-loop values $\beta'_{IR,4\ell}$ for some values of $r$
toward the lower part of $I_{IRZ,r}$ supersede the corresponding entries in
Table II of \cite{dex}.)  Both $\beta'_{IR,n\ell}$ and $\beta'_{IR,\Delta_r^n}$
use, as inputs, the coefficients of the beta function up to loop order $n$,
although $\beta'_{IR,\Delta_r^n}$ does this in a scheme-independent manner. We
see that, especially for $r$ values in the upper part of the interval
$I_{IRZ,r}$, the results are rather close, and, furthermore, that, as expected,
for a given $r$, the higher the loop level $n$ and the truncation order $p$ in
the respective calculations of $\beta'_{IR,n\ell}$ in the $\overline{\rm MS}$
scheme and the scheme-independent $\beta'_{IR,\Delta_r^p}$, the better the
agreement between these two results.  Toward the lower end of the interval
$I_{IRZ,r}$, both the conventional expansion of $\beta'_{IR}$ and the
scheme-independent expansion of $\beta'_{IR}$ in powers of $\Delta_r$ become
less reliable, and hence it is understandable that the results differ from each
other in this lower part of $I_{IRZ,r}$.

% ======================================================================

\subsection{$\beta'_{IR,\Delta_f^5}$ for $G={\rm SU}(N_c)$ and $R=adj$}
\label{betaprime_adj_section}

Here we calculate the $d_j$ and hence $\beta'_{IR,\Delta_f^j}$ for $j$ up to
$j=5$ in the SU($N_c$) gauge theory with fermion representation $R=adj$.  As
was discussed above, in this case, the interval $I_{IRZ}$ contains the single
Dirac value, $N_f=2$.  For this value of $N_f$, Eq. (\ref{Delta_adj}) yields 
$\Delta_f = 3/4$.  We recall that the $d_j$ for $2 \le j \le 4$ are
\cite{dex}
\beq
d_{2,adj} = \bigg ( \frac{2}{3} \bigg )^4 = 0.19753 \ ,  
\label{d2_adj}
\eeq
\beq
d_{3,adj} = \frac{2^8}{3^7} = 0.11706 \ ,  
\label{d3_adj}
\eeq
and
\beqs
d_{4,adj} & = & \frac{46871}{2^2 \cdot 3^{12}} + 
\frac{2368}{3^{10} N_c^2} \cr\cr
& = & 0.022049 + 0.040102N_c^{-2} \ .
\label{d4_adj}
\eeqs
Here, from our new general result (\ref{d5}) for $d_5$, we obtain the next
coefficient for this case of the adjoint representation:
\beqs
d_{5,adj} & = & -\frac{7141205}{2^3 \cdot 3^{16}} 
+ \frac{5504}{3^{12}}\zeta_3 \cr\cr
&-& \bigg ( \frac{30928}{3^{14}} + 
\frac{465152}{3^{13}}\zeta_3 \bigg )N_c^{-2} \cr\cr
& = & -(0.828739 \times 10^{-2}) - 0.357173N_c^{-2} \ . \cr\cr
& &  \ . 
\label{d5_adj}
\eeqs
While the $d_{j,adj}$ with $2 \le j \le 4$ are positive-definite,
we thus find that $d_{5,adj}$ is negative-definite.  These
results on signs are listed in Table \ref{dj_signs}. 
In the $N_c \to \infty$ (LN) limit of Eq. (\ref{ln}), the 
values of $\hat d_{j,adj}$ can be read off directly from our general
results in Eqs. (\ref{d2_adj})-(\ref{d5_adj}); for example, $\hat d_{4,adj} =
46871/(2^2 \cdot 3^{12})$, etc.

With these coefficients, one can again compute ratios to obtain a crude idea of
the region over which the small-$\Delta_f$ series expansion is reliable.  We
have 
\beq
\frac{d_{2,adj}}{d_{3,adj}} = \frac{3^3}{2^4} = 1.687
\label{d2_over_d3_adj}
\eeq
and, taking the large-$N_c$ limit for simplicity, 
\beq
\lim_{N_c \to \infty} 
\frac{d_{3,adj}}{d_{4,adj}} = \frac{3^5 \cdot 2^{10}}{46871} = 5.309
\label{d3_over_d4_adj}
\eeq
\beq
\lim_{N_c \to \infty} 
\frac{d_{4,adj}}{|d_{5,adj}|} = \frac{7593102}{7141205-3566592\zeta_3} 
= 2.6606 \ . 
\label{d4_over_d5_adj}
\eeq
Since $\Delta_f=0.75$ for $N_f=2$, these ratios indicate that 
the small-$\Delta_f$ expansion should be reasonably accurate here.

% ==========================================================================

\subsection{$\beta'_{IR,\Delta_f^5}$ for $G={\rm SU}(N_c)$ and 
$R=S_2, \ A_2$}
\label{betaprime_tensor_section}

Here we present our results for the $d_j$ coefficients 
and hence $\beta'_{IR,\Delta_f^j}$ with $j$ up to 5 for $G={\rm SU}(N_c)$ 
and $N_f$ fermions in the symmetric and antisymmetric rank-2 tensor 
representations, $S_2$ and $A_2$. As before with
$\gamma_{\bar\psi\psi,IR,\Delta_f^p}$, since many formulas for these two 
cases are simply related to each other by sign reversals in certain terms, 
it is convenient to treat these two cases together, denoting them collectively
as $T_2$.  We recall that for $R=A_2$, we restrict to $N_c \ge 3$.  

From our general formulas (\ref{d1})-(\ref{d5}), we obtain the following, where
the upper and lower signs refer to the $S_2$ and $A_2$ special cases of $T_2$,
respectively, and $F_\pm$ was defined in Eq. (\ref{fplusminus}):
\beq
d_{2,T_2} = \frac{2^3(N_c \pm 2)^2}{3^2 F_\pm}
\label{d2_tensor}
\eeq
\beq
d_{3,T_2} = \frac{2^4(N_c \pm 2)^3(8N_c^2 \pm 3N_c - 6)}
{3^3 N_c F_\pm^2}
\label{d3_tensor}
\eeq
\begin{widetext}
\beqs
d_{4,T_2} & = & \frac{(N_c \pm 2)^3}
{2 \cdot 3^5 N_c^2 F_\pm^5} \bigg [ 
\Big ( 1265517N_c^9 \pm 6305850N_c^8 + 8455112N_c^7 \mp 18825808N_c^6
-47225264N_c^5 \cr\cr
&\pm& 61021088N_c^4 + 70598528N_c^3 \mp 72131840N_c^2 
+ 3066624N_c \mp 2044416 \Big ) \cr\cr
& \pm &  
8448N_c^2(N_c \mp 2)(18N_c^2\pm 11N_c-22)(12N_c^3 \mp 9N_c^2 \pm 308)
\zeta_3 \ \bigg ]
\label{d4_tensor}
\eeqs
and
\beqs
d_{5,T_2} & = & \frac{(N_c \pm 2)^4}{2 \cdot 3^6 N_c^3 F_\pm^7} \Bigg [
\bigg (-578437605N_c^{13}\mp2353001022N_c^{12}-1643220810N_c^{11} 
\pm1685855300N_c^{10} \cr\cr
&+&12567177608N_c^9\pm29240054768N_c^8-75390007296N_c^7\mp70417381376N_c^6
+243309040128N_c^5 \cr\cr
&\mp&27199484928N_c^4-228577603584N_c^3\pm143780184064N_c^2
-38053396480N_c\pm15221358592 \bigg ) \cr\cr
&+&2^7F_\pm \bigg (125388N_c^{11}\pm372762N_c^{10}-7324047N_c^9\mp9682414N_c^8
+52934332N_c^7\mp12735976N_c^6 \cr\cr
&-&192234240N_c^5\pm112670976N_c^4+164609280N_c^3\mp111598080N_c^2+2973696N_c\mp1486848 \bigg )\zeta_3 \cr\cr
&+&2^{10} \cdot 55 N_c^2(N_c \mp 2)F_\pm^2\bigg (\mp 87N_c^5 + 259N_c^4
\pm1134N_c^3-3600N_c^2\mp 5016N_c+10032\bigg )\zeta_5 \ \Bigg ] \ . 
\label{d5_tensor}
\eeqs
\end{widetext} 
We find that, in addition to the manifestly positive $d_{2,T_2}$, the
coefficient $d_{3,T_2}$ is also positive for all relevant $N_c$. Here, by
``relevant $N_c$'', we mean $N_c \ge 2$ for $S_2$ and $N_c \ge 3$ for $A_2$.
In contrast, while $d_{4,S_2}$ is positive for all relevant $N_c$, we find that
$d_{4,A_2}$ is negative for $N_c=3, \ 4, \ 5$, passes through zero at $N_c =
5.515$, and is positive for $N_c \ge 6$. Further, we find that $d_{5,S_2}$ and
$d_{5,A_2}$ are both negative for their respective physical ranges, 
$N_c \ge 2$ and $N_c \ge 3$. These sign properties 
are listed in Table \ref{dj_signs}.

Some general comments are in order concerning these $d_{j,T_2}$ expressions.
These are analogous to the comments that we made for the $\kappa_{j,T_2}$
coefficients. The property that all of the $d_{j,A_2}$ coefficients contain an
overall factor of $(N_c-2)$ (possibly raised to a power higher than 1), and
hence vanish for $N_c=2$, is a consequence of the fact that for $N_c=2$, the
$A_2$ representation is a singlet, so for SU(2), fermions in the $A_2=$ singlet
representation have no gauge interactions and do not contribute to the beta
function or $\beta'_{IR}$.

Furthermore, if $N_c=2$, then the $S_2$ representation is the same as the
adjoint representation, so the coefficients must satisfy the equality
$d_{j,S_2}=d_{j,adj}$ for this SU(2) case, and we have checked that
they do.  This equality requires (i) that the term proportional to
$\zeta_3$ in $d_{4,S_2}$ must be absent if $N_c=2$, since $d_{4,adj}$
does not contain any $\zeta_3$ term, and this is accomplished by the
factor of $(N_c-2)$ multiplying the $\zeta_3$ term in $d_{4,S_2}$; and (ii)
the term proportional to $\zeta_5$ in $d_{5,S_2}$ must be absent if
$N_c=2$, since $d_{5,adj}$ does not contain any $\zeta_5$ term, and this
is accomplished by the factor $(N_c-2)$ multiplying this $\zeta_5$ term in
$d_{5,S_2}$. Similarly, as observed before, if $N_c=3$, then the $A_2$
representation is the same as the conjugate fundamental representation, $\bar
F$, so the coefficients must satisfy the equality $d_{j,A_2}=d_{j,F}$ 
for this SU(3) case, and we have checked that they do.

In the LN limit (\ref{ln}), as discussed above in the case of the anomalous
dimension $\gamma_{IR,T_2}$, the upper ends of the interval $I_{IRZ}$ for the
$S_2$ and $A_2$ theories approach the same value, $N_{u,T_2}$, given in
Eq. (\ref{nfb1z_tensor_nc_inf}), and similarly the lower ends of this interval
for these $S_2$ and $A_2$ theories approach the same value, $N_{\ell,T_2}$,
given in Eq. (\ref{nfb2z_tensor_nc_inf}).  We denote
\beq
\hat d_{j,T_2} = \lim_{LN} d_{j,T_2} \ , 
\label{dhat_tensor_ln}
\eeq
and we find 
\beq
\hat d_{j,S_2} = \hat d_{j,A_2} \ , 
\label{dhat_s2a2}
\eeq
which we denote simply as $\hat d_{j,T_2}$. Hence, 
\beq
\lim_{LN} \beta'_{IR,S_2} = \lim_{LN} \beta'_{IR,A_2} \ . 
\label{betaprime_ir_s2a2}
\eeq

Further, again in analogy with 
Eq. (\ref{kappahat_tensor_adj}) and for the same reasons concerning group
invariants in the LN limit, we have 
\beq
\hat d_{j,T_2} = 2^{-j} \hat d_{j,adj}
\label{dhat_tensor_adj}
\eeq

From our general expressions, we calculate 
\beq
\hat d_{2,T_2} = \frac{2^2}{3^4} = 0.049383
\label{d2_tensor_Ncinf}
\eeq
\beq
\hat d_{3,T_2} = \frac{2^5}{3^7} = 1.46319 \times 10^{-2}
\label{d3_tensor_Ncinf}
\eeq
\beq
\hat d_{4,T_2} = \frac{46871}{2^6 \cdot 3^{12}} = 1.37806 \times 10^{-3} 
\label{d4_tensor_Ncinf}
\eeq
and
\beqs
\hat d_{5,T_2} &=& -\frac{7141205}{2^8 \cdot 3^{16}} +
\frac{172}{3^{12}}\zeta_3 \cr\cr
&=& -(2.58981 \times 10^{-4}) \ . 
\label{d5_tensor_Ncinf}
\eeqs

To estimate the region over which the $\Delta_f$ expansion converges, we
calculate the ratios of adjacent coefficients.  We have 
\beq
\frac{d_{2,T_2}}{d_{3,T_2}} = \frac{3N_c(18N_c^2 \pm 11N_c - 22)}
{(N_c \pm 2)(8N_c^2 \pm 3N_c - 6)} \ .  
\label{d2_over_d3_tensor}
\eeq
and similarly for the ratios $d_{j-1,T_2}/d_{j,T_2}$ for $j=4, \ 5$. 
For the LN limit, 
\beq
\frac{\hat d_{2,T_2}}{\hat d_{3,T_2}} = \bigg ( \frac{3}{2} \bigg )^3 = 3.375
\label{d2_over_d3_tensor_Ncinf}
\eeq
\beq
\frac{\hat d_{3,T_2}}{\hat d_{4,T_2}} = \frac{497664}{46871} 
= 10.618
\label{d3_over_d4_tensor_Ncinf}
\eeq
and 
\beq
\frac{\hat d_{4,T_2}}{|\hat d_{5,T_2}|} = 5.321 \ . 
\label{d4_over_d5_tensor_Ncinf}
\eeq
Since formally, $(\Delta_f)_{max}=3.375$ from Eq.
(\ref{Deltaf_ir_max_tensor_nc_inf}) and $\Delta_f=5.5$ for $N_f=2$, these
ratios indicate that the $\Delta_f$ expansion for the LN limit of this $R=T_2$
case should be reasonably accurate in the interval $I_{IRZ}$ for this case.

% ========================================================================

% section V 
\section{IR Zero of $\beta_\xi$ in the LNN Limit}
\label{xiir_section}

In this section we analyze the zeros of the rescaled five-loop beta function
in the LNN limit.  This elucidates further the result that
we first found for a finite value of $N_c$, namely $N_c=3$, in \cite{flir},
namely that for SU(3), the five-loop beta function only has a physical IR zero
in the upper range of the interval $I_{IRZ}$.  We denote the $n$-loop rescaled
beta function (\ref{betaxi}) in this LNN limit as $\beta_{\xi,n\ell}$, and its
IR zero (if such a zero exists) as $\xi_{IR,n\ell} = 4\pi x_{IR,n\ell}$. The
analytic expressions of $\xi_{IR,2\ell}$ and $\xi_{IR,3\ell}$ were given in
\cite{lnn}, together with numerical values of $\xi_{IR,n\ell}$ for $1 \le n \le
4$.  Here we extend these results to the five-loop level, using the coefficient
$\hat b_5$ in Eq. (\ref{b5hat}). As noted before, we use the $\hat b_n$ with $3
\le n \le 5$ calculated in the $\overline{\rm MS}$ scheme. The reader is
referred to \cite{lnn} for analysis of these zeros up to the four-loop level.

In general, the IR zero of the $n$-loop beta function, $\beta_{\xi,n\ell}$, is
the positive real root closest to the origin (if such a root exists) of the
equation
\beq
\sum_{\ell=1}^n \hat b_\ell \, x^{\ell-1} = 0 \ , 
\label{beta_reduced_nloop_lnn}
\eeq
of degree $n-1$ in the variable $x$. The roots of
Eq. (\ref{beta_reduced_nloop_lnn}) depend on the $n-1$ ratios 
$\hat b_\ell/{\hat b_1}$ for $2 \le \ell \le n$.  In particular, at the
five-loop level, Eq. (\ref{beta_reduced_nloop_lnn}) is the quartic equation 
\beq
\hat b_1 + \hat b_2 x + \hat b_3 x^2 + \hat b_4 x^3 + \hat b_5 x^4 = 0 \ . 
\label{beta_reduced_5loop_lnn}
\eeq
To analyze the roots of this equation, it is natural to start with $r$ in the
vicinity of $r_u=11/2$, where $\hat b_1 \to 0$ and hence one solution of
Eq. (\ref{beta_reduced_5loop_lnn}) approaches zero, matching the behavior of
$x_{IR,n\ell}$ for $2 \le n \le 4$ in this limit. As we reduce $r$ from the
value $r_u$ in the interval $I_{IRZ,r}$, we can thus calculate how the physical
IR root, $x_{IR,5\ell} = \xi_{IR,5\ell}/(4\pi)$, 
changes.  We find that, in contrast to the behavior of the IR zero of the
lower-loop beta functions $\beta_{\xi,n\ell}$ with $2 \le n \le 4$, here at the
five-loop level, as $r$ decreases past a certain value $r_{cx}$, Eq.
(\ref{beta_reduced_5loop_lnn}) (with $\hat b_n$, $n=3, \ 4, \ 5$ calculated in
the $\overline{\rm MS}$ scheme) ceases to have a physical IR zero.  We find
that the value of $r_{cx}$ is
\beq
r_{cx} = 4.32264 \ , 
\label{rc}
\eeq
to the indicated floating-point accuracy.  This is determined as the relevant
root of the discriminant of Eq. (\ref{beta_reduced_5loop_lnn}), which is a
polynomial of degree 15 in the variable $r$.  (The discriminants of the
corresponding equations at loop levels 3 and 4 are polynomials of degree 3 and
8 in $r$.)  For example, for the illustrative value $r=5$, near to the upper
end of the interval $I_{IRZ,r}$, Eq. (\ref{beta_reduced_5loop_lnn}) has the
solutions in $x$, expressed in terms of $\xi = 4\pi x$:
$\xi = 0.36300$, $1.69540$, and $-1.48884 \pm 1.08446 i$. 
Of these, we identify the first as the IR zero, $\xi_{IR,5\ell}$.  As $r$
decreases and approaches $r_{cx}$ from above, the two real roots approach a
common value, $\xi \simeq 1.312$ and as $r$ decreases below $r_{cx}$,
Eq. (\ref{beta_reduced_5loop_lnn}) has only two complex-conjugate pairs of
solutions, roots, but no real positive solution. In Table
\ref{xiir_nloop_values} we list our new results for $\xi_{IR,5\ell}$, in
comparison with the previously calculated values of $\xi_{IR,n\ell}$ in the LNN
limit with $2 \le n \le 4$ from Table III of \cite{lnn}. 
Although we list $\xi_{IR,n\ell}$ values extending
to the lower part of the interval $I_{IRZ,r}$ for completeness, it is clear
that a number of these values are too large for the perturbative calculations
to be reliable.  For values of $r$ where the five-loop beta function
(calculated in the $\overline{\rm MS}$ scheme) has no physical IR zero, we
denote this as unphysical (u).

We note that the absence of a physical IR zero
in the five-loop beta function (calculated in the $\overline{\rm MS}$ scheme)
for $N_f$ values in the lower portion of the interval $I_{IRZ}$ does not
necessarily imply that higher-loop calculations would yield similarly
unphysical results.  We gave an example of this in Section VIII of the second
paper in \cite{sch}, using an illustrative exact beta function.  In this
example, it was shown that a certain order of truncation of the Taylor
series expansion in powers of $\alpha$ for this beta function did not yield any
physical IR zero, but higher orders did converge toward this zero.

% ========================================================================

% section VI 
\section{$\Delta_f$ Expansion for $\alpha_{IR}$ to $O(\Delta_f^4)$ }
\label{alfir_Deltasection}

\subsection{General $G$ and $R$}
\label{alfir_Deltageneral}

Since the exact $\alpha_{IR}$ (and also the $n$-loop approximation to this
exact $\alpha_{IR}$) vanishes as functions of $\Delta_f$, it follows that one
can expand it as a power series in this variable. This expansion was given
above as Eq. (\ref{alfir_Deltaseries}), and it was noted that the calculation
of the coefficient $a_j$ requires, as input, the $\ell$-loop beta function
coefficients $b_\ell$ with $1 \le \ell \le j+1$.  We denote the truncation of
this infinite series (\ref{alfir_Deltaseries}) to maximal power $j=p$ as
$\alpha_{IR,\Delta_f^p}$.  Here we present a calculation of this series to
$O(\Delta_f^4)$, which is the highest order to which it has been calculated.
Since $\alpha_{IR}$ is scheme-dependent, it follows that the $a_j$ coefficients
in Eq. (\ref{alfir_Deltaseries}) are also scheme-dependent, in contrast to the
scheme-independent coefficients $\kappa_j$ and $d_j$ in
Eqs. (\ref{gamma_ir_Deltaseries}) and (\ref{betaprime_ir_Deltaseries}).
Nevertheless, it is still worthwhile to calculate these coefficients $a_j$ and
the resultant finite-order approximations $\alpha_{IR,\Delta_f^p}$, for several
reasons.  First, this method has the advantage that $\alpha_{IR,\Delta_f^p}$ is
always physical and thus avoids the problem that we found in \cite{flir} and
have further studied above, that the five-loop beta function calculated in the
$\overline{\rm MS}$ scheme does not have a physical IR zero in the lower part
of the interval $I_{IRZ}$.  In \cite{gsi}, for the special case $G={\rm SU}(3)$
and $R=F$, we presented the $a_j$ (denoted $\tilde a_j$ there) for $1 \le j \le
4$.

Here, as a new result, we present the expressions for the $a_j$ for arbitrary
$G$ and $R$, for $1 \le j \le 4$.  For this purpose, we use the $n$-loop beta
function coefficients $b_n$ with $3 \le n \le 5$ calculated in the
$\overline{\rm MS}$ scheme.  In particular, our result for $a_4$ makes use of
the recently calculated five-loop beta function for general $G$ and $R$
\cite{b5}. 

For general $G$ and $R$, recalling the definition of the denominator
factor $D=7C_A+11C_f$ in Eq. (\ref{d}), we find 
\beq
a_1 = \frac{4T_f}{ 3C_A D}
\label{a1}
\eeq
\beq
a_2 = \frac{2T_f^2(-287C_A^2+1208C_AC_f+924C_f^2)}{3^3 C_A^2 D^3} 
\label{a2}
\eeq
\begin{widetext}
\beqs
a_3 &=& \frac{2T_f}{3^5C_A^4 D^5} \Bigg [ C_AT_f^2 \bigg ( -71491C_A^4
+ 372680C_A^3C_f + 2102252C_A^2C_f^2 + 835560C_AC_f^3+836352C_f^4\bigg ) 
\cr\cr
&-& 2560 T_f^2 D \frac{d_A^{abcd}d_A^{abcd}}{d_A}
+45056 C_A T_f D \frac{d_R^{abcd}d_A^{abcd}}{d_A}
-170368C_A^2T_fD \frac{d_R^{abcd}d_R^{abcd}}{d_A} \cr\cr
&+&4224D\bigg [ 3C_A^2T_f^2D(C_A-C_f) 
+ 16T_f^2 \frac{d_A^{abcd}d_A^{abcd}}{d_A}
-104 C_AT_f \frac{d_R^{abcd}d_A^{abcd}}{d_A}
+88C_A^2 \frac{d_R^{abcd}d_R^{abcd}}{d_A} \bigg ]\zeta_3 \Bigg ]
\label{a3}
\eeqs
and
\beqs
a_4 &=& \frac{T_f^2}{2 \cdot 3^7 C_A^5 D^7} \Bigg [ 
C_AT_f^2 \bigg ( 194849725C_A^6-684457480C_A^5C_f
+4175949036C_A^4C_f^2+13292017040C_A^3C_f^3 \cr\cr
&+&2617931536C_A^2C_f^4+8758858944C_AC_f^5+85865472C_f^6 \bigg ) 
\cr\cr
&+&2^{10}T_f^2 D \frac{d_A^{abcd}d_A^{abcd}}{d_A}
\Big (21287C_A^2-5504C_AC_f-19140C_f^2\Big ) \cr\cr
&+&2^{10}C_AT_f D \frac{d_R^{abcd}d_A^{abcd}}{d_A}
\Big (-194005C_A^2+253231C_AC_f+136488C_f^2\Big ) \cr\cr
&+&2^8 \cdot 11^2 C_A^2 D \frac{d_R^{abcd}d_R^{abcd}}{d_A}
\Big (917C_A^2-40412C_AC_f+26796C_f^2\Big ) \cr\cr
&-&2304 D \bigg [ C_A T_f^2 D\bigg ( 15456C_A^4-75039C_A^3C_f
+45716C_A^2C_f^2+23848C_AC_f^3+2112C_f^4\bigg ) \cr\cr
&+&16T_f^2 \frac{d_A^{abcd}d_A^{abcd}}{d_A} \Big (8610C_A^2
-15037C_AC_f-14036C_f^2\Big ) 
-8C_AT_f \frac{d_R^{abcd}d_A^{abcd}}{d_A} \Big (95984C_A^2-190355C_AC_f
-135036C_f^2 \Big ) \cr\cr
&+&88C_A^2 \frac{d_R^{abcd}d_R^{abcd}}{d_A} \Big (3199C_A^2-26004C_AC_f
-17908C_f^2 \Big ) \bigg ]\zeta_3 \cr\cr
&+&337920C_A D^2 \bigg [ -9C_AT_f^2D(C_A-C_f)(C_A+2C_f) 
-160T_f^2 \frac{d_A^{abcd}d_A^{abcd}}{d_A} \cr\cr
&+&80T_f(10C_A+3C_f)\frac{d_R^{abcd}d_A^{abcd}}{d_A}
+440C_A(C_A-3C_f)\frac{d_R^{abcd}d_R^{abcd}}{d_A} \bigg ] \zeta_5 \ \Bigg ] 
\ . 
\label{a4}
\eeqs
\end{widetext}
We next specialize to the case $G={\rm SU}(N_c)$ and give explicit reductions
of these general formulas for the representations of interest here. 

% ====================== R=fund ======================================

\subsection{$R=F$} 

For $R=F$, our general results (\ref{a1})-(\ref{a4}) reduce to the following
expressions: 
\beq
a_{1,F} = \frac{4}{3(25N_c^2-11)}
\label{a1_fund}
\eeq
\beq
a_{2,F} = \frac{4(548N_c^4-1066N_c^2+231)}{3^3N_c(25N_c^2-11)^3}
\label{a2_fund}
\eeq
\begin{widetext} 
\beqs
a_{3,F} & = & 
\frac{2^3}{3^5 N_c^2(25N_c^2-11)^5} \bigg [ \Big (730529N_c^8-1105385N_c^6
-719758N_c^4+389235N_c^2 + 52272 \Big ) \cr\cr
&+&1584 N_c^2(25N_c^2-11)\Big (25N_c^4-18N_c^2+77 \Big )\zeta_3 
\ \bigg ] 
\label{a3_fund}
\eeqs
and
\beqs
a_{4,F}&=&\frac{2^2}{3^7 N_c^3(25N_c^2-11)^7}\bigg [\Big ( 2783259085 N_c^{12} 
- 7278665930N_c^{10} + 4578046419 N_c^8 - 1719569282 N_c^6 \cr\cr
&+& 2905511455N_c^4 - 1137735654N_c^2 + 1341648 \Big ) \cr\cr 
&+& 288(25N_c^2-11) \Big ( 548025N_c^{10} - 1857036 N_c^8 + 4694107 N_c^6 
- 5482510 N_c^4 + 1098130 N_c^2 + 2904 \Big )\zeta_3 \cr\cr
&-& 190080N_c^2(25N_c^2-11)^2
\Big (40N_c^6-27N_c^4+124N_c^2-209\Big )\zeta_5 \ \bigg ] \ . 
\label{a4_fund}
\eeqs
\end{widetext}
We have checked that setting $N_c=3$ in our new $a_4$ coefficient in
Eq. (\ref{a4_fund}) yields agreement with the value that we obtained previous
for this special case in (Eq. (14) of) Ref. \cite{gsi}. 

We comment next on the signs of these coefficients.  The coefficient $a_1$ is
manifestly positive for arbitrary group $G$ and fermion representation $R$.  We
find that $a_{2,F}$ and $a_{3,F}$ are also positive for all physical $N_c \ge
2$.  In contrast, we find that $a_{4,F}$ is negative for $N_c=2$ and positive
for $N_c \ge 3$.  With $N_c$ generalized from positive integers to positive
real numbers in the range $N_c \ge 2$, we calculate that as $N_c$ increases
through the value $N_c=2.1184$ (given to the indicated accuracy), $a_{4,F}$
passes through zero with positive slope. 

We list below the explicit numerical expressions for $\alpha_{IR}$ to order
$\Delta_f^4$, for $N_c=2, \ 3, \ 4$ and $R=F$, denoted , the indicated
floating-point precision: 
\begin{widetext}
\beqs
{\rm SU}(2): \quad \alpha_{IR,F,\Delta_f^4} & =& \Delta_f \Big [ 
  (0.18826 
+ (0.62521 \times 10^{-2})\Delta_f 
+ (0.70548 \times 10^{-2})\Delta_f^2
- (0.45387 \times 10^{-4})\Delta_f^3 \ \Big ] \cr\cr
& &  
\label{alfir_sunf_p4_su2}
\eeqs
\beqs
{\rm SU}(3): \quad \alpha_{IR,F,\Delta_f^4} & =& \Delta_f \Big [ 
  (0.078295  
+ (2.2178 \times 10^{-3})\Delta_f 
+ (1.1314 \times 10^{-3})\Delta_f^2
+ (2.1932 \times 10^{-5})\Delta_f^3 \ \Big ] \cr\cr
& &  
\label{alfir_sunf_p4_su3}
\eeqs
and
\beqs
{\rm SU}(4): \quad \alpha_{IR,F,\Delta_f^4} & =& \Delta_f \Big [ 
  (0.043072
+ (0.97619 \times 10^{-3})\Delta_f 
+ (0.33823 \times 10^{-3})\Delta_f^2
+ (0.71999 \times 10^{-5})\Delta_f^3 \ \Big ] \ . \cr\cr
& &  
\label{alfir_sunf_p4_su4}
\eeqs
\end{widetext}

\begin{figure}
  \begin{center}
    \includegraphics[height=6cm]{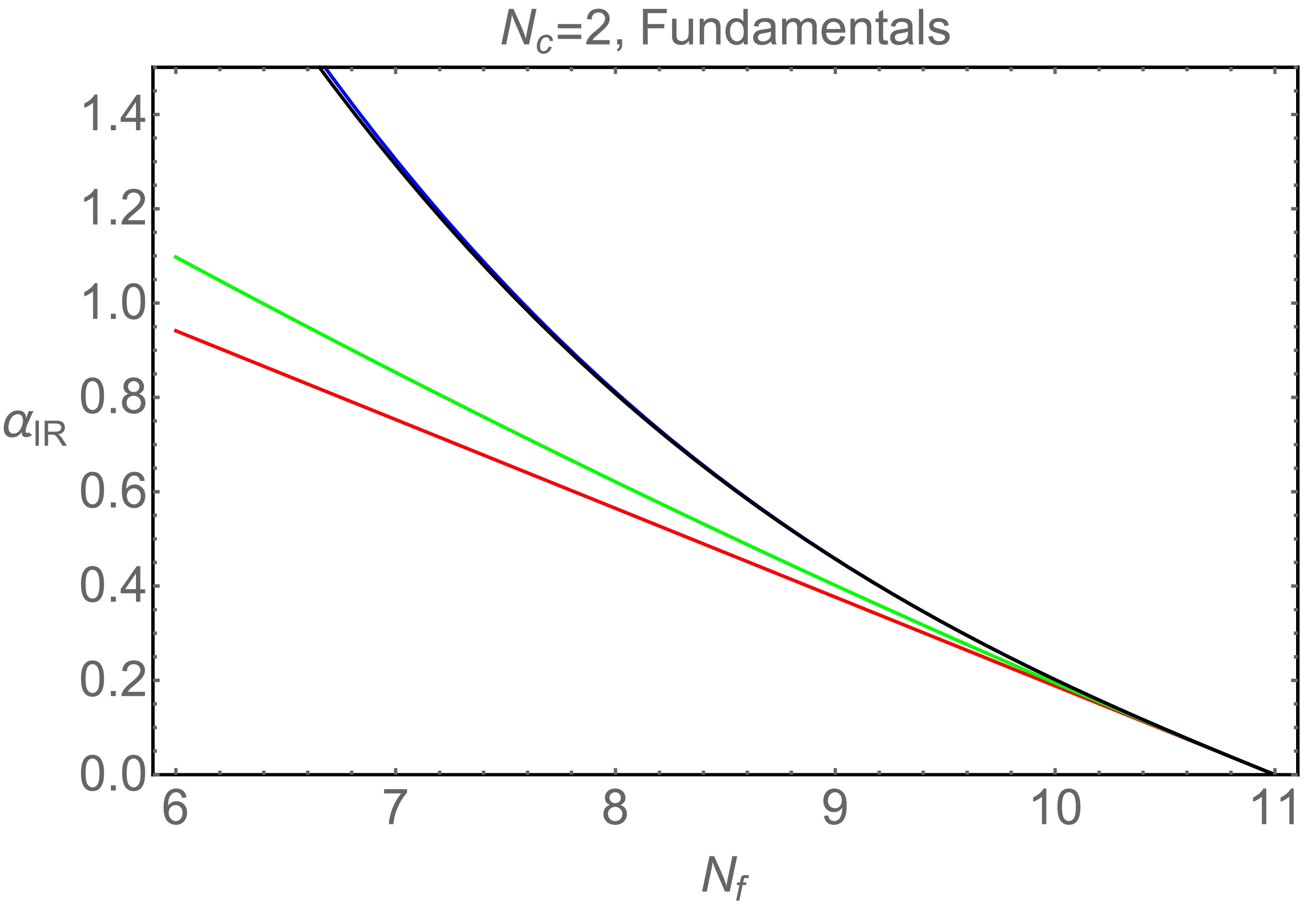}
  \end{center}
\caption{Plot of $\alpha_{IR,F,\Delta_f^p}$ 
(denoted as $\alpha_{IR}$ on the vertical axis) 
with $1 \le p \le 4$ for $G={\rm SU}(2)$, as functions of $N_f \in I_{IRZ}$. 
From bottom to top, the curves (with colors online) refer to 
$\alpha_{IR,F,\Delta_f}$ (red),
$\alpha_{IR,F,\Delta_f^2}$ (green),
$\alpha_{IR,F,\Delta_f^3}$ (blue),
$\alpha_{IR,F,\Delta_f^4}$ (black). Note that the curves for 
$\alpha_{IR,F,\Delta_f^3}$ and $\alpha_{IR,\Delta_f^4}$ are so close as to be
indistinguishable in this figure.}
\label{alphaNc2fund_plot}
\end{figure}

\begin{figure}
  \begin{center}
    \includegraphics[height=6cm]{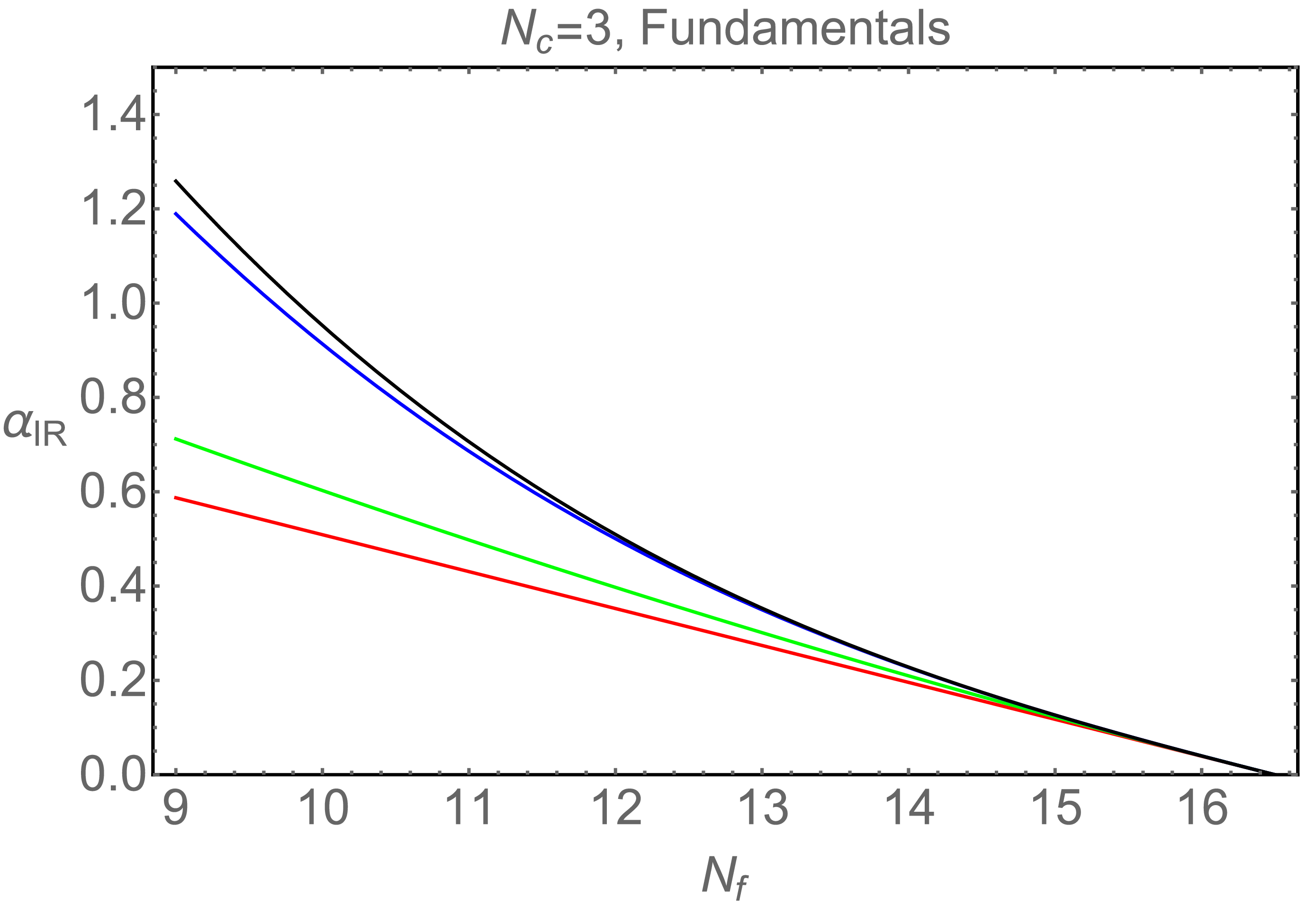}
  \end{center}
\caption{Plot of $\alpha_{IR,F,\Delta_f^p}$ 
with $1 \le p \le 4$ for $G={\rm SU}(3)$, as functions of $N_f \in I_{IRZ}$. 
From bottom to top, the curves (with colors online) refer to 
$\alpha_{IR,F,\Delta_f}$ (red),
$\alpha_{IR,F,\Delta_f^2}$ (green),
$\alpha_{IR,F,\Delta_f^3}$ (blue),
$\alpha_{IR,F,\Delta_f^4}$ (black).}
\label{alphaNc3fund_plot}
\end{figure}

\begin{figure}
  \begin{center}
    \includegraphics[height=6cm]{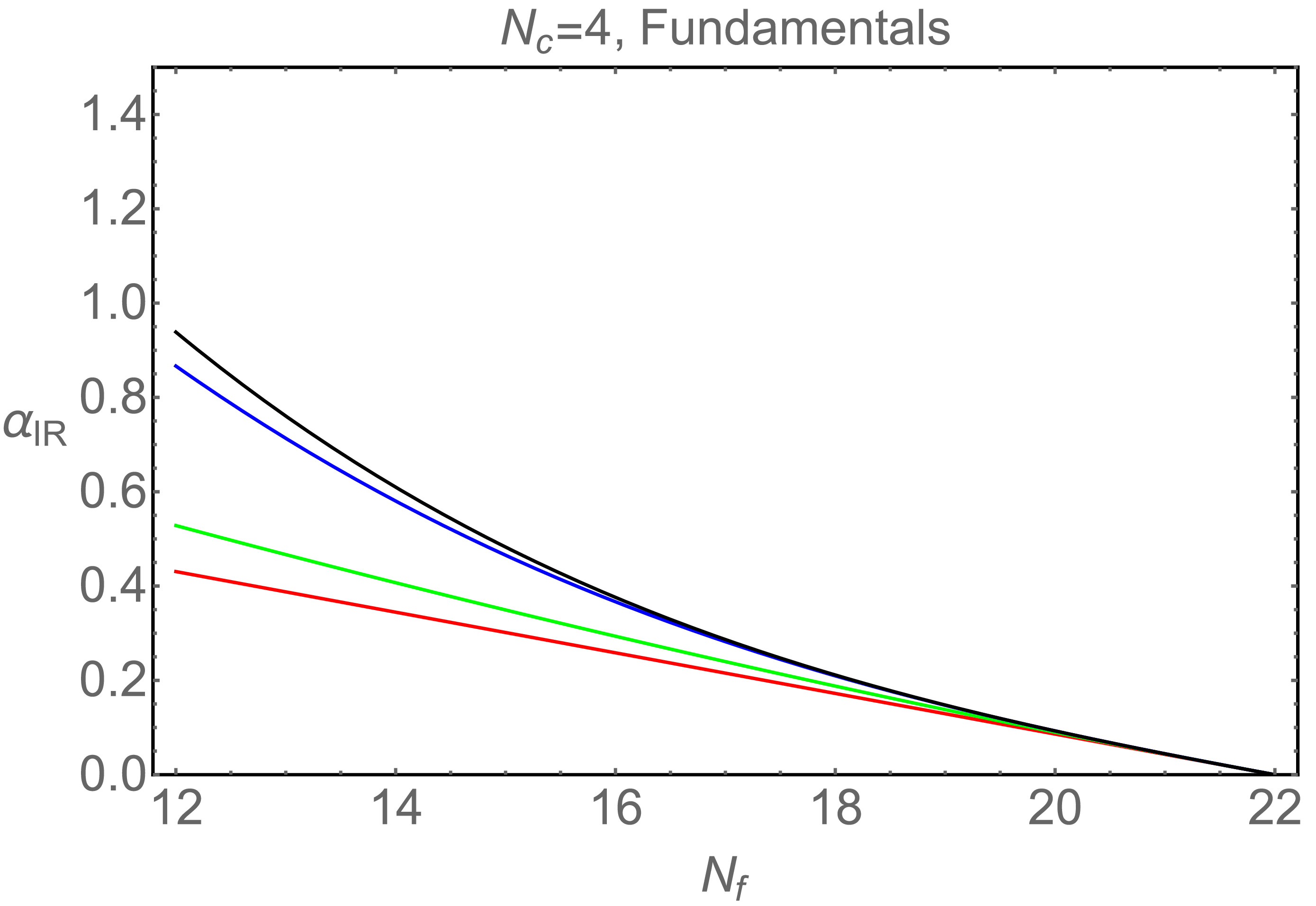}
  \end{center}
\caption{Plot of $\alpha_{IR,F,\Delta_f^p}$ 
with $1 \le p \le 4$ for $G={\rm SU}(4)$, as functions of $N_f \in I_{IRZ}$. 
From bottom to top, the curves (with colors online) refer to 
$\alpha_{IR,F,\Delta_f}$ (red),
$\alpha_{IR,F,\Delta_f^2}$ (green),
$\alpha_{IR,F,\Delta_f^3}$ (blue),
$\alpha_{IR,F,\Delta_f^4}$ (black).}
\label{alphaNc4fund_plot}
\end{figure}

In Figs. \ref{alphaNc2fund_plot}-\ref{alphaNc4fund_plot} we show 
$\alpha_{IR,F,\Delta_f^p}$ for $N_c=2, \ 3, \ 4$ and $1 \le p \le 4$ as a
function of $N_f$. Note that
in Fig. \ref{alphaNc2fund_plot} the curves for $p=3$ and $p=4$ are so close
as to be indistinguishable for this  this range of $N_f$. 

In Table \ref{alfir_nloop_Delta} we compare the values of the IR zero of the
$n$-loop beta function for $1 \le n \le 4$ from \cite{bvh} with our values of
$\alpha_{IR,F,\Delta_f^p}$ for $1 \le p \le 4$ and $N_c=2, \ 3, \ 4$.  
Since the calculation of $\alpha_{IR,n\ell}$ uses the $\ell$-loop
beta function coefficients $b_\ell$ with $1 \le \ell \le n$, while the
calculation of $\alpha_{IR,\Delta_f^p}$ uses the $b_\ell$ for 
$1 \le \ell \le p+1$, the closest comparison is of $\alpha_{IR,n\ell}$ with
$\alpha_{IR,\Delta_f^{n-1}}$, which both use $n$-loop information from the beta
function.  Although, for completeness, we include values of $\alpha_{IR,2\ell}$
for $N_f$ extending down to the lower end of the respective intervals $I_{IRZ}$
for each value of $N_c$, we caution that in a number of cases, including
$N_f=6$ for SU(2), $N_f=9$ for SU(3), and $10 \le N_f \le 12$ for SU(4), these
values of $\alpha_{IR,2\ell}$ are too large for the perturbative $n$-loop
calculations to be reliable.  Concerning the comparison of the
higher-order $n$-loop values of $\alpha_{IR,n\ell}$ with our values of
$\alpha_{IR,F,\Delta_f^p}$, we see that for a given $N_c$ and $N_f$, at the
upper end of the non-Abelian Coulomb phase, the values of
$\alpha_{IR,\Delta_f^{n-1}}$ and $\alpha_{IR,n\ell}$ are quite close to each
other, but $N_f$ decreases in this NACP in in the interval $I_{IRZ}$,
$\alpha_{IR,\Delta_f^{n-1}}$ becomes slightly larger than $\alpha_{IR,n\ell}$.

In the LNN limit, for the IR zero of the rescaled beta function, we write 
\beq
\xi_{IR} = 4 \pi \sum_{j=1}^\infty \hat a_{j,F} \Delta_r^j \quad (
{\rm LNN \ limit}) \ , 
\label{xi_Deltaseries_lnn}
\eeq
where
\beq
\hat a_{j,F} = \lim_{LNN} N_c^{j+1} a_{j,F} \ . 
\label{ajhat}
\eeq
From our results for $a_{j,F}$, we calculate
\beq
\hat a_{1,F} = \frac{4}{3 \cdot 5^2} = 0.053333
\label{a1hat}
\eeq
\beq
\hat a_{2,F} = \frac{2192}{3^3 \cdot 5^6} = 0.519585 \times 10^{-2} 
\label{a2hat}
\eeq
\beq
\hat a_{3,F} = \frac{5844232}{3^5 \cdot 5^{10}} + 
\frac{1408}{3^3 \cdot 5^6}\zeta_3 = 0.647460 \times 10^{-2} 
\label{a3hat}
\eeq
and
\beqs
\hat a_{4,F} &=& \frac{2226607268}{3^7 \cdot 5^{13}} +
\frac{935296}{3^4 \cdot 5^{10}}\zeta_3 - \frac{45056}{3^4 \cdot 5^8}\zeta_5 
\cr\cr
&=& 0.778770 \times 10^{-3} \ . 
\label{a4hat}
\eeqs
Thus, in the LNN limit, the expansion of $\xi_{IR}$, to $O(\Delta_r^4)$, is 
\beqs
&&\xi_{IR,\Delta_r^4} =
 4\pi \Delta_r \bigg [ 0.053333 + (0.519585 \times 10^{-2})\Delta_r
\cr\cr
&+&(0.647460 \times 10^{-2})\Delta_r^2 + (0.778770 \times 10^{-3})\Delta_r^3 
\bigg ] \ . \cr\cr
& & 
\label{xiir_lnn}
\eeqs
%

% ========================== R = adj =======================================

\subsection{$R=adj$}

For $R=adj$, our general results (\ref{a1})-(\ref{a4}) reduce to the following
expressions:
\beq
a_{1,adj} = \frac{2}{3^3N_c} = \frac{0.074747}{N_c} 
\label{a1_adj}
\eeq
\beq
a_{2,adj} = \frac{205}{2^2 \cdot 3^7 N_c} = \frac{0.023434}{N_c} 
\label{a2_adj}
\eeq
\beqs
a_{3,adj}&=& \frac{49129}{2^4 \cdot 3^{11}N_c} - \frac{296}{3^9 N_c^3} 
\cr\cr
&=& \frac{0.017333}{N_c} - \frac{0.015038}{N_c^3} 
\label{a3_adj}
\eeqs
and
\beqs
a_{4,adj}&=&\bigg ( \frac{38811689}{2^8 \cdot 3^{15}} - \frac{40}{3^9}\zeta_3
\bigg )\frac{1}{N_c} \cr\cr
&+&\bigg (-\frac{3157}{3^{13}} + \frac{25616}{3^{12}} \bigg )\frac{1}{N_c^3}
\cr\cr
&=&\frac{0.0081230}{N_c} + \frac{0.055960}{N_c^3} \ . 
\label{a4_adj}
\eeqs
The coefficients $a_{j,adj}$ with $j=1, \ 2, \ 4$ are manifestly positive, and
we find that $a_{3,adj}$ is also positive for all $N_c \ge 2$.  

Since for the adjoint representation, $R=adj$, the upper and lower boundaries
of the interval $I_{IRZ}$, $N_{u,T_2}=11/2$ in Eq. (\ref{nfb1z_adj}) and 
$N_{\ell,adj}=17/16$ in (\ref{nfb2z_adj}), are independent of $N_f$, it follows
that $\Delta_f=N_u-N_f$ is also independent of $N_c$.  From the general
formula (\ref{alfir_Deltaseries}), in the LN limit of a theory with fermions in
a two-index representation $R_2$, including the adjoint and symmetric and
antisymmetric tensors, we can write 
\beq
\quad \xi_{IR} = 4 \pi \sum_{j=1}^\infty \hat a_{j,R_2} \Delta_f^j \quad (
{\rm LN \ limit}) \ , 
\label{xi_Deltaseries_ln}
\eeq
where
\beq
\hat a_{j,R_2} = \lim_{LN} N_c a_{j,R_2} \ . 
\label{ajhat_r2}
\eeq

From our calculations above, setting $R_2=adj$, we have 
\beq
\hat a_{1,adj} = \frac{2}{3^3} = 0.074747
\label{a1hat_adj}
\eeq
\beq
\hat a_{2,adj} = \frac{205}{2^2 \cdot 3^7} = 0.023434
\label{a2hat_adj}
\eeq
\beq
\hat a_{3,adj} =  \frac{49129}{2^4 \cdot 3^{11}} = 0.017333
\label{a3hat_adj}
\eeq
and
\beq
\hat a_{4,adj}= \frac{38811689}{2^8 \cdot 3^{15}} - \frac{40}{3^9}\zeta_3
=0.0081230 \ . 
\label{hata4_adj}
\eeq
%

% ========================== R = T_2 ======================================

\subsection{$R=S_2, \ A_2$ }

For $R$ equal to the symmetric or antisymmetric rank-2 tensor representations,
$S_2$ and $A_2$, 
we give the reductions of our general results (\ref{a1})-(\ref{a4}) next. 
As before, it is convenient to consider these together, since many terms differ
only by sign reversal.  As above, the upper and lower signs refer to the $S_2$
and $A_2$ representations, respectively.  Also, as before, for $A_2$, we
require that $N_c \ge 3$.  Recalling the definition of the denominator factor
$F_\pm$ in Eq. (\ref{fplusminus}), we have
\beq
a_{1,T_2} = \frac{2(N_c \pm 2)}{3F_\pm} 
\label{a1_tensor}
\eeq
\begin{widetext}
\beq
a_{2,T_2} = \frac{(N_c \pm 2)^2(1845N_c^4 \pm 3056N_c^3 - 5188N_c^2 \mp3696N_c
+ 3696)}{2 \cdot 3^3 N_c F_\pm^3}
\label{a2_tensor}
\eeq
\beqs
a_{3,T_2} &=& \frac{(N_c \pm 2)^2}{2^2 \cdot 3^5 N_c^2 F_\pm^5} \Bigg [ 
\bigg ( 3979449N_c^9 \pm 16999002N_c^8 + 761444N_c^7 \mp 52233472N_c^6 
-3099440N_c^5 \cr\cr
&\pm& 11578144N_c^4 - 16368000N_c^3 \pm 36440448N_c^2
-40144896N_c \pm 26763264 \bigg ) \cr\cr
&\mp& 12672N_c^2(N_c \mp2)F_\pm(12N_c^3 \mp 9N_c^2 \pm308)\zeta_3 \ \Bigg ] 
\label{a3_tensor} 
\eeqs
and
\beqs
& & a_{4,T_2} = \frac{(N_c \pm 2)^3}{2^5 \cdot 3^7 N_c^3 F_\pm^7} \Bigg [ 
\bigg ( 28293721281N_c^{13} \pm 156860406306N_c^{12} + 13832572748N_c^{11}
\mp 547968555432N_c^{10} \cr\cr
&-&929147053664N_c^9 \pm428226859968N_c^8
+2279581786496N_c^7 \pm586028410624N_c^6 - 4633121830656N_c^5 \cr\cr
&\pm& 143588589056N_c^4 +4686268342272N_c^3 \mp2321839534080N_c^2
-27476951040N_c \pm10990780416 \bigg ) \cr\cr
&-&2304F_\pm \bigg ( 131220N_c^{11} \pm695898N_c^{10}-6916683N_c^9
\mp 10687114N_c^8 +60333108N_c^7 \mp12100440N_c^6 \cr\cr
&-& 239418432N_c^5
\pm 140804928N_c^4 + 208053120N_c^3 \mp 140560640N_c^2 +2973696N_c
\mp 1486848 \bigg )\zeta_3 \cr\cr
&+& 1013760 N_c^2 (N_c \mp 2)F_\pm^2\bigg ( \pm 87N_c^5-259N_c^4 \mp1134N_c^3
+ 3600N_c^2 \pm 5016N_c -10032 \bigg )\zeta_5 \ \Bigg ] \ . 
\label{a4_tensor}
\eeqs
\end{widetext}

The same general comments that we made before concerning factors in the
$\kappa_{j,T_2}$ and $d_{j,T_2}$ coefficients also apply here.  Thus, for
arbitrary $j$, the $a_{j,A_2}$ coefficients contain at least one overall factor
of $(N_c-2)$ and hence vanish for $N_c=2$, as a result of the fact that for
$N_c=2$, the $A_2$ representation is a singlet, so for SU(2), fermions in the
$A_2=$ singlet representation are free fields and hence make no
contribution to the beta function. Moreover, if $N_c=2$, then the $S_2$
representation is the same as the adjoint representation, so the $a_j$
coefficients must satisfy the equality $a_{j,S_2}=a_{j,adj}$ for this SU(2)
case, and we have checked that they do.  Similarly, if $N_c=3$, then the $A_2$
representation is the same as the conjugate fundamental representation, $\bar
F$, so these coefficients must satisfy the equality $a_{j,A_2}=a_{j,F}$ for
this SU(3) case, and we have checked that they do.

We next consider the LN limit of the theory with fermions in the $S_2$ or $A_2$
representations.  Using the definition (\ref{ajhat_r2}) with $R_2=S_2$ and
$R_2=A_2$, we find that 
\beq
\hat a_{j,S_2} = \hat a_{j,A_2}
\label{ajhat_s2a2}
\eeq
so we denote these simply as $\hat a_{j,T_2}$.  In general, for the same
group-theoretical reasons as led to the LN relation $\hat
\kappa_{j,T_2}=2^{-j}\hat\kappa_{j,adj}$ in Eq. (\ref{kappahat_tensor_adj}) and
the LN relation $\hat d_{j,T_2}=2^{-j} \hat d_{j,adj}$ in
Eq. (\ref{dhat_tensor_adj}),we have, in the LN limit,
\beq
\hat a_{j,T_2} = 2^{-j} \hat a_{j,adj} \ . 
\label{ahat_tensor_adj}
\eeq
Explicitly, we calculate 
\beq
\hat a_{1,T_2} = \frac{1}{3^3} = 0.05333
\label{a1hat_tensor}
\eeq
\beq
\hat a_{2,T_2} = \frac{205}{2^4 \cdot 3^7} = 0.58585 \times 10^{-2} 
\label{a2hat_tensor}
\eeq
\beq
\hat a_{3,T_2} = \frac{49129}{2^7 \cdot 3^{11}} = 2.16668 \times 10^{-3} 
\label{a3hat_tensor}
\eeq
and
\beq
\hat a_{4,T_2} = \frac{38811689}{2^{12} \cdot 3^{15}} 
- \frac{5}{2 \cdot 3^9}\zeta_3 = 0.50769 \times 10^{-3} \ . 
\label{a4hat_tensor}
\eeq
%

% ==========================================================================

% Section VII 
\section{Conclusions}
\label{conclusion_section}

In conclusion, in this paper we have presented a number of new results on
scheme-independent calculations of various quantities in an asymptotically free
vectorial gauge theory having an IR zero of the beta function.  We have
presented scheme-independent series expansions of the anomalous dimension
$\gamma_{\bar\psi\psi,IR}$ to $O(\Delta_f^4)$ and the derivative of the beta
function, $\beta'_{IR}$, to $O(\Delta_f^5)$ for a theory with a general gauge
group $G$ and $N_f$ fermions in a representation $R$ of $G$.  We have given
reductions of our general formulas for theories with $G={\rm SU}(N_c)$ and $R$
equal to the fundamental, adjoint, and symmetric and antisymmetric rank-2
tensor representations. We have compared our scheme-independent calculations of
$\gamma_{\bar\psi\psi,IR}$ and $\beta'_{IR}$ with previous $n$-loop values of
these quantities calculated via series expansions in powers of the coupling.
For a number of specific theories we have also compared our new
scheme-independent calculations of $\gamma_{\bar\psi\psi,IR}$ and $\beta'_{IR}$
with lattice measurements.  We have shown that for all of the representations
we have studied, and for the full range $1 \le p \le 4$ for which we have
performed calculations, $\gamma_{\bar\psi\psi,IR}$ calculated to
$O(\Delta_f^p)$, denoted $\gamma_{\bar\psi\psi,IR,\Delta_f^p}$, increases
monotonically with decreasing $N_f$ (i.e., increasing $\Delta_f$) and, for a
fixed $N_f$, $\gamma_{\bar\psi\psi,IR,\Delta_f^p}$, increases monotonically
with the order $p$.  For the representation $R=F$, we have presented results
for the limit $N_c \to \infty$ and $N_f \to \infty$ with $N_f/N_c$ fixed.
These higher-order results have been applied to obtain estimates of the lower
end of the (IR-conformal) non-Abelian Coulomb phase.  We have confirmed and
extended our earlier finding that our expansions in powers of $\Delta_f$ should
be reasonably accurate throughout a substantial portion of the non-Abelian
Coulomb phase.  We have also given expansions for $\alpha_{IR}$ calculated to
$O(\Delta_f^4)$ which provide a useful complementary approach to calculating
$\alpha_{IR}$.  Our scheme-independent calculations of physical quantities at a
conformal IR fixed point yield new information about the properties of a
conformal field theory.

% =======================================================================

\begin{acknowledgments}

This research was supported in part by the Danish National
Research Foundation grant DNRF90 to CP$^3$-Origins at SDU (T.A.R.) and 
by the U.S. NSF Grant NSF-PHY-16-1620628 (R.S.) 

\end{acknowledgments}

% =======================================================================

\begin{appendix}

\section{Series Coefficients for $\beta_\xi$ and 
$\gamma_{\bar\psi\psi}$ in the LNN Limit}
\label{bellhatappendix} 

For reference, we list here the rescaled series coefficients for 
$\beta_\xi$ and $\gamma_{\bar\psi\psi}$ in the LNN limit (\ref{lnn}).  
From the (scheme-independent) one-loop and two-loop coefficients in the beta
function \cite{b1,b2},
it follows that in the LNN limit the $\hat b_\ell$ with $\ell=1,2$ are 
\beqs
\hat b_1 & = & \frac{1}{3}(11-2r) \cr\cr
         & = & 3.667-0.667r
\label{b1hat}
\eeqs
and
\beqs
\hat b_2 & = & \frac{1}{3}(34-13r)  \cr\cr
         & = & 11.333 - 4.333r \ . 
\label{b2hat}
\eeqs
The coefficients $b_3$ and $b_4$ have been calculated in the 
$\overline{\rm MS}$ scheme \cite{b3,b4}. With these inputs, one has \cite{lnn}
\beqs
\hat b_3 & = & \frac{1}{54}(2857-1709r+112r^2) \cr\cr
         & = & 52.907 - 31.648r + 2.074r^2 
\label{b3hat}
\eeqs
and 
\beqs
\hat b_4 & = & \bigg ( \frac{150473}{486} + \frac{44}{9}\zeta_3 \bigg ) 
- \bigg (\frac{485513}{1944} + \frac{20}{9}\zeta_3 \bigg ) r \cr\cr
& + & \bigg ( \frac{8654}{243} + \frac{28}{3}\zeta_3 \bigg ) r^2 
+ \bigg ( \frac{130}{243}  \bigg ) r^3 \cr\cr
& = & 315.492 - 252.421r + 46.832r^2 + 0.5350r^3 \ . \cr\cr
& & 
\label{b4hat}
\eeqs
The behavior of these coefficients $\hat b_\ell$ as functions of $r$ was
discussed in \cite{lnn} for $1 \le \ell \le 4$.  The positivity of $\hat b_1$
is equivalent to the asymptotic freedom of the theory, and requires $r$ to lie
in the interval $0 \le r < 11/2$.  The existence of an IR zero in the two-loop
beta function is equivalent to the condition that $\hat b_2 < 0$, which, in
turn, is equivalent to the condition that $r \in I_{IRZ,r}$ as given in
Eq. (\ref{intervalr}) . In this interval,
$\hat b_3$ is negative-definite, while $\hat b_4$ is negative for for $2.615 <
r < 3.119$ and positive for $3.119 < r < 5.5$ \cite{lnn}.

For the coefficients $\hat c_\ell$ in Eq. (\ref{gamma_ir_lnn}), from 
\cite{c4} and references therein, one has \cite{lnn}  
\beq
\hat c_1 = 3 \ , 
\label{chat1}
\eeq
\beq
\hat c_2 = \frac{203}{12} - \frac{5}{3} r \ , 
\label{chat2}
\eeq
\beq
\hat c_3 = \frac{11413}{108}-\bigg ( \frac{1177}{54} + 12\zeta_3 \bigg ) r
- \frac{35}{27}r^2 \ , 
\label{chat3}
\eeq
and
\beqs
\hat c_4 & = & \frac{460151}{576}-\frac{23816}{81}r+\frac{899}{162}r^2
-\frac{83}{81}r^3 \cr\cr
&+& \bigg (\frac{1157}{9}-\frac{889}{3}r+20r^2+\frac{16}{9}r^3 \bigg )\zeta_3
\cr\cr 
&+& r\Big (66-12r \Big )\zeta_4 + \Big (-220+160r \Big )\zeta_5 \ . 
\label{chat4}
\eeqs

\end{appendix}

% =====================================================================

% ========================================================================

\newpage

\begin{table}
  \caption{\footnotesize{
Values of the anomalous dimension $\gamma_{\bar\psi\psi,IR,F}$ calculated to
$O(\Delta_f^p)$, i.e., $\gamma_{\bar\psi\psi,IR,F,\Delta_f^p}$, with
$1 \le p \le 4$, for $G={\rm SU}(N_c)$, as a 
function of $N_c$ and $N_f$ for $2 \le N_c \le 4$ and $N_f$ in the respective
intervals $I_{IRZ}$ for each $N_c$.  For comparison, we also include the
$n$-loop values $\gamma_{\bar\psi\psi,IR,F,n\ell}$ with $2 \le n \le 4$ 
from Table VI of \cite{bvh}. Values that exceed the bound
$\gamma_{\bar\psi\psi,IR} \le 2$ in Eq. (\ref{gamma_upperbound}) are marked as
unphysical (u). For notational brevity in this and successive tables, we 
omit the subscript $\bar\psi\psi$. See text for further details.}}
\begin{center}
\begin{tabular}{|c|c|c|c|c|c|c|c|c|} \hline\hline
$N_c$ & $N_f$ & $\gamma_{IR,F,2\ell}$&$\gamma_{IR,F,3\ell}$ &
$\gamma_{IR,F,4\ell}$ &
$\gamma_{IR,F,\Delta_f}$  &
$\gamma_{IR,F,\Delta_f^2}$ &
$\gamma_{IR,F,\Delta_f^3}$ &
$\gamma_{IR,F,\Delta_f^4}$ 
\\ \hline
2& 6& u & u        & u      & 0.337  & 0.520   & 0.596  & 0.698 \\
2& 7& u & u        & u      & 0.270  & 0.387   & 0.426  & 0.467 \\
2& 8& 0.752& 0.272 & 0.204  & 0.202  & 0.268   & 0.285  & 0.298 \\
2& 9& 0.275& 0.161 & 0.157  & 0.135  & 0.164   & 0.169  & 0.172 \\
2&10& 0.0910& 0.0738& 0.0748& 0.0674 & 0.07475 & 0.07535& 0.0755 \\
\hline
3& 9& u     & u     & u     & 0.374  & 0.587  & 0.687  & 0.804  \\
3&10& u     & u     & u     & 0.324  & 0.484  & 0.549  & 0.615  \\
3&11& 1.61  & 0.439 & 0.250 & 0.274  & 0.389  & 0.428  & 0.462  \\
3&12& 0.773 & 0.312 & 0.253 & 0.224  & 0.301  & 0.323  & 0.338  \\
3&13& 0.404 & 0.220 & 0.210 & 0.174  & 0.221  & 0.231  & 0.237  \\
3&14& 0.212 & 0.146 & 0.147 & 0.125  & 0.148  & 0.152  & 0.153  \\
3&15& 0.0997& 0.0826& 0.0836& 0.0748 & 0.0833 & 0.0841 & 0.0843 \\
3&16& 0.0272& 0.0258& 0.0259& 0.0249 & 0.0259 & 0.0259 & 0.0259 \\
\hline
4&11& u     & u     & u     & 0.424  & 0.694  & 0.844 & 1.029     \\
4&12& u     & u     & u     & 0.386  & 0.609  & 0.721 & 0.8475    \\
4&13& u     & u     & u     & 0.347  & 0.528  & 0.610 & 0.693     \\
4&14& u     & u     & u     & 0.308  & 0.451  & 0.509 & 0.561     \\
4&15& 1.32  & 0.420 & 0.281 & 0.270  & 0.379  & 0.418 & 0.448     \\
4&16& 0.778 & 0.325 & 0.269 & 0.231  & 0.312  & 0.336 & 0.352     \\
4&17& 0.481 & 0.251 & 0.234 & 0.193  & 0.249  & 0.263 & 0.2705    \\
4&18& 0.301 & 0.189 & 0.187 & 0.154  & 0.190  & 0.197 & 0.200     \\
4&19& 0.183 & 0.134 & 0.136 & 0.116  & 0.136  & 0.139 & 0.140     \\
4&20& 0.102 & 0.0854& 0.0865& 0.0771 & 0.0860 & 0.0869 & 0.0871   \\
4&21& 0.0440& 0.0407& 0.0409& 0.0386 & 0.0408 & 0.0409& 0.0409    \\
\hline\hline
\end{tabular}
\end{center}
\label{gamma_values}
\end{table}

% ======================================================================

\begin{table}
  \caption{\footnotesize{Values of the scheme-independent
$\gamma_{IR,F,\Delta_r^p}$ in the LNN limit (\ref{lnn}) 
for $1 \le p \le 4$, together with $\gamma_{IR,F,n\ell}$ with 
$n=2, \ 3, \ 4$ from Table V of \cite{lnn} for
comparison, as a function of $r$ for $r \in I_{IRZ,r}$. Values
that exceed the bound $\gamma_{IR} \le 2$ are marked as unphysical (u) 
or placed in parentheses. We also list the extrapolated estimate 
$\gamma_{IR,F,ex234}$ of $\gamma_{IR,F,\Delta_r^\infty}$ and, 
in the last column, the ratio 
$\gamma_{IR,F,ex234}/\gamma_{IR,F,\Delta_r^4}$.}}
\begin{center}
\begin{tabular}{|c|c|c|c|c|c|c|c|c|c|} \hline\hline
$r$ & $\gamma_{_{IR,F,2\ell}}$ & $\gamma_{_{IR,F,3\ell}}$ & 
$\gamma_{_{IR,F,4\ell}}$
& $\gamma_{IR,F,\Delta_r}$ 
& $\gamma_{IR,F,\Delta_r^2}$ 
& $\gamma_{IR,F,\Delta_r^3}$
& $\gamma_{IR,F,\Delta_r^4}$
& $\gamma_{IR,F,ex234}$
& $\frac{\gamma_{IR,F,ex234}}{\gamma_{IR,F,\Delta_r^4}}$
\\ \hline
2.8& u    & 1.708 & 0.190 & 0.432 & 0.706 & 0.870  & 1.064  &(2.09)&1.96 \\
3.0& u    & 1.165 & 0.225 & 0.400 & 0.635 & 0.765  & 0.908  &1.645 &1.82 \\
3.2& u    & 0.854 & 0.264 & 0.368 & 0.567 & 0.668  & 0.770  &1.28  &1.66 \\
3.4& u    & 0.656 & 0.293 & 0.336 & 0.502 & 0.579  & 0.650  &0.993 &1.53 \\
3.6&1.853 & 0.520 & 0.308 & 0.304 & 0.440 & 0.497   & 0.5445 &0.763 &1.40 \\
3.8&1.178 & 0.420 & 0.306 & 0.272 & 0.381 & 0.422   & 0.452  &0.584 &1.29 \\
4.0&0.785 & 0.341 & 0.288 & 0.240 & 0.325 & 0.353   & 0.371  &0.444 &1.20 \\
4.2&0.537 & 0.277 & 0.257 & 0.208 & 0.272 & 0.290   & 0.300  &0.337 &1.12 \\
4.4&0.371 & 0.222 & 0.217 & 0.176 & 0.2215& 0.233   & 0.238  &0.253 &1.06 \\
4.6&0.254 & 0.1735 & 0.1745 & 0.144&0.1745& 0.1805  & 0.183  &0.188 &1.03 \\
4.8&0.170 & 0.129 & 0.131 & 0.112 & 0.130 & 0.133   & 0.134  &0.135 &1.01 \\
5.0&0.106 & 0.0889& 0.0900& 0.0800& 0.0894& 0.09045 & 0.0907 &0.0905&1.00 \\
5.2&0.0562& 0.0512& 0.0516& 0.0480& 0.0514& 0.0516  & 0.0516 &0.0516&1.00 \\
5.4&0.0168& 0.0164& 0.0164& 0.0160& 0.0164& 0.0164  & 0.0164 &0.0164&1.00 \\
\hline\hline
\end{tabular}
\end{center}
\label{gamma_values_lnn}
\end{table}

% ========================================================================

%
\begin{table}
  \caption{\footnotesize{Values of the anomalous dimension
$\gamma_{IR,adj,\Delta_f^p}$ with 
$1 \le p \le 4$, for $N_f=2$ and $G={\rm SU}(N_c)$ with 
$N_c=2, \ 3$. For comparison, we also list our $n$-loop values, 
$\gamma_{IR,adj,n\ell}$ for this theory from Table VIII of Ref. \cite{bvh}.}}
\begin{center}
\begin{tabular}{|c|c|c|c|c|c|c|c|c|c|} \hline\hline
$N_c$ &  
$\gamma_{IR,adj,2\ell}$  &
$\gamma_{IR,adj,3\ell}$  & 
$\gamma_{IR,adj,4\ell}$  &
$\gamma_{IR,adj,\Delta_f}$ & 
$\gamma_{IR,adj,\Delta_f^2}$ &
$\gamma_{IR,adj,\Delta_f^3}$ &
$\gamma_{IR,adj,\Delta_f^4}$
\\ \hline
2 & 0.820 & 0.543 & 0.500 & 0.333 & 0.465 & 0.511 & 0.556  \\ \hline
3 & 0.820 & 0.543 & 0.523 & 0.333 & 0.465 & 0.516 & 0.553  \\ \hline
\hline\hline
\end{tabular}
\end{center}
\label{gamma_ir_adj_values}
\end{table}

% ========================================================================     

%
\begin{table}
\caption{\footnotesize{Values of the anomalous dimension
$\gamma_{IR,S_2,\Delta_f^p}$ with $1 \le p \le 4$, for 
$G={\rm SU}(N_c)$ with $N_c=3, \ 4$ and 
$N_f = 2, \ 3$ (so $N_f \in I_{IRZ}$). For comparison, 
we also include values of $\gamma_{IR,S_2,n\ell}$ with
$2 \le n \le 4$ for this theory from Table XI in our Ref. \cite{bvh}}.
Values that exceed the upper bound $\gamma_{IR} < 2$ are marked as 
unphysical (u).}
\begin{center}
\begin{tabular}{|c|c|c|c|c|c|c|c|c|} \hline\hline
$N_c$ & $N_f$ & 
$\gamma_{IR,S_2,2\ell}$  &
$\gamma_{IR,S_2,3\ell}$  &
$\gamma_{IR,S_2,4\ell}$  &
$\gamma_{IR,S_2,\Delta_f}$  &
$\gamma_{IR,S_2,\Delta_f^2}$ &
$\gamma_{IR,S_2,\Delta_f^3}$ &
$\gamma_{IR,S_2,\Delta_f^4}$
\\ \hline
3 & 2 & u      & 1.28  & 1.12  & 0.501 & 0.789 & 0.960 & 1.132  \\
3 & 3 & 0.144  & 0.133 & 0.133 & 0.116 & 0.131 & 0.133 & 0.1335  \\
\hline 
4 & 2 & u      & u     & 1.79  & 0.581 & 0.966 & 1.242 & 1.536  \\
4 & 3 & 0.381  & 0.313 & 0.315 & 0.232 & 0.294 & 0.312 & 0.319  \\
\hline\hline
\end{tabular}
\end{center}
\label{gamma_ir_sym_values}
\end{table}

% =======================================================================

%
\begin{table}
\caption{\footnotesize{Values of the anomalous dimension
$\gamma_{IR,A_2,\Delta_f^p}$ calculated to order $1 \le p \le 4$, for 
$G={\rm SU}(4)$ and $N_f \in I_{IRZ}$. For comparison, we also include values 
of $\gamma_{IR,A_2,n\ell}$ with $2 \le n \le 4$ for this theory from 
Table XII in \cite{bvh}. Values that exceed the upper bound $\gamma_{IR} < 2$ 
are marked as unphysical (u).}}
\begin{center}
\begin{tabular}{|c|c|c|c|c|c|c|c|c|} \hline\hline
$N_c$ & $N_f$ & 
$\gamma_{IR,A_2,2\ell}$  &
$\gamma_{IR,A_2,3\ell}$  &
$\gamma_{IR,A_2,4\ell}$  &
$\gamma_{IR,A_2,\Delta_f}$  &
$\gamma_{IR,A_2,\Delta_f^2}$ &
$\gamma_{IR,A_2,\Delta_f^3}$ &
$\gamma_{IR,A_2,\Delta_f^4}$
\\ \hline
4 & 5 & u      & u     & u     & 0.5405& 0.941 & 1.287 & 1.671  \\
4 & 6 & u      & 1.38  & 0.293 & 0.450 & 0.728 & 0.928 & 1.114  \\
4 & 7 & u      & 0.695 & 0.435 & 0.360 & 0.538 & 0.641 & 0.717  \\
4 & 8 & 0.802  & 0.402 & 0.368 & 0.270 & 0.370 & 0.4135& 0.438  \\
4 & 9 & 0.331  & 0.228 & 0.232 & 0.180 & 0.225 & 0.237 & 0.242  \\
4 &10 & 0.117  & 0.101 & 0.103 & 0.0901& 0.101 & 0.103 & 0.103  \\
\hline\hline
\end{tabular}
\end{center}
\label{gamma_ir_asym_values}
\end{table}

% ========================================================================     

%
\begin{table}
\caption{\footnotesize{Values of the anomalous dimension
$\gamma_{IR,T_2,\Delta_f^p}$ for $T_2=S_2$ or $T_2=A_2$, 
calculated to order $1 \le p \le 4$, in the limit $N_c
\to \infty$ with $N_f \in I_{IRZ}$ for this limit, namely 
$3 \le N_f \le 5$.}}
\begin{center}
\begin{tabular}{|c|c|c|c|c|} \hline\hline
$N_f$ & 
$\gamma_{IR,T_2,\Delta_f}$  &
$\gamma_{IR,T_2,\Delta_f^2}$ &
$\gamma_{IR,T_2,\Delta_f^3}$ &
$\gamma_{IR,T_2,\Delta_f^4}$
\\ \hline
3 & 0.5555 & 0.921 & 1.177 & 1.408 \\
4 & 0.333  & 0.465 & 0.520 & 0.550 \\
5 & 0.111  & 0.126 & 0.128 & 0.128 \\
\hline\hline
\end{tabular}
\end{center}
\label{gamma_ir_tensor_ncinf_values}
\end{table}

% ========================================================================     

\begin{table}
  \caption{\footnotesize{Signs of the $d_{j,R}$ coefficients for $2 \le j \le
      5$ 
      for gauge group $G={\rm SU}(N_c)$ and fermion representations $R$ equal
      to $F$ (fundamental), $adj$ (adjoint), $S_2$, and $A_2$ (symmetric and
      antisymmetric rank-2 tensor). Note that $d_1=0$ for all $G$ and $R$.
      In the case $R=A_2$, we restrict to $N_c \ge 3$.}}
\begin{center}
\begin{tabular}{|c|c|c|c|c|} \hline\hline 
$j$ & $d_{j,F}$ & $d_{j,adj}$ & $d_{j,S_2}$ & $d_{j,A_2}$ 
\\ \hline
 2  & $+$  &  $+$  & $+$  & $+$         \\
\hline
 3  & $+$  &  $+$  & $+$  & $+$         \\
\hline
 4  & $-$  &  $+$  & $+$  & $-$ for $N_c=3,4,5$  \\
    &      &       &      & $+$ for $N_c \ge 6$ \\
\hline
 5  & $-$  &  $-$  & $-$  & $-$  \\    
\hline\hline
\end{tabular}
\end{center}
\label{dj_signs}
\end{table}

% ==========================================================================

\begin{table}
\caption{\footnotesize{ Scheme-independent values of
$\beta'_{IR,F,\Delta_f^p}$ with $2 \le p \le 4$ for $G={\rm SU}(2)$, SU(3), 
and SU(4), as functions of $N_f$ in the respective intervals $I_{IRZ}$. 
For comparison, we list the $n$-loop values of $\beta'_{IR,F,n\ell}$ with 
$2 \le n \le 5$, where $\beta'_{IR,F,n\ell}$ with $n=3,\ 4,\ 5$ are computed
in the $\overline{\rm MS}$ scheme. 
The notation $a$e-$n$ means $a \times 10^{-n}$.}}
\begin{center}
\begin{tabular}{|c|c|c|c|c|c|c|c|c|} \hline\hline
$N_c$ & $N_f$ 
    & $\beta'_{IR,F,2\ell}$
    & $\beta'_{IR,F,3\ell,{\overline{\rm MS}}}$
    & $\beta'_{IR,F,4\ell,{\overline{\rm MS}}}$
    & $\beta'_{IR,F,\Delta_f^2}$
    & $\beta'_{IR,F,\Delta_f^3}$
    & $\beta'_{IR,F,\Delta_f^4}$
    & $\beta'_{IR,F,\Delta_f^5}$ 
\\ \hline
 2 & 6 & 6.061  & 1.620 & 0.975 & 0.499   & 0.957   & 0.734  & 0.6515 \\
 2 & 7 & 1.202  & 0.728 & 0.677 & 0.320   & 0.554   & 0.463  & 0.436  \\
 2 & 8 & 0.400  & 0.318 & 0.300 & 0.180   & 0.279   & 0.250  & 0.243  \\
 2 & 9 & 0.126  & 0.115 & 0.110 & 0.0799  & 0.109   & 0.1035 & 0.103  \\
 2 &10 & 0.0245 & 0.0239& 0.0235& 0.0200  & 0.0236  & 0.0233 & 0.0233 \\
 \hline
 3 & 9 & 4.167  & 1.475  & 1.464 & 0.467   & 0.882   & 0.7355  & 0.602 \\
 3 &10 & 1.523  & 0.872  & 0.853 & 0.351   & 0.621   & 0.538   & 0.473 \\
 3 &11 & 0.720  & 0.517  & 0.498 & 0.251   & 0.415   & 0.3725  & 0.344 \\
 3 &12 & 0.360  & 0.2955 & 0.282 & 0.168   & 0.258   & 0.239   & 0.228 \\
 3 &13 & 0.174  & 0.1556 & 0.149 & 0.102   & 0.144   & 0.137   & 0.134 \\
 3 &14 & 0.0737 & 0.0699 & 0.678 & 0.0519  & 0.0673  & 0.0655  & 0.0649 \\
 3 &15 & 0.0227 & 0.0223 & 0.0220& 0.0187  & 0.0220  & 0.0218  & 0.0217 \\
 3 &16 & 2.21e-3& 2.20e-3& 2.20e-3& 2.08e-3 & 2.20e-3 & 2.20e-3& 2.20e-3 \\
\hline
 4 &11 &16.338  & 2.189 & 2.189  & 0.553   & 1.087   & 0.898  & 0.648  \\
 4 &12 & 3.756  & 1.430 & 1.429  & 0.457   & 0.858   & 0.729  & 0.574  \\
 4 &13 & 1.767  & 0.965 & 0.955  & 0.370   & 0.663   & 0.578  & 0.486  \\
 4 &14 & 0.984  & 0.655 & 0.639  & 0.292   & 0.498   & 0.445  & 0.394  \\
 4 &15 & 0.581  & 0.440 & 0.424  & 0.224   & 0.362   & 0.331  & 0.3045 \\
 4 &16 & 0.348  & 0.288 & 0.276  & 0.1645  & 0.251   & 0.234  & 0.222  \\
 4 &17 & 0.204  & 0.180 & 0.1725 & 0.114   & 0.164   & 0.156  & 0.1515 \\
 4 &18 & 0.113  & 0.105 & 0.101  & 0.0731  & 0.0988  & 0.0955 & 0.0939 \\
 4 &19 & 0.0558 & 0.0536& 0.0522 & 0.0411  & 0.0520  & 0.0509 & 0.0505 \\
 4 &20 & 0.0222 & 0.0218& 0.0215 & 0.0183  & 0.0215  & 0.0213 & 0.0212 \\
 4 &21 & 5.01e-3&4.99e-3&4.96e-3 & 4.57e-3 & 4.97e-3 & 4.96e-3& 4.96e-3 \\
\hline\hline
\end{tabular}
\end{center}
\label{betaprime_values}
\end{table}

% ========================================================================

\begin{table}
\caption{\footnotesize{ Scheme-independent values of
$\beta'_{IR,\Delta_r^p}$ for $2 \le p \le 5$
in the LNN limit (\ref{lnn}) as functions of $r=5.5-\Delta_r$. For comparison,
we also list the $n$-loop values $\beta'_{IR,n\ell}$ with $2 \le n \le 5$,
where $\beta'_{IR,n\ell}$ with $n=3, \ 4, \ 5$ are computed in the 
$\overline{\rm MS}$ scheme. The notation $a$e-$n$ means $a \times 10^{-n}$. }}
\begin{center}
\begin{tabular}{|c|c|c|c|c|c|c|c|} \hline\hline
$r$ & $\beta'_{IR,2\ell}$
    & $\beta'_{IR,3\ell}$
    & $\beta'_{IR,4\ell}$
    & $\beta'_{IR,\Delta_r^2}$
    & $\beta'_{IR,\Delta_r^3}$
    & $\beta'_{IR,\Delta_r^4}$
    & $\beta'_{IR,\Delta_r^5}$ 
\\ \hline
 2.8 & 8.100    & 1.918  & 1.913  &  0.518  & 1.004  & 0.851 & 0.583    \\
 3.0 & 3.333    & 1.376  & 1.379  &  0.444  & 0.830  & 0.717 & 0.535    \\
 3.2 & 1.856    & 1.006  & 1.003  &  0.376  & 0.676  & 0.596 & 0.4755   \\
 3.4 & 1.153    & 0.7395 & 0.729&  0.314  & 0.542  & 0.486 & 0.410    \\
 3.6 & 0.752    & 0.542  & 0.527  &  0.257  & 0.426  & 0.388 & 0.342    \\
 3.8 & 0.500    & 0.393  & 0.378  &  0.2055 & 0.327  & 0.303 & 0.276    \\
 4.0 & 0.333    & 0.279  & 0.267  &  0.160  & 0.243  & 0.229 & 0.214    \\
 4.2 & 0.219    & 0.193  & 0.184  &  0.120  & 0.174  & 0.166 & 0.159    \\
 4.4 & 0.139    & 0.128  & 0.122  &  0.0860 & 0.119  & 0.115 & 0.112    \\
 4.6 & 0.0837   & 0.0792 & 0.0766 &  0.0576 & 0.0756 & 0.0737 & 0.0726  \\
 4.8 & 0.0460   & 0.0445 & 0.0435 &  0.0348 & 0.0433 & 0.0426 & 0.0423  \\
 5.0 & 0.0215   & 0.0212 & 0.0208 &  0.0178 & 0.0209 & 0.0207 & 0.0206  \\
 5.2 & 0.714e-2 & 0.710e-2& 0.706e-2&0.640e-2&0.707e-2&0.704e-2& 0.704e-3 \\
 5.4 & 0.737e-3 & 0.736e-3& 0.7356e-3& 0.7111e-3 & 0.7358e-3 & 0.7355e-3 
& 0.7355e-3  \\
\hline\hline
\end{tabular}
\end{center}
\label{betaprime_values_lnn}
\end{table}

% ========================================================================

\begin{table}
\caption{\footnotesize{Values of the IR zero $\xi_{IR,n\ell}$ 
of the $\beta_{\xi,n\ell}$ function in the LNN limit 
for $2 \le n \le 5$ and $r \in I_r$.
Notation u (unphysical) means that there is no physical IR zero 
$\xi_{IR,5\ell}$ of the 5-loop beta function.}}
\begin{center}
\begin{tabular}{|c|c|c|c|c|} \hline\hline
$r$ & $\xi_{IR,2\ell}$ & $\xi_{IR,3\ell}$ & $\xi_{IR,4\ell}$ & 
$\xi_{IR,5\ell}$ 
\\ \hline
 2.8 &  28.274   &  3.573    &  3.323    & u   \\
 3.0 &  12.566   &  2.938    &  2.868    & u   \\
 3.2 &  7.606    &  2.458    &  2.494    & u   \\
 3.4 &  5.174    &  2.076    &  2.168    & u   \\
 3.6 &  3.731    &  1.759    &  1.873    & u   \\
 3.8 &  2.774    &  1.489    &  1.601    & u   \\
 4.0 &  2.095    &  1.252    &  1.349    & u   \\
 4.2 &  1.586    &  1.041    &  1.115    & u   \\
 4.4 &  1.192    &  0.8490   &  0.9003   & 1.0353   \\
 4.6 &  0.8767   &  0.6725   &  0.7038   & 0.7439   \\
 4.8 &  0.6195   &  0.5083   &  0.5244   & 0.5364   \\
 5.0 &  0.4054   &  0.3538   &  0.3603   & 0.3630   \\
 5.2 &  0.2244   &  0.2074   &  0.2089   & 0.2093   \\
 5.4 &  0.06943  &  0.06769  &  0.06775  & 0.06776  \\
\hline\hline
\end{tabular}
\end{center}
\label{xiir_nloop_values}
\end{table}

% =======================================================================

\begin{widetext}
\begin{table}
  \caption{\footnotesize{Values of $\alpha_{IR,\Delta_f^p}$ with 
$1 \le p \le 4$ for $N_c=2, \ 3, \ 4$ and $R=F$, as functions of 
$N_f \in I_{IRZ}$, together with $\alpha_{IR,2\ell}$ and $\overline{\rm MS}$ 
values of $n$-loop $\alpha_{IR,n\ell}$ with $3 \le n \le 4$ from \cite{bvh}, 
for comparison.}}
\begin{center}
\begin{tabular}{|c|c|c|c|c|c|c|c|c|} \hline\hline
$N_c$ & $N_f$ & $\alpha_{IR,2\ell}$ & $\alpha_{IR,3\ell}$ & $\alpha_{IR,4\ell}$
& $\alpha_{IR,\Delta_f}$
& $\alpha_{IR,\Delta_f^2}$
& $\alpha_{IR,\Delta_f^3}$
& $\alpha_{IR,\Delta_f^4}$
\\ \hline
2 & 6  &11.42  & 1.645 & 2.395 & 0.941 & 1.098 & 1.979 & 1.951  \\
2 & 7  & 2.83  & 1.05  & 1.21  & 0.753 & 0.853 & 1.305 & 1.293  \\
2 & 8  & 1.26  & 0.688 & 0.760 & 0.565 & 0.621 & 0.8115& 0.808  \\
2 & 9  & 0.595 & 0.418 & 0.444 & 0.377 & 0.402 & 0.458 & 0.457  \\
2 & 10 & 0.231 & 0.196 & 0.200 & 0.188 & 0.1945& 0.202 & 0.2015 \\
\hline
3 & 9  & 5.24  & 1.028 & 1.072 & 0.587 & 0.712 & 1.19  & 1.26  \\
3 & 10 & 2.21  & 0.764 & 0.815 & 0.509 & 0.603 & 0.913 & 0.952 \\
3 & 11 & 1.23  & 0.578 & 0.626 & 0.431 & 0.498 & 0.686 & 0.706 \\
3 & 12 & 0.754 & 0.435 & 0.470 & 0.352 & 0.397 & 0.500 & 0.509 \\
3 & 13 & 0.468 & 0.317 & 0.337 & 0.274 & 0.301 & 0.350 & 0.353 \\
3 & 14 & 0.278 & 0.215 & 0.224 & 0.196 & 0.210 & 0.227 & 0.228 \\
3 & 15 & 0.143 & 0.123 & 0.126 & 0.117 & 0.122 & 0.126 & 0.126 \\
3 & 16 & 0.0416& 0.0397& 0.0398& 0.0391 & 0.0397 & 0.0398 & 0.0398 \\
\hline
4  & 11  & 14.00   & 0.972  & 0.943 & 0.474 & 0.592 & 1.042 & 1.1475 \\
4  & 12  &  3.54   & 0.754  & 0.759 & 0.431 & 0.528 & 0.867 & 0.939  \\
4  & 13  &  1.85   & 0.6035 & 0.628 & 0.388 & 0.467 & 0.713 & 0.7605 \\
4  & 14  &  1.16   & 0.489  & 0.521 & 0.345 & 0.407 & 0.580 & 0.610 \\
4  & 15  &  0.783  & 0.397  & 0.428 & 0.3015& 0.349 & 0.465 & 0.483 \\
4  & 16  &  0.546  & 0.320  & 0.345 & 0.258 & 0.294 & 0.367 & 0.376 \\
4  & 17  &  0.384  & 0.254  & 0.271 & 0.215 & 0.240 & 0.282 & 0.2865\\
4  & 18  &  0.266  & 0.194  & 0.205 & 0.172 & 0.188 & 0.210 & 0.211 \\
4  & 19  &  0.175  & 0.140  & 0.145 & 0.129 & 0.138 & 0.147 & 0.148 \\
4  & 20  &  0.105  & 0.091  & 0.092 & 0.0861& 0.09005&0.0928& 0.0929 \\
4  & 21  &  0.0472 & 0.044  & 0.044 & 0.0431& 0.04405&0.0444& 0.0444 \\
\hline\hline
\end{tabular}
\end{center}
\label{alfir_nloop_Delta}
\end{table}
\end{widetext}


\begin{thebibliography}{99}

% 1 
\bibitem{rg}
Some early discussions of the renormalization group in quantum field theory 
include
E. C. G. Stueckelberg and A. Peterman, Helv. Phys. Acta {\bf 26}, 499 (1953);
M. Gell-Mann and F. Low, Phys. Rev. {\bf 95}, 1300 (1954);
N. N. Bogolubov and D. V. Shirkov, Doklad. Akad. Nauk SSSR {\bf 103}, 391
(1955); C. G. Callan, Phys. Rev. D {\bf 2}, 1541 (1970);
K. Symanzik, Commun. Math. Phys. {\bf 18}, 227 (1970); K. Wilson,
Phys. Rev. D {\bf 3}, 1818 (1971).

% 2 
\bibitem{scalecon}
Some early analyses of connections between scale and
  conformal invariance include A. Salam, Ann. Phys. (NY) {\bf 53}, 174 (1969);
  A. M. Polyakov, JETP Lett. {\bf 12}, 381 (1970); 
  D. J. Gross and J. Wess, Phys. Rev. D {\bf 2}, 753 (1970); C. G. Callan,
  S. Coleman, and R. Jackiw, Ann. Phys. (NY) {\bf 59}, 42 (1970).  More recent
  works include J. Polchinski, Nucl. Phys. B {\bf 303}, 226 (1988);
  J.-F. Fortin, B. Grinstein and A. Stergiou, JHEP 01 (2013) 184 (2013);
  A. Dymarsky, Z. Komargodski, A. Schwimmer, and S. Thiessen, JHEP {\bf 10},
  171 (2015) and references therein.

% 3 
\bibitem{fm}
Note that our assumption that the fermions are massless does not produce any
loss of generality, since if a fermion had a nonzero mass $m_0$, it would be
integrated out of the low-energy effective field theory at scales $\mu <
m_0$, and hence would not affect the IR limit $\mu \to 0$ under consideration
here. 

% 4 
\bibitem{nacp_comment}
Fully nonperturbative evidence for the non-Abelian Coulomb phase comes from
lattice simulations, as discussed below.  Furthermore, in the case where 
$G={\rm SU}(N_c)$ and the fermions are in the fundamental representation, 
one can take the limit in Eq. (\ref{lnn}), namely $N_c \to \infty$ and 
$N_f \to \infty$ with $r=N_f/N_c$ fixed and finite.  In this case, 
$\alpha_{IR}$ can be made arbitrarily small so that strength of the gauge
coupling at the IR fixed point approaches arbitrarily close to zero.

% 5
\bibitem{gross75}
D. J. Gross, in R. Balian and J. Zinn-Justin, eds. 
{\it Methods in Field Theory}, Les Houches 1975 
(North Holland, Amsterdam, 1976), p. 141.

% 6 
\bibitem{casimir}
%
The Casimir invariants $C_2(R)$ and $T(R)$ are defined as
$\sum_a \sum_j {\cal D}_R(T_a)_{ij} {\cal D}_R(T_a)_{jk}\equiv 
C_2(R)\delta_{ik}$ and
$\sum_{i,j} {\cal D}_R(T_a)_{ij} {\cal D}_R(T_b)_{ji} \equiv T(R) \delta_{ab}$,
where $R$ is the
representation, $T_a$ are the generators of $G$, normalized according to 
${\rm Tr}(T_aT_b)=(1/2)\delta_{ab}$ and ${\cal D}_R$ is the matrix
representation ({\it Darstellung}) of $R$.  For the adjoint representation, we
denote $C_2(adj) \equiv C_A$, and for fermions transforming according
to the representation $R$, we denote $C_2(R) \equiv C_f$ and
$T(R) \equiv T_f$. 

% 7
\bibitem{b1}
D. J. Gross and F. Wilczek, Phys. Rev. Lett. {\bf 30}, 1343 (1973);
H. D. Politzer, Phys. Rev. Lett. {\bf 30}, 1346 (1973); G. 't Hooft,
unpublished.

% 8
\bibitem{b2}
W. E. Caswell, Phys. Rev. Lett. {\bf 33}, 244 (1974);
D. R. T. Jones, Nucl. Phys. B {\bf 75}, 531 (1974).

% 9 
\bibitem{gw2}
D. J. Gross and F. Wilczek, Phys. Rev. D {\bf 8}, 3633 (1973). 

% 10
\bibitem{bz}
T. Banks and A. Zaks, Nucl. Phys. B {\bf 196}, 189 (1982).

% 11
\bibitem{grunberg92}
G. Grunberg, Phys. Rev. D {\bf 46}, 2228 (1992). 

% 12
\bibitem{gtr}
T. A. Ryttov, Phys. Rev. Lett. {\bf 117}, 071601 (2016) [arXiv:1604.00687].
Our $b_\ell = 2^\ell \beta_{\ell-1}$ in this paper. 

% 13 
\bibitem{dex}
T. A. Ryttov and R. Shrock, Phys. Rev. D {\bf 94}, 125005 (2016)
[arXiv:1610.00387].

% 14
\bibitem{gsi}
T. A. Ryttov and R. Shrock, Phys. Rev. D {\bf 94}, 105014 (2016)
[arXiv:1608.00068].

% 15
\bibitem{flir}
T. A. Ryttov and R. Shrock, Phys. Rev. D {\bf 94}, 105015 (2016)
[arXiv:1607.06866]. 

% 16
\bibitem{dexs}
T. A. Ryttov  and R. Shrock, arXiv:1701.06083. 

% 17 
\bibitem{b5}
F. Herzog, B. Ruijl, T. Ueda, J. A. M. Vermaseren, and A. Vogt,
JHEP 02 (2017) 090 [arXiv:1701.01404]. 
Our $b_\ell = \beta_{\ell-1}$ in this paper. 

% 18
\bibitem{b5su3}
P. A. Baikov, K. G. Chetyrkin, and J. H. K\"uhn, 
Phys. Rev. Lett. {\bf 118}, 082002 (2017) [arXiv:1606.08659].  
Our $b_\ell = 4^\ell \beta_{\ell-1}$ in this paper. 

% 19
\bibitem{bvh} 
T. A. Ryttov, R. Shrock, Phys. Rev. D {\bf 83}, 056011 (2011) 
[arXiv:1011.4542]. 

% 20
\bibitem{bc}
R. Shrock, Phys. Rev. D {\bf 87}, 105005 (2013).

% 21
\bibitem{lnn}
R. Shrock, Phys. Rev. D {\bf 87}, 116007 (2013).

% 22
\bibitem{lgtreviews}
See, e.g., talks in the
CP3 Workshop at http://cp3-origins.dk/events/meetings/mass2013;
Lattice-2014 at https://www.bnl.gov/lattice2014;
SCGT15 at http://www.kmi.nagoya-u.ac.jp/workshop/SCGT15; 
Lattice-2015 at http://www.aics.riken.jp/sympo/lattice2015;
Lattice-2016 at http://www.southampton.ac.uk/lattice2016; see also
T. Degrand, Rev. Mod. Phys. {\bf 88}, 015001 (2016).

% 23
\bibitem{cft_bootstrap}
Some recent reviews include S. Rychkov, 
{\it EPFL Lectures on Conformal Field Theory in $D \ge 3$ Dimensions} 
[arXiv:1601.05000];
D. Simmons-Duffin, 2015 TASI Lectures on the Conformal Bootstrap
[arXiv:1602.07982], and D. Poland, Nature Phys. {\bf 12}, 535 (2016).

% 24
\bibitem{wtc}
B. Holdom, Phys. Lett. B {\bf 150}, 301 (1985);
K. Yamawaki, M. Bando, and K. Matumoto, Phys. Rev. Lett. {\bf 56}, 1335
(1986); T. Appelquist, D. Karabali, and
L. C. R. Wijewardhana, Phys. Rev. Lett. {\bf 57}, 957 (1986).

% 25
\bibitem{ps}
C. Pica, F. Sannino, Phys. Rev. D {\bf 83}, 035013 (2011) [arXiv:1011.5917].

% 26
\bibitem{gammaconvention}
%
Some authors use the opposite sign convention for the anomalous dimension,
writing $D_{\cal O} = D_{\cal O,{\rm free}} + \gamma_{\cal O}$. Our sign
convention is the same as the one used in \cite{wtc} and 
the lattice gauge theory literature.

% 27
\bibitem{b3}
O. V. Tarasov, A. A. Vladimirov, and A. Yu. Zharkov, Phys. Lett. B {\bf 93},
429 (1980); S. A. Larin and J. A. M. Vermaseren, Phys. Lett. B {\bf 303}, 334
(1993).

% 28
\bibitem{b4}
T. van Ritbergen, J. A. M. Vermaseren, and S. A. Larin, Phys. Lett. B 
{\bf 400}, 379 (1997). 

% 29
\bibitem{b4p}
M. Czakon, Nucl. Phys. B {\bf 710}, 485 (2005). 

% 30
\bibitem{msbar}
W. A. Bardeen et al., Phys. Rev. D {\bf 18}, 3998 (1978).

% 31
\bibitem{c4}
K. G. Chetyrkin, Phys. Lett. B {\bf 404}, 161 (1997);
J. A. M. Vermaseren, S. A. Larin, and T. van Ritbergen, Phys. Lett. B 
{\bf 405}, 327 (1997). 

% 32
\bibitem{nfintegral}
%
Here and below, if an expression for $N_f$
formally evaluates to a non-integral real value, it is understood
implicitly that one infers an appropriate integral value from it.

% 33
\bibitem{phasenote}
%
$N_u$ and $N_\ell$ were denoted $N_{f,b1z}$ and $N_{f,b2z}$ in \cite{bvh} and
some of our subsequent works.  In principle, for an appropriate $G$ and $R$, 
between the confining phase with spontaneous chiral symmetry breaking at 
small $N_f$ and the (deconfined) non-Abelian Coulomb phase at larger $N_f \lsim
N_u$, there could be an intermediate phase with confinement but no chiral
symmetry breaking, provided that the 't Hooft anomaly-matching conditions are
satisfied (a necessary but not sufficient condition for such a phase). 

% 34
\bibitem{gkgg}
E. Gardi and M. Karliner, Nucl. Phys. B {\bf 529}, 383 (1998);
E. Gardi and G. Grunberg, JHEP 03, 024 (1999).

% 35
\bibitem{elias}
F. A. Chishtie, V. Elias, V. A. Miransky, and T. G. Steele, 
Prog. Theor. Phys. {\bf 104}, 603 (2000). 

% 36
\bibitem{bfs}
T. A. Ryttov and R. Shrock, Phys. Rev. D {\bf 85}, 076009 (2012);

% 37
\bibitem{bfss}
R. Shrock, Phys. Rev. D {\bf 91}, 125039 (2015); G. Choi and R. Shrock,  
Phys. Rev. D {\bf 93}, 065013 (2016).

% 38
\bibitem{sch}
T. A. Ryttov and R. Shrock, Phys. Rev. D {\bf 86}, 065032 (2012);
T. A. Ryttov and R. Shrock, Phys. Rev. D {\bf 86}, 085005 (2012).

% 39
\bibitem{sch2}
R. Shrock, Phys. Rev. D {\bf 88}, 036003 (2013); 
R. Shrock, Phys. Rev. D {\bf 90}, 045011 (2014);
R. Shrock, Phys. Rev. D {\bf 91}, 125039 (2015);
G. Choi and R. Shrock, Phys. Rev. D {\bf 90}, 125029 (2014);
G. Choi and R. Shrock, Phys. Rev. D {\bf 94}, 065038 (2016). 

% 40
\bibitem{tr}
T. A. Ryttov, Phys. Rev. D {\bf 89}, 016013 (2014);
T. A. Ryttov, Phys. Rev. D {\bf 89}, 056001 (2014);
T. A. Ryttov, Phys. Rev. D {\bf 90}, 056007 (2014).

% 41
\bibitem{gracey2015}
J. A. Gracey and R. M. Simms, Phys. Rev. D {\bf 91}, 085037 (2015). 

% 42
\bibitem{gracey_gammatensor}
J. A. Gracey, Phys. Lett. B {\bf 488}, 175 (2000).

% 43
\bibitem{c5su3}
P. A. Baikov, K. G. Chetyrkin, and J. H. K\"uhn, JHEP 10, 076 (2014) 
[arXiv:1402.6611]. Our $c_\ell = 2^{1+2\ell}\gamma_{\ell-1}$ in this 
paper.

% 44
\bibitem{gammabound}
G. Mack, Commun. Math. Phys. {\bf 55}, 1 (1977);
B. Grinstein, K. Intriligator, and I. Rothstein, Phys. Lett. B {\bf 662}, 367
(2008); Y. Nakayama, Phys. Repts. {\bf 569}, 1 (2015).

% 45
\bibitem{nsvz}
V. A. Novikov, M. A. Shifman, A. I. Vainshtein, and V. I. Zakharov (NSVZ),
Nucl. Phys. B {\bf 229}, 381, 407 (1983).

% 46
\bibitem{seiberg}
N. Seiberg, Phys. Rev. D {\bf 49}, 6857 (1994).

% 47
\bibitem{othergroups}
F. Sannino, Phys. Rev. D {\bf 79}, 096007 (2009); 
M. Mojaza, C. Pica, T. A. Ryttov, and F. Sannino, Phys. Rev. D {\bf 86}, 
076012 (2012). 

% --------------------------- SU(3), R=F, N_f=12 -------------------------

% 47a
\bibitem{lsd}
T. Appelquist et al. (LSD Collab.), Phys. Rev. D {\bf 84}, 054501 (2011).

% 48
\bibitem{degrand}
T. DeGrand, Phys. Rev. D {\bf 84}, 116901 (2011).

% 49
\bibitem{latkmi}
Y. Aoki et al. (LatKMI Collab.), Phys. Rev. D {\bf 86}, 054506 (2012).

% 50
\bibitem{ah1}
A. Hasenfratz, A. Cheng, G. Petropoulos, and D. Schaich, in
{\it Proc. Lattice-2012} [arXiv:1207.7162].

% 51
\bibitem{ah2}
A. Hasenfratz, A. Cheng, G. Petropoulos, and D. Schaich, in 
{\it Proc. Lattice-2013} [arXiv:1310.1124].

% 52
\bibitem{ah3}
A. Hasenfratz and D. Schaich, arXiv:1610.10004 and private communications. 

% 53
\bibitem{lmnp}
M. P. Lombardo, K. Miura, T. J. Nunes da Silva, and E. Pallante,
JHEP 12, 183 (2014). See also 
A. Deuzeman, M. P. Lombardo, T. Nunes Da Silva, and E. Pallante, 
Phys. Lett. B {\bf 720}, 358 (2013); 
K. Miura, M. P. Lombardo, and E. Pallante, 
Phys. Lett. B {\bf 710}, 676 (2012). 

% 54
\bibitem{kuti}
Z. Fodor, K. Holland, J. Kuti, D. Nogradi, C. Schroeder, and C.H. Wong, in
{\it Proc. Lattice-2012} [arXiv:1211.4238]; Z. Fodor, K. Holland, J. Kuti,
S. Mondal, D. Nogradi, and C. H. Wong, Phys. Rev. D {\bf 94} 091501 (2016). 

% ------------------------- SU(3), Nf=10 ---------------------------------

% 55
\bibitem{lsdnf10}
T. Appelquist et al. (LSD Collab.), arXiv:1204.6000.

% ------------------------ SU(3), Nf=8 --------------------------------

% 56
\bibitem{latkminf8}
Y. Aoki et al. (LatKMI Collab.), Phys. Rev. D {\bf 87}, 094511 (2013);
Y. Aoki et al. (LatKMI Collab.), Phys. Rev. D {\bf 89}, 111502 (2014);
Y. Aoki et al. (LatKMI Collab.), Nuclear and Particle Physics Proc. {\bf
270-272}, 242 (2016). 

% 57
\bibitem{lsdnf8}
T. Appelquist et al. (LSD Collab.), Phys. Rev. D {\bf 90}, 114502 (2014);
T. Appelquist et al. (LSD Collab.), Phys. Rev. D {\bf 93}, 114514 (2016) 
and references therein. 

% 58
\bibitem{appelquist_eft}
T. Appelquist, J. Ingoldby, and M. Piai, arXiv:1702.04410.  

% 59
\bibitem{sannino_eft} 
M. Hansen, K. Lang\ae ble, and F. Sannino, Phys. Rev. D {\bf 95},
036005 (2017). 

\bibitem{fleming_eft}
A. D. Gasbarro and G. T. Fleming, arXiv:1702.00480. 

% --------------------------- SU(2), R=F, Nf=8 -------------------------

% 60
\bibitem{ckm}
T. Appelquist and R. Shrock, Phys. Lett. B {\bf 548}, 204 (2002);
T. Appelquist and R. Shrock, Phys. Rev. Lett. {\bf 90}, 201801 (2003);
T. Appelquist, M. Piai, and R. Shrock, Phys. Rev. D {\bf 69}, 015002 (2004);
N. C. Christensen and R. Shrock, Phys. Rev. Lett. {\bf 94}, 241801 (2005) and
references therein. 

% 61
\bibitem{su2nf8}
H. Ohki et al., PoS Lattice2010, 066, arXiv:1011.0373;
C. Y.-H. Huang et al., PoS Lattice2015, arXiv:1511.01968.

% 62
\bibitem{tuominen2017}
V. Leino et al., arXiv:1701.04666.

% ---------------------------- SU(2), R=F, Nf=6 --------------------------

% 63
\bibitem{su2nf6}
F. Bursa et al., Phys. Lett. {\bf B696}, 374 (2011);
T. Karavirta et al., JHEP 1205 (2012) 003;
M. Tomii et al., arXiv:1311.0099.
T. Appelquist et al. (LSD Collab.), Phys. Rev. Lett. {\bf 112}, 111601 (2014).

% 64
\bibitem{yamada_su2nf6}
M. Hayakawa, K.-I. Ishikawa, S. Takeda, and N. Yamada, 
Phys. Rev. D {\bf 88}, 094504, 094506 (2013);

% 65
\bibitem{tuominen_su2nf6}
J. M. Suorsa et al., arXiv:1611.02022; see also 
V. Leino et al., arXiv:1610.09989. 

% -----------------------------------------------------------------------

% 66
\bibitem{stevenson2016}
P. M. Stevenson, Mod. Phys. Lett. A {\bf 31}, 1650226 (2016).

% ----------------------- SU(2), Nf=2 adjoint ----------------------

% 67
\bibitem{sannino_su2adj}
D. D. Dietrich and F. Sannino, Phys. Rev. D {\bf 75}, 085018 (2007);
S. Catterall and F. Sannino, Phys. Rev. D {\bf 76}, 034504 (2007).

% 68
\bibitem{catterall2010}
S. Catterall, L. Del Debbio, J. Giedt, and L. Keegan,
PoS Lattice2010, 057 (2010) [arXiv:1010.5909].

% 69
\bibitem{deldebbio2010}
L. Del Debbio, B. Lucini, A. Patella, C. Pica, and A. Rago,
Phys. Rev. D {\bf 82}, 014510 (2010).

% 70
\bibitem{degrand2011}
T. DeGrand, Y. Shamir, and B. Svetitsky, Phys. Rev. D {\bf 83}, 074507 (2011).

% 71
\bibitem{lsd2011}
T. Appelquist et al. (LSD Collab.), Phys. Rev. D {\bf 84}, 054501 (2011).

% 72
\bibitem{deldebbio2016}
L. Del Debbio, B. Lucini, A. Patella, C. Pica, and A. Rago,
Phys. Rev. D {\bf 93}, 054505 (2016). 

% 73
\bibitem{tuominen2016}
J. Rantaharju, T. Rantalaiho, K. Rummukainen, and K. Tuominen, Phys. Rev.
D {\bf 93}, 094509 (2016).

% 74
\bibitem{giedt2016}
J. Giedt, Int. J. Mod. Phys. A {\bf 31}, 1630011 (2016).

% 75
\bibitem{montvay}
G. Bergner, P. Giudice, G. M\"unster, I. Montvay, and S. Piemonte,
arXiv:1610.01576. 

% ---------------------- SU(3), R=sym2 ------------------------------

% 76
\bibitem{degrand_sextet}
T. DeGrand, Y. Shamir, and B. Svetitsky, Phys. Rev. D {\bf 87}, 074507 (2013).

% 77
\bibitem{kuti_sextet}
Z. Fodor, K. Holland, J. Kuti, D. Nogradi, C. Schroeder, and C. H. Wong,
Phys. Lett. B {\bf 718}, 657 (2012).

% 78
\bibitem{kogut_sinclair_sextet}
J. B. Kogut and D. K. Sinclair, Phys. Rev. D {\bf 81}, 114507 (2010);
Phys. Rev. D {\bf 92}, 054508 (2015).

% ------------------------- betaprime ----------------------------------

% 79
\bibitem{traceanomaly}
S. L. Adler, J. C. Collins, and A. Duncan, Phys. Rev. D {\bf 15}, 1712 (1977);
J. C. Collins, A. Duncan, and S. Joglekar, Phys. Rev. D {\bf 16}, 438
(1977); N. K. Nielsen, Nucl. Phys. B {\bf 120}, 212 (1977); see also
H. Kluberg-Stern and J.-B. Zuber, Phys. Rev. D {\bf 12}, 467 (1975).

% 80
\bibitem{tracerel}
See, e.g., S. S. Gubser, A. Nellore, S. S. Pufu, and E. D. Rocha,
Phys. Rev. Lett. {\bf 101}, 131601 (2008); see also M. Kurachi, S. Matsuzaki, 
and K. Yamawaki, Phys. Rev. D {\bf 90}, 055028 (2014); 
R. J. Crewther and L. C. Tunstall, Phys. Rev. D {\bf 91}, 034016 (2015). See
also T. Nunes da Silva, E. Pallante, and L. Robroek, arXiv:1609.06298, 
arXiv:1506.06396.

% 81
\bibitem{hasenfratz_betaprime}
A. Hasenfratz and D. Schaich, arXiv:1610.10004. 

\end{thebibliography}
\end{document}